\documentclass{ieeeaccess}
\usepackage{cite}
\usepackage{amsmath,amssymb,amsfonts}
\usepackage{pifont}
\newcommand{\cmark}{\ding{51}}%
\newcommand{\xmark}{\ding{55}}%
\usepackage{algorithmic}
\usepackage{graphicx}
\usepackage{textcomp}

\usepackage{bm}
\makeatletter
\AtBeginDocument{\DeclareMathVersion{bold}
\SetSymbolFont{operators}{bold}{T1}{times}{b}{n}
\SetSymbolFont{NewLetters}{bold}{T1}{times}{b}{it}
\SetMathAlphabet{\mathrm}{bold}{T1}{times}{b}{n}
\SetMathAlphabet{\mathit}{bold}{T1}{times}{b}{it}
\SetMathAlphabet{\mathbf}{bold}{T1}{times}{b}{n}
\SetMathAlphabet{\mathtt}{bold}{OT1}{pcr}{b}{n}
\SetSymbolFont{symbols}{bold}{OMS}{cmsy}{b}{n}
\renewcommand\boldmath{\@nomath\boldmath\mathversion{bold}}}
\makeatother

\def\BibTeX{{\rm B\kern-.05em{\sc i\kern-.025em b}\kern-.08em
    T\kern-.1667em\lower.7ex\hbox{E}\kern-.125emX}}

\usepackage{amsthm}
\newtheorem{definition}{Definition}

\usepackage{tabularx}
\usepackage{booktabs}
\usepackage{makecell}

\usepackage{microtype}
\usepackage{colortbl}
\definecolor{lightgray}{gray}{0.9} 
\definecolor{lightaccessblue}{cmyk}{0.10, 0.05, 0, 0} 
\newcommand{\contentfont}{\fontfamily{phv}\selectfont} 

\usepackage{url}

\usepackage[hidelinks]{hyperref}

\begin{document}
\history{Date of publication xxxx 00, 0000, date of current version xxxx 00, 0000.}
\doi{10.1109/ACCESS.2024.0429000}

\title{High-Performance Serverless Computing: A Systematic Literature Review on Serverless for HPC, AI, and Big Data}

\author{
\uppercase{Valerio Besozzi}\authorrefmark{1}\authorrefmark{2}, 
\uppercase{Matteo Della Bartola}\authorrefmark{1}, 
\uppercase{Patrizio Dazzi}\authorrefmark{1}, and 
\uppercase{Marco Danelutto}\authorrefmark{1}
}

\address[1]{Department of Computer Science, University of Pisa, Pisa, 56127, Italy}
\address[2]{ISTI – National Research Council, Pisa, 56124, Italy}

\tfootnote{This work has been partially funded by the NOUS (A catalyst for EuropeaN ClOUd Services in the era of data spaces, high-performance and edge computing) HORIZON-CL4-2023-DATA-01-02 project, G.A. n. 101135927.}

\corresp{Corresponding author: Valerio Besozzi (e-mail: valerio.besozzi@phd.unipi.it).}

\markboth
{Besozzi and Della Bartola \headeretal: High-Performance Serverless Computing}
{Besozzi and Della Bartola \headeretal: High-Performance Serverless Computing}

\corresp{Corresponding author: Valerio Besozzi (e-mail: \url{valerio.besozzi@phd.unipi.it}).}

\begin{abstract}
The widespread deployment of large-scale, compute-intensive applications such as high-performance computing, artificial intelligence, and big data is leading to convergence between cloud and high-performance computing infrastructures. Cloud providers are increasingly integrating high-performance computing capabilities in their infrastructures, such as hardware accelerators and high-speed interconnects, while researchers in the high-performance computing community are starting to explore cloud-native paradigms to improve scalability, elasticity, and resource utilization. In this context, serverless computing emerges as a promising execution model to efficiently handle highly dynamic, parallel, and distributed workloads. This paper presents a comprehensive systematic literature review of 122 research articles published between 2018 and early 2025, exploring the use of the serverless paradigm to develop, deploy, and orchestrate compute-intensive applications across cloud, high-performance computing, and hybrid environments.
%
From these, a taxonomy 
comprising eight primary research directions and nine targeted use case domains is 
proposed, alongside an analysis of recent publication trends and collaboration networks among authors, highlighting the growing interest and interconnections 
within this emerging research field.
Overall, this work aims to offer a valuable foundation for both new researchers and experienced practitioners, guiding the development of next-generation serverless solutions for parallel compute-intensive~applications.
\end{abstract}


\titlepgskip=-21pt

\maketitle

\section{Introduction}
\label{sec:introduction}
\IEEEPARstart{H}{igh-performance} computing infrastructures are increasingly being used for computationally intensive computations across a wide range of diverse domains. However, a major problem in current high-performance computing systems is that they are characterized by rigid resource allocation systems and programming models that fail to adapt to dynamic and bursty workloads, leading to inefficient resource utilization and prolonged job queuing times~\cite{copik_software_2024}.

Historically, high-performance computing infrastructures have been distinguished from general IT and cloud infrastructures due to the specialized software, hardware characteristics, and their typical on-premises nature~\cite{buyya_high_1999,ALI201387}. High-performance computing systems are specifically designed to support workloads that require massive parallelism, accelerator support, high-throughput communication, and low-latency data access. In contrast, cloud infrastructures have been developed with the objective of providing elastic, scalable, on-demand, and affordable access to resources~\cite{BUYYA20133}.

Nevertheless, an intersection is emerging, driven by the proliferation of applications such as big data, artificial intelligence, machine learning, and scientific simulations, which increasingly blur the lines between high-performance computing and cloud computing infrastructures. 
Consequently, modern cloud infrastructures have been increasingly converging towards high-performance computing systems, mirroring their performance capabilities and hardware characteristics. In that sense, the recently introduced Microsoft Azure Eagle cloud supercomputer is a notable example of this trend\footnote{It was also recently included in the TOP500 supercomputer list. More information at: \url{https://www.top500.org/system/180236/}.}. In addition, cloud providers are introducing HPC-tailored solutions into their offerings. This is exemplified by AWS Parallel Computing Service~\cite{aws_pcs}, a fully managed service that simplifies the deployment and scaling of HPC workloads in their cloud infrastructure.
Modern cloud infrastructures have been starting to rely on specialized hardware, such as GPUs and FPGAs, and leverage high-speed networking to support compute-intensive tasks that were previously the exclusive domain of high-performance computing environments.

This convergence is unlikely to remain a mere one-way process. As cloud computing infrastructures evolve by incorporating characteristics and capabilities increasingly resembling those traditionally associated with high-performance computing systems, the reverse trend is also likely to emerge. 
Namely, to meet the growing demand for computational capacity tailored to support compute-intensive applications while simultaneously improving flexibility and resource efficiency, high-performance computing systems can adopt strategies and approaches originating from cloud computing, particularly in terms of scalability and dynamic resource management.
Novel cloud computing paradigms, such as serverless computing, have the potential to make high-performance computing systems more efficient at handling dynamic and bursty parallel and distributed workloads, thereby reducing idle time, optimizing overall performance, and improving server consolidation.

In this regard, a new research area is emerging at the intersection of these two paradigms. We refer to it as \textit{High-Performance Serverless Computing}, which seeks to adapt the serverless execution model to support HPC, AI, and big data workloads in a more elastic and efficient way. This involves rethinking aspects such as resource provisioning and management, accelerator integration, and support for heterogeneous and disaggregated computing resources. Recent efforts highlight this trend. For instance, Meta's hyperscale infrastructure~\cite{10.1145/3701296} demonstrates how serverless architectures can effectively support both general-purpose and high-performance workloads. The rise in popularity of serverless computing over the past seven years makes it a promising area for research.

In this paper, we present a comprehensive and systematic review of the literature on research at the intersection of serverless and high-performance computing. To this end, we analyze a total of 122 studies, providing a snapshot of the current state of research in this emerging domain. We hereby propose a taxonomy to classify existing research directions in high-performance serverless computing, with the aim of identifying both well-explored areas and current research gaps in the field. In addition, we examine major use cases, investigate the publication landscape, and analyze emerging research trends to demonstrate the increasing attention this topic is receiving. Finally, we conclude by outlining key insights and opportunities for future research and development in high-performance serverless computing.

The remainder of this paper is structured as follows. Section~\ref{sec:related_work} provides an overview of related work, focusing on secondary studies in the serverless
computing literature. Section~\ref{sec:background} introduces background concepts related to serverless computing. Section~\ref{sec:methodology} presents the review protocol and methodology adopted, including the process used to collect the selected research articles. Section~\ref{sec:results} reports the results of our analysis of current high-performance serverless computing literature. Specifically, Section~\ref{subsec:taxonomy} presents the taxonomy developed to classify research directions; Section~\ref{subsec:research_directions} discusses how current research
addresses these directions; Section~\ref{subsec:use_cases}
analyzes the use cases and application domains targeted by the selected works; Section~\ref{subsec:trends} examines research trends, influential contributions, and publication venues; and Section~\ref{subsec:bibliometric} provides a bibliometric
analysis of collaboration patterns and influential authors within the high-performance serverless computing research community. Finally, Section~\ref{sec:discussion} discusses our findings and presents open problems and research gaps, while Section~\ref{sec:conclusion}
provides concluding remarks.

\section{Related Work}
\label{sec:related_work}
Since the introduction of serverless computing, numerous systematic and non-systematic literature reviews have addressed various aspects of this emerging paradigm~\cite{9756233,10.1145/3510611, hassan_survey_2021, 10.1145/3579643, 10895318,10.1145/3700875,ghorbian_survey_2024}. Among these, Wen et al.~\cite{10.1145/3579643} presented a comprehensive review and proposed a taxonomy of research directions within the broader serverless computing landscape. Similarly, Hassan et al.~\cite{hassan_survey_2021} and Shafiei et al.~\cite{10.1145/3510611} provided extensive overviews of trends, use cases and open challenges in serverless computing, focusing primarily on typical cloud workloads and event-driven applications.
Other surveys have focused on specific subdomains within the field. For instance, Alam et al.~\cite{10895318} concentrated on serverless architectures, Golec et al.~\cite{10.1145/3700875} reviewed the literature on cold start mitigation techniques and proposed a taxonomy, and Ghorbian et al.~\cite{ghorbian_survey_2024} surveyed scheduling mechanisms across various computing environments. Although these studies offer valuable insights into specific aspects and challenges, they do not explore the intersection between serverless computing and high-performance~workloads.

In contrast, our work, while adopting the same methodological rigor as the aforementioned surveys, narrows the focus to systematically examine how serverless computing has been employed to support HPC, AI, and big data applications. 
To the best of our knowledge, no prior literature review has investigated this intersection. Hence, our goal is to address this gap by surveying and classifying research efforts that target compute-intensive applications on serverless platforms, thereby providing a current and focused overview of the state of the art in this emerging field.

\section{Background}
\label{sec:background}
Serverless computing is a cloud execution model that promotes full abstraction of the underlying computing infrastructure. It enables developers to focus solely on implementing business logic, without having to address operational logic. This section explores the fundamentals of serverless, including key definitions, characteristics, and service models. 

\subsection{Definition and Characteristics of Serverless Computing}
\label{subsec:characteristics_of_serverless}
First popularized by Amazon in 2014 with AWS Lambda,\footnote{Although the foundational ideas of serverless computing can be traced back to 2008 with the introduction of Google App Engine.} serverless computing is a cloud computing execution model that allows users to deploy and execute \textit{fine-grained billed} and \textit{automatically scaled} applications, without having to address the underlying operational logic. 
Within the cloud computing reference model, serverless computing is often positioned alongside \textit{Platform-as-a-Service (PaaS)}.
However, 
it is distinguished by its complete abstraction of infrastructure management, automatic resource scaling, and, in certain service models, the \textit{separation of computation from storage} and \textit{event-driven} architecture.
While these latter aspects were initially regarded as intrinsic to serverless computing, recent literature acknowledges 
that they are specific to certain service models,
particularly \textit{Function-as-a-Service (FaaS)}, rather than 
general attributes 
of the entire serverless paradigm~\cite{10.1145/3587249}.

Although research publications on serverless computing have surged since its introduction and the subject is extensively covered in the literature, 
no universally accepted formal definition of this model exists.
Nonetheless, current literature concurs that serverless computing is characterized by the following core traits:
\begin{itemize}    
    \item \textit{No operational logic (NoOps).} 
    Developers are not required to handle the provisioning and management of cloud execution environments, including virtual machines or containers, nor their operational aspects, such as resource management, container/VM lifecycle, fault tolerance, security, monitoring, and accounting~\cite{10.1145/3579643,10.1145/3587249,10.1145/3510611,10.1145/3368454, 8481652, Baldini2017, hellerstein2018serverlesscomputingstepforward}. The term ``serverless'' is derived from this very concept, not because servers are absent, but because all these responsibilities are fully abstracted from the customer and managed by the cloud provider.

    \item \textit{Utilization-based billing.}
    The billing mechanism accounts solely for the resources actually used by the customer and the duration of their usage, following a \textit{pay-as-you-go} billing model~\cite{10.1145/3579643,10.1145/3587249,10.1145/3510611,10.1145/3368454, 8481652, Baldini2017, hellerstein2018serverlesscomputingstepforward}.
    This differs significantly from conventional cloud billing,
    where users incur costs for allocated resources even when idle.

    \item \textit{Auto-scaling.}
        The cloud provider has the responsibility of automatically allocating and scaling the resources associated with the service on demand, with the ability to scale from zero to infinity~\cite{Baldini2017, Jonas:EECS-2019-3, 10.1145/3368454, hellerstein2018serverlesscomputingstepforward, 10.1145/3579643, 10.1145/3510611}.
        This means that resources are provisioned and deallocated as needed, ensuring an efficient match to workloads and optimizing utilization when applications are idle. Although automatic scaling exists in some PaaS offerings, in serverless computing it specifically denotes the capacity to scale from zero to infinity, and vice versa, which is essential for event-driven and variable workloads.
\end{itemize}

Together, these properties distinguish serverless computing from other cloud paradigms, making it particularly suited for elastic, event-driven workloads.

\subsection{Serverless Service Models}
\label{subsec:service_models}
Similar to cloud computing, serverless services can be categorized into distinct service models, each representing a specific type of service or product. Current literature identifies three primary types of serverless service models: \textit{Function-as-a-Service (FaaS)}, \textit{Backend-as-a-Service (BaaS)}, and \textit{Container-as-a-Service (CaaS)}. The following are the most widely accepted definitions of these models:

\begin{definition}[]
    Function-as-a-Service (FaaS) is a form of serverless computing in which the cloud provider manages the resource requirements, life-cycle, and event-driven triggering of user-provided functions~\cite{8481652}.
\end{definition}
\begin{definition}[]
    Backend-as-a-Service (BaaS) is a form of serverless computing focused on providing specialized serverless frameworks that cater to specific application requirements (i.e., object storage, databases, or messaging services) \cite{Jonas:EECS-2019-3}.
\end{definition}
\begin{definition}[]
    Container-as-a-Service (CaaS) is a cloud service model that allows users to deploy and manage containers in the cloud \cite{VARGHESE2018849}.  Depending on the level of abstraction and automation it provides, CaaS can also be considered a form of serverless computing \cite{10.1145/3587249}.
\end{definition}

Going into details, FaaS represents the most widely utilized form of serverless computing. It allows developers to create and run event-driven applications, where the unit of computation is one or more serverless functions that are activated and executed in response to specific events~\cite{10.1145/3368454}. 
Moreover, FaaS is distinguished by its separation of computation from storage. To enable easy and elastic auto-scaling, serverless functions should be stateless, not retaining any data between executions. Instead, they should rely on external storage solutions, such as cloud storage services (e.g., AWS S3, Azure Blob Storage) or Backend-as-a-Service (BaaS) offerings, to handle data storage and retrieval. Examples of FaaS solutions include AWS Lambda, Azure Functions, and Google Cloud Run functions.

In contrast, BaaS represents a more tailored, vendor-specific approach to serverless computing compared to FaaS, offering services that simplify the development of backend functionalities~\cite{10.1145/3579643}. These typically include specialized serverless frameworks for object storage, databases, and authentication components~\cite{10.1145/3587249,9190031}. Examples of BaaS offerings include AWS DynamoDB, AWS Amplify, and Google Cloud Firestore.


Finally, the classification of CaaS as a serverless service model remains a debated topic. Container orchestration services like AWS Elastic Container Service, Google Kubernetes Engine, and Azure Kubernetes Service manage containers partially but do not fully abstract the underlying infrastructure, making their classification as serverless counterintuitive. However, recent CaaS platforms like Google Cloud Run, AWS Fargate, and Azure Container Apps offer a higher level of abstraction, managing the container lifecycle with minimal user intervention. These platforms can be considered \textit{serverless-enabled} CaaS, depending on the level of abstraction provided by the specific container platform~\cite{10.1145/3587249}.

\begin{table*}[ht]
\footnotesize
\centering
\caption{Comparison of PaaS with serverless service models.}
\label{tab:comp}
\rowcolors{2}{white}{lightaccessblue}
\begin{tabularx}{\textwidth}{>{\raggedright\arraybackslash}p{2.75cm}| *{4}{>{\raggedright\arraybackslash}X}}
\toprule
{\footnotesize\sffamily\bfseries\textcolor{accessblue}{Feature}} & 
\multicolumn{1}{c}{\footnotesize\sffamily\bfseries\textcolor{accessblue}{PaaS}} & 
\multicolumn{1}{c}{\footnotesize\sffamily\bfseries\textcolor{accessblue}{CaaS}} & 
\multicolumn{1}{c}{\footnotesize\sffamily\bfseries\textcolor{accessblue}{FaaS}} & 
\multicolumn{1}{c}{\footnotesize\sffamily\bfseries\textcolor{accessblue}{BaaS}} \\
\midrule
{\contentfont \textbf{Infrastructure abstraction}} &
{\contentfont Medium: abstracts hardware and OS; exposes runtime and middleware} &
{\contentfont Medium-High: abstracts hardware, OS, and container orchestration} &
{\contentfont High: abstracts servers, execution environment, and runtime} &
{\contentfont Very high: abstracts entire infrastructure} \\

{\contentfont \textbf{Degree of control}} &
{\contentfont Medium: user manages app deployment and configuration} &
{\contentfont Medium: user manages container images} &
{\contentfont Low: provider handles most operations; user provides application code} &
{\contentfont Very low: provider manages all backend} \\

{\contentfont \textbf{Scaling}} & 
{\contentfont Dynamic scaling available through manually or often predefined rules; scaling to zero generally not supported} & 
{\contentfont Dynamic scaling available through predefined rules; scaling to zero only with serverless extension} & 
{\contentfont Auto-scaling on-demand, with the ability to scale to zero} & 
{\contentfont Hidden from users, depends on pricing and QoS} \\

{\contentfont \textbf{Unit of work}} & 
{\contentfont Single deployed application or service} & 
{\contentfont Single container or pod} & 
{\contentfont Single function} & 
{\contentfont Single request or API call} \\

{\contentfont \textbf{Granularity of billing}} & 
{\contentfont Medium to large: minutes to hours per resource} & 
{\contentfont Medium-Fine: seconds to minutes per resource} & 
{\contentfont Very fine: hundreds of milliseconds of function execution time} & 
{\contentfont Very fine: pay per request or operation} \\
\bottomrule
\end{tabularx}
\end{table*}

\subsection{Comparison of Serverless with PaaS}

\begin{figure}[ht]
    \centering
     \includegraphics[width=0.95\columnwidth]{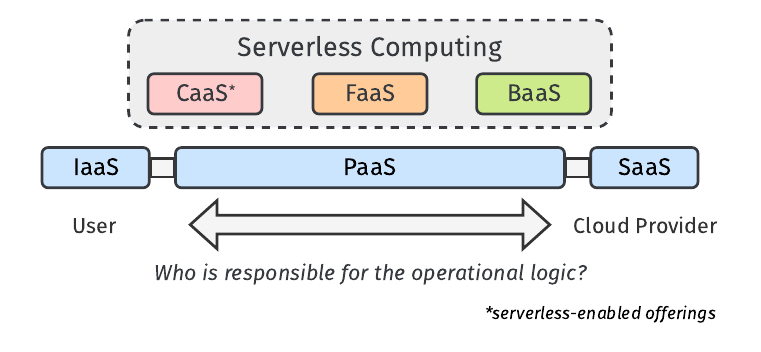}
    \caption{Position of serverless computing and its service models within the cloud computing reference model.}
    \label{fig:ref_serverless}
\end{figure}

Given the level of control provided to developers and the degree of abstraction of computing resources and underlying infrastructure, serverless computing can be positioned within the cloud computing reference model, overlapping with Platform-as-a-Service (PaaS) \cite{10.1145/3154847.3154848, 10.1145/3368454}. However, the degree of control and abstraction offered by a serverless service model determines whether it is more closely aligned with \textit{Infrastructure-as-a-Service (IaaS)} or \textit{Software-as-a-Service (SaaS)}, as illustrated in Figure~\ref{fig:ref_serverless}.
While serverless and PaaS share some similarities, they differ in key aspects, as summarized in Table~\ref{tab:comp}. Both offer developers a comparable degree of control, yet serverless services fully abstract infrastructure management tasks, offer automatic scaling of resources, and adopt a fine-grained billing model~\cite{10.1145/3368454, 10.1145/3579643, 10.1145/3133850.3133855}.

Among the serverless service models, serverless-enabled CaaS offers developers the highest level of control, requiring them to create and manage container images and, in some cases, to influence scaling behavior through configurable policies. However, unlike container orchestration platforms typically associated with PaaS or traditional CaaS, these offerings abstract away the operational aspects of management and configuration associated with the underlying infrastructure.

FaaS, on the other hand, occupies an intermediate position. It typically adheres to a microservice-like development style comparable to that advocated in certain PaaS products, but with finer granularity.
FaaS relies on an event-driven execution model, where functions are triggered by events and billed only for the duration of their execution.
This results in a more cost-effective billing model. As a downside, FaaS usually does not support long-running tasks or stateful applications, which are instead supported by PaaS~\cite{10.1145/3579643}.

Finally, at the other end of the spectrum, BaaS provides the highest level of abstraction, but the least amount of control. 
BaaS offerings typically consist of prebuilt serverless components or frameworks designed for specific use cases. In this regard, BaaS partially bridges the gap between PaaS and SaaS~\cite{10.1145/3368454, 10.1145/3154847.3154848}.

\subsection{Serverless Architecture}
\label{subsec:arch}

\begin{figure*}[ht]
    \centering
    \includegraphics[width=0.80\textwidth]{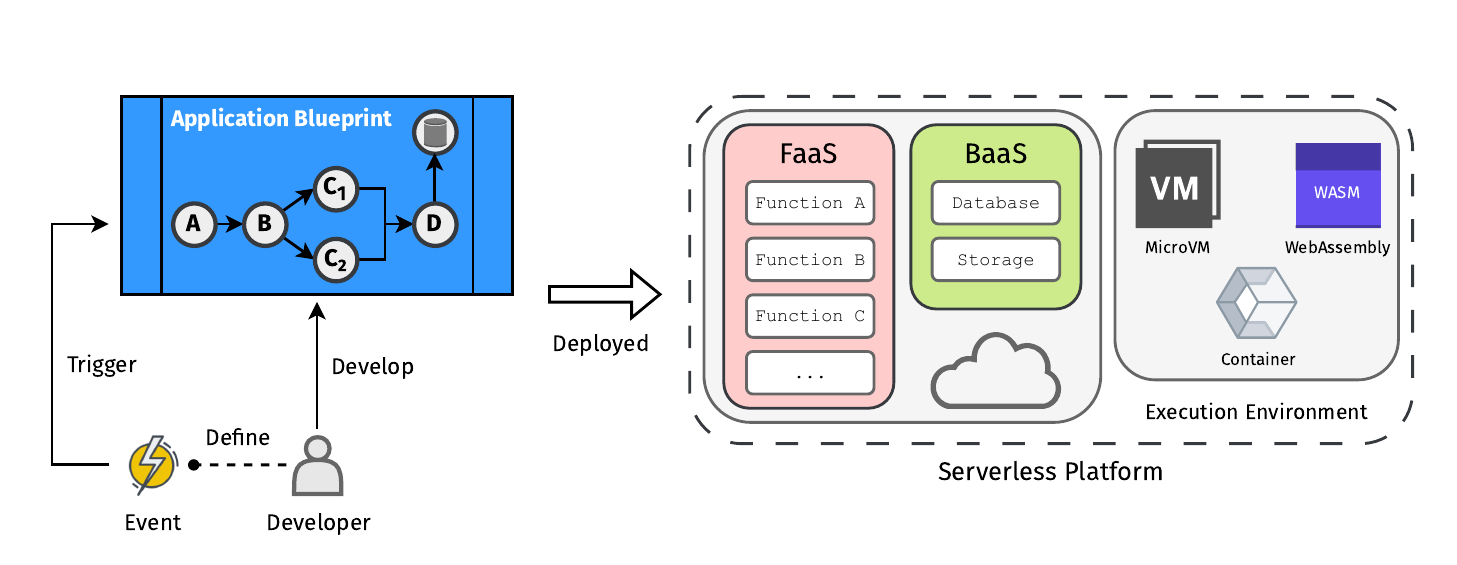}
    \caption{Overview of the high-level architecture and deployment workflow in a serverless application.}
    \label{fig:serv_arch}
\end{figure*}

Serverless computing introduces a distinct architectural paradigm compared to traditional and cloud-native software development. In contrast to monolithic or microservice architectures, which typically characterize cloud-native applications, serverless architectures prioritize a more modular and distributed design.
%
FaaS is commonly used as the primary service model in serverless architectures, allowing applications to be decomposed into smaller, independent components, commonly known as serverless functions. Each function is designed to perform a specific task or handle an individual event. This fine-grained decomposition usually leads to greater elasticity and allows cloud providers to improve fault tolerance, load balancing, and scalability~\cite{10.1145/3579643}.
In addition, serverless applications frequently leverage BaaS offerings, which simplify backend integration by delegating non-core functionalities to third-party managed services. This allows developers to focus on implementing business logic, rather than configuring and managing infrastructure components.

Bringing it all together, developers build serverless applications by decomposing them into multiple functions and deploying them on a serverless platform, either managed by a third-party cloud provider or self-hosted. The resulting application is often modeled as a function composition expressed through a workflow or directed acyclic graph (DAG). Figure~\ref{fig:serv_arch} provides a high-level overview of this workflow. The execution lifecycle of a serverless application typically involves:
\begin{enumerate}
    \item A predefined event triggers a serverless function that has been previously bound to it.
    \item 
    The serverless platform provisions the required resources and sets up the execution environment to run the function.
    This 
    includes 
    allocating a virtual machine or container, pulling in the function code, and performing any required application-specific startup tasks.
    \item 
    After execution, the provisioned resources remain idle for a predefined timeout, referred to as the \textit{keep-alive time}, while the platform awaits any additional requests for the instantiated function. If no further requests are received within that time, the platform releases the allocated resources.
\end{enumerate}

Serverless functions are typically instantiated within isolated execution environments, commonly referred to as \textit{function instances} in the literature~\cite{10.1145/3579643}. These instances rely on classical Infrastructure-as-a-Service (IaaS) technologies, such as containers or virtual machines. This concept is pivotal as it underpins the serverless computing architecture, enabling horizontal scaling and fine-grained utilization billing. However, it also introduces some challenges, most notably the \textit{cold start} problem~\cite{10.1145/3587249, 9191377, EBRAHIMI2024103115}. This issue arises due to serverless computing's ability to \textit{scale to zero}. 
When a function instance remains inactive and the keep-alive time expires, its resources are deallocated, causing subsequent requests to experience startup latency. This aspect is critical, particularly when designing applications with stringent latency requirements or unpredictable traffic patterns.

\subsection{Serverless Platforms}
\label{subsec:serverless_platforms}
As previously stated, serverless computing was first popularized by Amazon in 2014 with AWS Lambda, establishing its roots in commercial cloud platforms. Since then, it has gained significant traction, leading to the development of a wide range of both commercial and open-source FaaS platforms. The remainder provides an overview of notable examples from both categories, emphasizing their key features and characteristics.

\subsubsection{Commercial Platforms}
After the introduction of AWS Lambda~\cite{AWSLambda}, other major cloud providers have launched their own commercial FaaS platforms. Microsoft, Google, and IBM have all launched their own cloud functions services: Azure Functions~\cite{AzureFunctions}, Cloud Run functions~\cite{GoogleCloudFunctions} (formerly Cloud Functions), and IBM Cloud Code Engine~\cite{IBMCodeEngine} (formerly IBM Cloud Functions), respectively.
Most commercial FaaS platforms differ in their billing model, programming language support, and each one imposes its own platform-specific restrictions on memory usage, timeout limits, and so on.
Furthermore, like for other commercial cloud products, these FaaS platforms usually hide their architecture details and impose strict vendor lock-in. However, for companies and industries that don't have their private cloud infrastructure, these platforms integrate well within the cloud services ecosystem of their vendor.
In addition, similar to what Docker has done with Docker Hub, cloud providers such as Amazon and Microsoft offer repositories where users can upload and publish pre-built serverless functions. Amazon offers this service through AWS SAR~\cite{AWSServerlessRepo}, while Microsoft offers the Azure Serverless Community Library~\cite{ServerlessLibrary}.

\subsubsection{Open Source Platforms}
Although serverless has primarily been popularized by commercial cloud providers, it has also recently attracted interest from the open-source community.
The primary distinction between commercial FaaS platforms and open-source ones, aside from the freely available source code, is that open-source solutions can be instantiated on private cloud infrastructures.
Moreover, these solutions rely on open-source components and use Linux containers as execution environments~\cite{DBLP:journals/corr/abs-2106-03601}.
Notable open-source FaaS platforms include OpenFaaS~\cite{openfaas}, Apache OpenWhisk~\cite{OpenWhisk}, Knative~\cite{KnativeDocs}, and Nuclio~\cite{Nuclio}.
Among these, only Nuclio specifically targets compute-intensive workloads for data science.
Finally, open-source platforms such as OpenFaaS offer specific repositories or marketplaces that allow users to share pre-made serverless functions, similar to what is done in some commercial FaaS platforms~\cite{OpenFaaSStore}.

\subsection{Execution Environments}
One of the key enabling technologies of cloud computing is virtualization \cite{BUYYA20133}, and the same holds true for serverless computing. The deployment of function instances relies on virtualized environments, typically implemented via lightweight virtualization stacks, to enable rapid instantiation and mitigate cold start latency.
%
%
That said, it is clear that the choice of execution environment plays a significant role in the architecture of a serverless platform.
In practice, most serverless platforms rely on Linux containers, often instantiated via Docker~\cite{merkel2014docker} or other OCI-compliant runtimes, for sandboxing. Containers share the host kernel and provide efficient isolation, portability, and ease of deployment. Despite some security and stability concerns related to the shared kernel~\cite{8693491}, 
containerization remains the de facto standard for sandboxing in serverless platforms and, more broadly, in cloud infrastructures.
Specialized containerization solutions such as Singularity have been developed for high-performance computing, offering better integration with HPC software stacks while avoiding the need for elevated privileges~\cite{10.1371/journal.pone.0177459,10.1145/3332186.3332192}.\footnote{Singularity has since split into two different projects: Apptainer, maintained by the Linux Foundation, and Singularity-CE, maintained by Sylabs.} Singularity emphasizes integration over isolation, allowing seamless access to host resources such as GPUs and high-speed interconnects (e.g., InfiniBand, OmniPath). It also supports job schedulers like Slurm and PBS, facilitating the deployment of container-based workloads in HPC systems~\cite{10.1145/3332186.3332192}. These features position Singularity as a strong alternative for containerization in high-performance serverless applications.

\begin{figure}[ht]
    \centering
    \includegraphics[width=0.95\columnwidth]{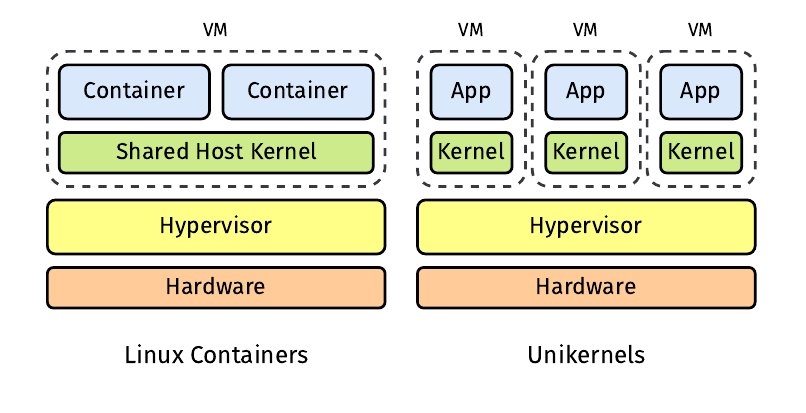}
    \caption{Comparison between containers and unikernels.}
    \label{fig:unikernel}
\end{figure}

Furthermore, recent literature has investigated the potential of alternative solutions, including WebAssembly and unikernels.
WebAssembly has gained attention as a lightweight, portable virtual ISA supporting multiple programming languages and safe, sandboxed execution~\cite{WebAssemblyCoreSpecification1,gackstatter_pushing_2022}. In addition, the \textit{WebAssembly System Interface (WASI)} extends WebAssembly use beyond just web applications, enabling applications to interact with the underlying operating system through \textit{syscall-like} APIs.
Unikernels represent another emerging option, packaging applications into single-purpose, single-address-space images built with library operating systems~\cite{10.1145/2557963.2566628}. Each unikernel runs on a virtual machine on a hypervisor, providing isolation through virtualization. 
Unlike containers, as shown in Figure~\ref{fig:unikernel}, unikernels do not share the same kernel. This architecture ensures that if a unikernel fails, only the affected application is impacted.
Since both application and kernel operate in kernel mode, the overhead of user-kernel transitions is minimized. These properties make unikernels well-suited for serverless, offering a compelling alternative to traditional containers~\cite{10749739,8567366}.
However, alternatives such as WebAssembly and unikernels present viable options to containers in serverless computing, balancing isolation, performance, and startup overhead.

\section{Research Methodology}
\label{sec:methodology}
To conduct a precise and structured review of the literature on high-performance serverless computing, we followed the guidelines proposed by Kitchenham et al.~\cite{kitchenham2007guidelines}, which are well-established and widely adopted within software engineering for systematic literature reviews.
This section outlines the methodology adopted in our study.

\subsection{Research Questions}
In order to achieve our objectives and provide a guide for the analysis of current research on the suitability of the serverless execution model for supporting HPC, AI, big data, and other parallel  compute-intensive workloads, we have formulated the following five research questions:

\paragraph{\textbf{RQ1}} \textit{Which research directions may be identified in the current literature on high-performance serverless computing?} The objective of this research question is to identify and construct a taxonomy of current research directions in the domain of high-performance serverless computing literature.

\paragraph{\textbf{RQ2}} \textit{How does the existing literature address specific problems for each research direction identified in the current literature?} The objective of this research question is to understand, for each of the previously identified research directions, the different solutions and approaches proposed in the literature to address specific problems.

\paragraph{\textbf{RQ3}} \textit{What specific use cases are addressed by the solutions proposed in the current literature on high performance serverless computing?} This research question aims to identify and classify the different use cases in which serverless computing is applied to generic and domain-specific HPC, AI, big data, and other parallel compute-intensive workloads.

\paragraph{\textbf{RQ4}} \textit{Is high-performance serverless computing an emerging research area? Which studies are most influential and where are they published?} This research question aims to analyze the growing interest in high-performance serverless computing over time, identify the most impactful papers through citation analysis, and determine the primary publication venues for related research.

\paragraph{\textbf{RQ5}} \textit{Who are the main contributors and what are the collaboration patterns in the current high-performance serverless computing research community?} The objective of this research question contributes to identify the leading authors and institutions, examine co-authorship networks and highlight collaboration patterns, providing an overview of the structure and impact of the community.

\subsection{Review Protocol}
\begin{figure*}[ht]
    \centering
    \includegraphics[width=0.85\textwidth]{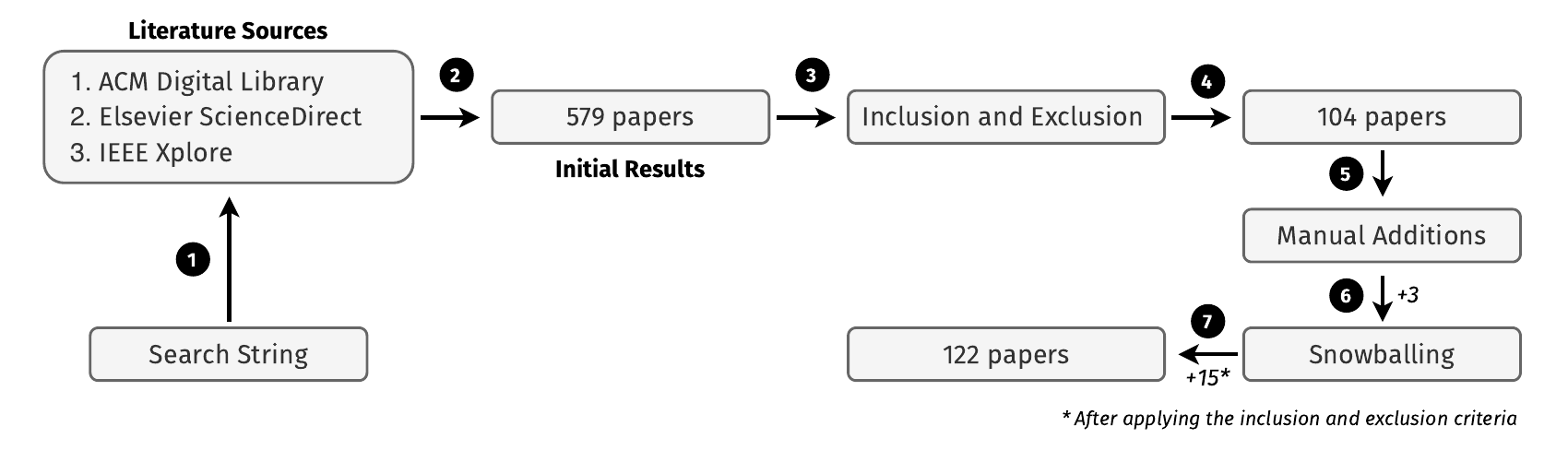}
    \caption{Overview of the research methodology adopted in this study for the article selection process.}
    \label{fig:methodology}
\end{figure*}

A defining characteristic of a systematic literature review is the implementation of a rigorous review protocol, which differentiates it from conventional review methodologies~\cite{kitchenham2007guidelines}. This is imperative to ensure the identification of a comprehensive set of primary studies relevant to the research questions, thus mitigating biases in study retrieval and selection. The following describes the review protocol we implemented, while Figure~\ref{fig:methodology} provides an overview of its execution.

\subsubsection{Literature Sources}
To perform an exhaustive search of the relevant literature, we selected three digital libraries as literature sources: (1) \textit{ACM Digital Library}, (2) \textit{Elsevier ScienceDirect}, and (3) \textit{IEEE Xplore}.

\subsubsection{Search String}
To gather relevant articles on high-performance serverless computing, we defined and applied a search string across the three selected digital libraries.
We initially defined a search string, which we then refined through trial and error to better capture studies on serverless computing for HPC, AI, big data workloads, and accelerator integration. This approach led to the following search string:
\begin{quote}
    \textit{((serverless OR "function-as-a-service" OR faas OR "high-performance serverless") AND (("high-performance computing" OR HPC) OR (accelerators OR GPUs OR FPGAs OR TPUs OR "heterogeneous computing")))}
\end{quote}

We applied the chosen search string to the selected digital libraries. For ScienceDirect, due to its limit of eight Boolean operators per query, we adapted the search string by splitting it into multiple queries.

\subsubsection{Inclusion and Exclusion Criteria}
To determine the relevance of a publication to the scope of this literature review, we established the following criteria. 

A publication was retained if it met all the inclusion criteria. The adopted inclusion criteria were as follows: (1) Publications that focused on serverless computing within the context of high-performance computing, AI inference, big data, or other parallel compute-intensive workloads; (2) Publications that proposed use cases, design solutions, algorithms, optimization approaches, or conceptual frameworks for such applications, or compared their performance against traditional IaaS or on-premises infrastructure; (3) Publications published between January 2017 and February~2025. 

A publication was excluded if it met any of the following criteria. The exclusion criteria were: (1) Secondary or tertiary studies, e.g., empirical studies, literature reviews, and surveys; (2) Bachelor, master, or doctoral dissertations; (3) Pre-printed studies and other publications not formally peer-reviewed; (4) Publications not written in English; (5)~Publications focused exclusively on AI training, which remains impractical on serverless platforms due to runtime and resource demands.\footnote{While there are some studies on the use of serverless architectures for AI/DL training~\cite{9888122}, we considered their impact on the goals of this survey to be minimal. One exception was retained~\cite{kang_towards_2024} for its generalizable scheduling and data placement strategy.}

\subsubsection{Search Process}  
Once the search string and inclusion/exclusion criteria were established, we initiated the search process. We applied the search string to the selected digital libraries, with results restricted to research papers published between January 1, 2017 and February 28, 2025, as preliminary searches did not reveal relevant studies prior to 2017. This initial search yielded a total of 579 publications.

We conducted the selection process iteratively. Two authors independently reviewed the entire list of retrieved articles, designating each for inclusion or exclusion. During this phase, we evaluated most of the articles based on their titles and abstracts. When a decision could not be made from these, we reviewed the full content of the paper. In cases where consensus could not be reached, a third author was consulted to make the final determination. Following this, two authors reviewed the included papers, further refining the selection by analyzing the full content.

After applying the inclusion and exclusion criteria, we reduced the number of selected articles to 104. Additionally, we manually added three relevant research articles~\cite{shillaker_faasm_2020,petrosyan_serverless_2022,du_serverless_2022} that were not retrieved by the initial search string but were considered seminal works directly relevant to our scope. This brought the total to 107 publications.

To expand the set of relevant studies on high-performance serverless computing, we applied a \textit{snowballing} procedure following Claes Wohlin's guidelines~\cite{10.1145/2601248.2601268}. This process involved \textit{backward snowballing}, which analyzed the reference lists of included articles, and \textit{forward snowballing}, which examined papers citing them. We conducted both steps using Semantic Scholar\footnote{Available at \url{https://www.semanticscholar.org/}.} and repeated the process until no new relevant articles were found. The first round identified 41 additional studies, filtered down to 15. A second round yielded four more articles, but none met our relevance criteria. In total, we selected 122 relevant papers for this literature review.

\subsubsection{Data Extraction}
\label{subsubsection:data_extract}
To systematically collect relevant information from the 122 primary studies included in this literature review, we created and maintained a spreadsheet as the central repository for extracted data. 
For each paper, we gathered and stored relevant data in the spreadsheet, refining and updating it with each iteration of the article review.
The data collected included a concise description of the scope of each article, details on publication venues, citation counts, and an initial classification of the research directions addressed by the article. Additional bibliographic metadata (e.g., author names, titles, abstracts, references) were maintained in a separate RIS file. This facilitated cross-verification and export for bibliometric analysis.

Citation counts were retrieved as of February 28, 2025, using a combination of Scopus and Google Scholar databases, and the original digital libraries. To mitigate potential discrepancies due to indexing variations or self-citations, we cross-verified counts where possible.

The resulting dataset~\cite{1ed9-pc52-25}\footnote{The dataset is available at \url{https://github.com/Della97/survey-artifact/}.} was subsequently used to construct the taxonomy of research directions and use cases, presented in Section~\ref{subsec:taxonomy_explained}, as well as to perform the bibliometric analyses, reported in Section~\ref{subsec:bib_explained}.

\subsection{Taxonomy Construction}
\label{subsec:taxonomy_explained}
To answer RQ1 and provide a comprehensive overview of the current research directions in  literature regarding high-performance serverless computing, we adhered to established guidelines from both general literature and specific to software engineering, adopting an open coding approach~\cite{williams2019art,799955}.
The same methodology was applied for RQ3, where it was employed to construct a taxonomy of use cases addressed in the selected research papers.

The taxonomy construction process consisted of three phases. In the first phase, two authors assigned short phrases to each selected paper to represent the research directions and addressed use cases. In the second phase, the authors compared their assigned phrases, merging them and consulting a third author in case of disagreements to reach consensus. In the third phase, high-level categories were defined based on the most frequent terms in the assigned phrases. These phrases were then grouped into categories, forming a hierarchical taxonomy.  We iteratively refined the categories, assigning papers to the relevant categories. The conflicts were resolved with the help of a third author. This process continued until a consensus was reached, which resulted in the final labels for our taxonomy.


Finally, it is important to note that some research articles, such as Yang et al.~\cite{yang_infless_2022}, may span multiple research directions. For this reason, we classified certain works into more than one category, based on the relevance and depth of their contributions. For example, Yang et al.~\cite{yang_infless_2022} proposed a serverless framework for ML inference, addressing topics such as accelerator support, domain-specific frameworks, and resource scheduling. 
%
In such cases, we included the work in multiple research directions, but only when substantial contributions were made in each.

\subsection{Bibliometric Analysis Methodology}
\label{subsec:bib_explained}
To address RQ4 and RQ5, we conducted a bibliometric analysis of the 122 primary studies included in this review. This analysis builds on the dataset described in Section~\ref{subsubsection:data_extract}, using citation counts, publication metadata, and author information.

For RQ4, we aimed to capture the scientific impact and dissemination trends of research on high-performance serverless computing. We analyzed citation counts to identify the most influential papers and venue statistics (e.g., publication frequency by journal/conference, yearly trends). Citation data were aggregated per paper and venue using Python scripts.

For RQ5, we investigated collaboration patterns within the research community. To this end, we constructed an undirected graph with authors as nodes, $496$ in total, and edges representing co-authorship on at least one of the selected studies. Using VOSviewer~\cite{van_eck_software_2010}, we applied a modularity-based clustering method for community detection to partition the graph into collaboration clusters. To reduce sparsity and highlight recurring collaborations, we applied a filtering step retaining only authors with three or more publications, yielding a graph with $31$ nodes, which was used for our analysis.

The results of this bibliometric analysis are presented and discussed in Section~\ref{subsec:trends} and Section~\ref{subsec:bibliometric}.

\section{Results}
\label{sec:results}
The selected publications were thoroughly reviewed to answer previously identified research questions. This section presents and discusses the results obtained for each question.

\subsection{RQ1: Research Directions}
\label{subsec:taxonomy}

\begin{figure*}[ht]
    \centering
    \includegraphics[width=0.90\textwidth]{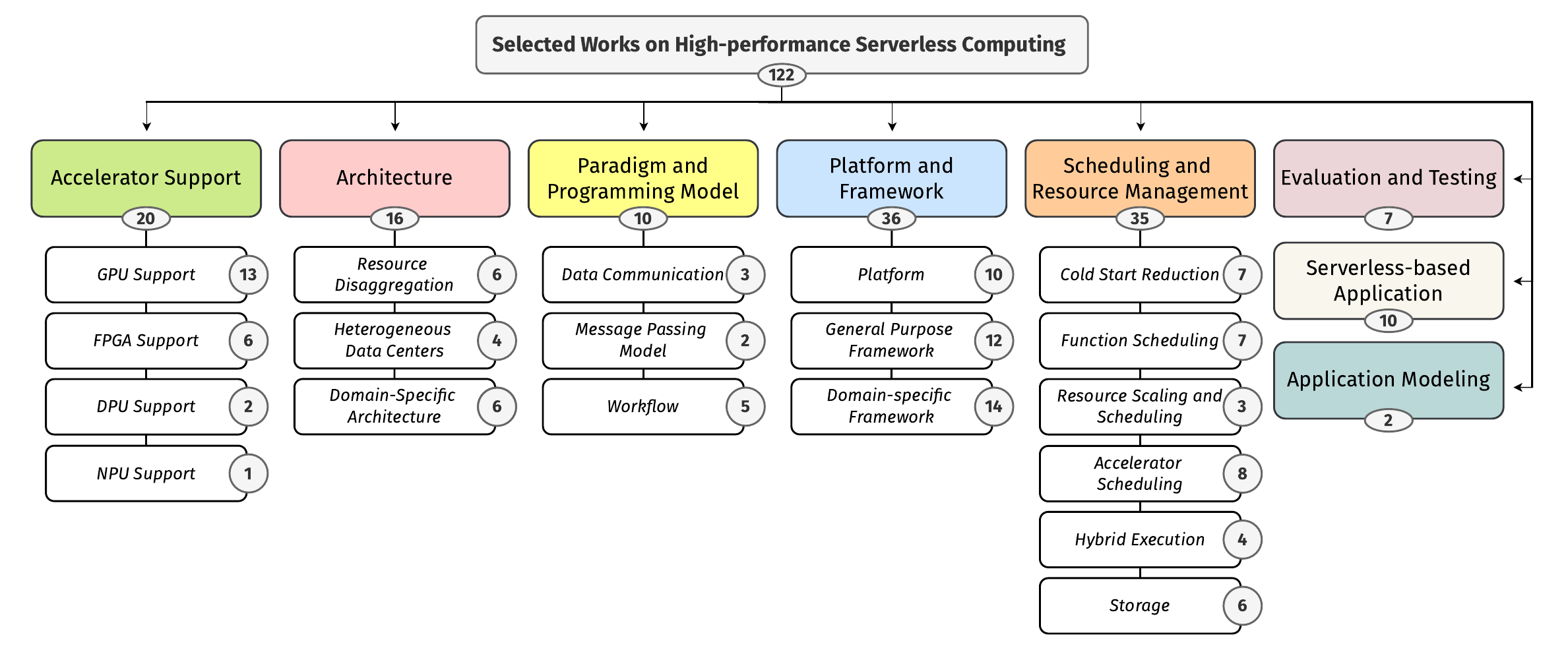}
    \caption{Proposed taxonomy of research directions within the high-performance serverless computing literature.}
    \label{fig:taxonomy}
\end{figure*}

In order to answer RQ1, we built a taxonomy of current research directions in the field of high-performance serverless computing. We followed the methodology presented in Section~\ref{subsec:taxonomy_explained}, which resulted in the taxonomy depicted in Figure~\ref{fig:taxonomy}. Each research direction in the figure is annotated with the number of articles classified under it and its related topics. 
Overall, our taxonomy comprises eight main research directions, each describing a specific aspect addressed in the context of high-performance serverless computing. For some of these directions, we defined subcategories that group works that fall within the same area and propose similar solutions. These subcategories, also illustrated in Figure~\ref{fig:taxonomy}, reflect how research articles are organized throughout this paper.

Focusing on the taxonomy in detail, studies classified as \textit{Accelerator Support} cover aspects related to the integration and efficient use of accelerators in serverless platforms and architectures. Depending on the type of accelerator addressed, we defined different subcategories to distinguish between their specific support strategies.
Studies under the \textit{Architecture} research direction present architectural designs or approaches that adopt the serverless execution model to address challenges in data center operations. This direction includes three subcategories: \textit{Resource Disaggregation}, comprising works that adopt serverless to support disaggregated data centers; \textit{Heterogeneous Data Centers}, focusing on architectural approaches for integrating serverless across heterogeneous resources; and \textit{Domain-specific Architecture}, collecting works that design serverless architectures tailored to specific domains or applications.
The \textit{Paradigm and Programming Model} research direction includes works that propose programming abstractions or execution paradigms based on serverless to serve HPC, AI, or big data applications. Subcategories include \textit{Data Communication}, grouping abstractions for efficient data and inter-function communication; \textit{Message Passing Model}, which adapts MPI-style models for serverless HPC applications; and \textit{Workflow}, covering paradigms and models that target the development of scientific workflows and AI or ML pipelines.
Works labeled as \textit{Platform and Framework} propose serverless platforms or frameworks specifically designed for high-performance workloads. This includes general-purpose serverless solutions and domain-specific frameworks. We grouped the contributions into three subcategories: \textit{Platform}, for full-stack high-performance serverless platforms; \textit{General-purpose Framework}, which includes frameworks targeting general parallel and distributed workloads; and \textit{Domain-specific Framework}, for frameworks targeting specific domains such as ML inference or stream processing.
The \textit{Scheduling and Resource Management} research direction addresses various techniques and strategies to manage resources and schedule workloads in high-performance serverless platforms. Subcategories include \textit{Cold Start Reduction}, focusing on techniques to mitigate cold start delays; \textit{Function Scheduling}, which covers adaptive scaling and function placement; \textit{Resource Scaling and Scheduling}, addressing general provisioning and orchestration; \textit{Accelerator Scheduling}, targeting the allocation and scheduling of accelerator resources; \textit{Hybrid Execution}, which combines on-premises infrastructure with serverless resources; and \textit{Storage}, covering approaches for data placement and efficient access in high-performance serverless platforms.
The \textit{Evaluation and Testing} category includes works that provide comparisons, benchmarking tools, simulation environments, and evaluation methodologies to evaluate the performance of serverless systems in the context of HPC, AI, and big data workloads.
The \textit{Serverless-based Applications} research direction collects early and exploratory efforts that apply serverless to specific use cases such as numerical computing, health monitoring, and scientific computing. These works typically focus on assessing feasibility and potential benefits.
Finally, works in the \textit{Application Modeling} research direction propose approaches and tools to model applications that are intended to be deployed in serverless environments. These are designed to help developers understand, optimize, or adapt applications for this paradigm.

From our taxonomy, the research direction with the highest number of contributions is \textit{Platform and Framework}, accounting for $29.5\%$ of the surveyed studies. This highlights the strong research interest in developing tailored serverless platforms and frameworks to serve domains that traditionally rely on HPC infrastructures, such as scientific computing, AI, and big data. This is followed by \textit{Scheduling and Resource Management}, which comprises $28.6\%$ of the selected studies. This reflects the importance of developing scheduling and resource management techniques for high-performance serverless systems, especially given the growing interest in applying serverless to HPC workloads and the distinct characteristics of HPC clusters compared to traditional cloud data centers.
Interestingly, within this research direction, the most represented subcategory is \textit{Accelerator Scheduling}, which accounts for $22.8\%$ of the works. Combined with the fact that the third most represented research direction is \textit{Accelerator Support}, accounting $16.4\%$ of the works, this suggests that the management of accelerators in increasingly heterogeneous HPC environments is of high relevance. This is particularly true in domains such as AI and ML, where accelerators play a critical role in improving performance and reducing execution time~\cite{10.1145/1553374.1553486,10.1117/12.2279088}.
Finally, in descending order by the number of contributions, the remaining research directions are \textit{Architecture} with $13.1\%$, followed by \textit{Paradigm and Programming Model} and \textit{Serverless-based Applications}, each accounting for $8.2\%$, and finally \textit{Application Modeling}, which represents only $1.6\%$ of the considered studies.

\subsection{RQ2: Current Solutions}
\label{subsec:research_directions}
To address RQ2, we investigated the solutions and approaches proposed in the literature for each research direction identified in the taxonomy presented in Section~\ref{subsec:taxonomy}. 
To this end, we re-read the selected studies, summarized their contributions, and classified each work into one or more research directions based on its primary focus.
The following presents our findings, organized by research direction and associated subtopics.

\begin{table*}[ht]
\footnotesize
\centering
\caption{Summary of studies addressing accelerator support in serverless computing.}
\label{tab:accelerator}
\rowcolors{2}{white}{lightaccessblue}
\begin{tabularx}{\textwidth}{>{\raggedright\arraybackslash}p{3.80cm} >{\raggedright\arraybackslash}X >{\raggedright\arraybackslash}l}
\toprule
{\footnotesize\sffamily\bfseries\textcolor{accessblue}{Study}} & 
{\footnotesize\sffamily\bfseries\textcolor{accessblue}{Proposed Approach}} & 
{\footnotesize\sffamily\bfseries\textcolor{accessblue}{Hardware}} \\
\midrule
{\contentfont Naranjo et al.~\cite{naranjo_accelerated_2020,naranjo_delgado_acceleration_2023}} & 
{\contentfont Provide access to centralized GPU clusters through rCUDA and NVIDIA-Docker.} & 
{\contentfont GPU} \\

{\contentfont Kim et al. \cite{kim_gpu_2018}} & 
{\contentfont Propose a GPU-supported serverless framework using IronFunction and NVIDIA-Docker.} & 
{\contentfont GPU} \\

{\contentfont Satzke et al.~\cite{satzke_efficient_2021}} &  
{\contentfont Provide access to vGPUs through partition of physical GPUs.}  & 
{\contentfont GPU} \\

{\contentfont Fingler et al.~\cite{fingler_dgsf_2022,fingler_disaggregated_2023}} & 
{\contentfont Provide access to centralized GPU clusters through custom APIs and implementation.} & 
{\contentfont GPU} \\

{\contentfont Related Studies~\cite{dhakal_fine-grained_2023,rattihalli_fine-grained_2023, hui_esg_2024, bhasi_paldia_2024, yang_infless_2022}} & 
{\contentfont Enable GPU resource partitioning and multiplexing through NVIDIA MIG and/or MPS.} & 
{\contentfont GPU} \\

{\contentfont Related Studies~\cite{garg_faaster_2021, prakash_optimizing_2021}} & 
{\contentfont Leverage kernel slicing of GPU tasks to improve resource usage.} & 
{\contentfont GPU} \\

{\contentfont Related Studies~\cite{maschi_serverless_2023, yang_exploring_2023, rattihalli_fine-grained_2023}} & 
{\contentfont Conduct exploratory studies on the feasibility of FPGA offloading in serverless platforms.} & 
{\contentfont FPGA} \\

{\contentfont Ringlein et al.~\cite{ringlein_case_2021}} & 
{\contentfont Use partial reconfiguration to reduce the provisioning time of FPGA resources.} & 
{\contentfont FPGA} \\

{\contentfont Bacis et al.~\cite{bacis_blastfunction_2020}} & 
{\contentfont Propose a distributed FPGA sharing system and a time-sharing mechanism for multi-tenant execution.} & 
{\contentfont FPGA} \\

{\contentfont Liu et al.~\cite{liu_fuyao_2024}} & 
{\contentfont Offload intermediate data transfers to DPUs by decoupling control and data flow.} & 
{\contentfont DPU} \\

{\contentfont Du et al.~\cite{du_serverless_2022}} & 
{\contentfont Introduce abstractions for DPU and FPGA support, enabling multi-tenant execution and direct function invocation across heterogeneous accelerators.} & 
{\contentfont FPGA, DPU} \\

{\contentfont Ma et al.~\cite{ma_widepipe_2021}} & 
{\contentfont Present a serverless framework that supports NPU clusters to accelerate DNN inference.} & 
{\contentfont NPU} \\
\bottomrule
\end{tabularx}
\end{table*}

\subsubsection{Accelerator Support}
Although CPU-centric architectures have traditionally dominated computing hardware, there has been a surge in workload-specific accelerators (e.g., GPUs, FPGAs, and DPUs). This trend is driven by the slowdown of Moore's Law~\cite{6307773}, advances in hardware technology, and increasing demand for specialized applications such as machine learning and video processing. Accelerators, with their parallel processing capabilities, are leading to a shift toward heterogeneous computing in both cloud and HPC infrastructures~\cite{VARGHESE2018849}.
%
%
Serverless platforms have primarily focused on CPU-based execution, limiting their applicability for compute-intensive workloads. Hence, extending serverless to support a diverse array of accelerators in increasingly heterogeneous data centers becomes crucial.
The following paragraphs present recent literature focused on different aspects of supporting accelerators in serverless architectures, while Table~\ref{tab:accelerator} provides a summary of these~works.

\paragraph{GPU Support}
\label{par:gpu-support}
Despite the popularity of GPUs for accelerating compute-intensive workloads, integrating them into serverless computing remains challenging. 
Recent research has explored diverse approaches to enable GPU support in serverless architectures, which can be broadly distinguished between approaches that provide access to entire GPU devices and techniques that enable fine-grained partitioning or sharing of GPU resources.

The former typically leverage centralized GPU clusters, allowing both public (i.e. AWS Lambda) and self-hosted serverless platforms (i.e. Apache OpenWhisk) to offload GPU workloads. These solutions rely on traditional cloud technologies, such as remote API calls via \textit{rCUDA} and containerization with \textit{NVIDIA-Docker}~\footnote{Superseded by the NVIDIA Container Toolkit in 2024.}~\cite{naranjo_accelerated_2020,naranjo_delgado_acceleration_2023,yang_infless_2022}. rCUDA creates virtual GPU devices that act as proxies for physical GPUs located on remote machines, effectively forming GPU clusters. Conversely, NVIDIA-Docker is a runtime that enables Docker containers to access GPUs seamlessly, thereby simplifying the development and deployment process. Other approaches instead rely on custom interfaces~\cite{fingler_dgsf_2022,fingler_disaggregated_2023,kim_gpu_2018}. 
Although these solutions provide full GPU access with relatively simple management, they cannot exploit heterogeneous or disaggregated GPU resources, lack support for advanced function scheduling, and are limited in multi-tenant environments~\cite{garg_faaster_2021}.

To address these limitations, fine-grained partitioning and multiplexing techniques have been proposed.
For example, Satzke et al.~\cite{satzke_efficient_2021} extended the KNIX high-performance serverless framework by integrating a GPU Manager that partitions physical GPUs into \textit{virtual GPUs (vGPUs)}, allowing multiple functions to access shared GPU cluster resources concurrently.
Further research has explored NVIDIA's \textit{Multi-Process Service (MPS)} and \textit{Multi-Instance GPUs (MIG)} multiplexing techniques to partition GPU resources more effectively in serverless environments. MPS enables fine-grained control over GPU allocation by allowing an application to specify the maximum percentage of streaming multiprocessors dedicated to its tasks. Similarly, MIG partitions a single GPU into isolated instances that can be independently assigned to different functions. Dhakal et al.~\cite{dhakal_fine-grained_2023} integrated support for GPU multiplexing into \textit{Parsl} (the parallel runtime system behind Globus Compute, formerly FuncX) by leveraging both MPS and MIG, demonstrating improved utilization and performance for deep-learning workloads.
While fine-grained partitioning improves multi-tenancy and scalability, allowing multiple functions to share GPU resources efficiently, it introduces additional management complexity.
Hence, advanced scheduling techniques have been proposed in the literature to support such multiplexing technologies~\cite{rattihalli_fine-grained_2023, hui_esg_2024, bhasi_paldia_2024, yang_infless_2022}.

Finally, dynamic kernel slicing has been proposed to minimize wasted GPU resources and better meet task deadlines~\cite{prakash_optimizing_2021, garg_faaster_2021}. It involves dividing a large GPU kernel into multiple smaller sub-kernels, enabling fine-grained scheduling and improved resource sharing.
Originally designed for concurrent task execution, kernel slicing can also support horizontal scaling across multiple GPUs~\cite{garg_faaster_2021}. In this sense, it complements multiplexing technologies such as vGPUs, MPS, and MIG, improving efficiency and resource management in multi-tenant, heterogeneous serverless environments.

\paragraph{FPGA, NPU, and DPU Support}
Recent research has also explored the integration of alternative accelerators, such as FPGAs, DPUs, and NPUs, into serverless architectures.
These devices can complement GPUs by addressing specific compute- or data-intensive workloads,
but their unique characteristics pose challenges in terms of portability, resource management, and multi-tenancy.

FPGAs, thanks to their flexibility and reconfigurable logic, can effectively accelerate a wide range of applications. 
However, their device- and vendor-specific toolchains often result in monolithic applications, limiting portability and usability in cloud environments.
Several research efforts have explored FPGA integration within serverless platforms, primarily through proof-of-concept and exploratory studies~\cite{maschi_serverless_2023, yang_exploring_2023, rattihalli_fine-grained_2023}.  
Ringlein et al.~\cite{ringlein_case_2021} presented \textit{Mantle}, an architecture for disaggregated FPGA resources designed to simplify integration in FaaS platforms that leverages partial reconfiguration to reduce cold-boot and provisioning times of new FPGA instances.
%
Similarly, Bacis et al.~\cite{bacis_blastfunction_2020} proposed \textit{BlastFunction}, a distributed FPGA sharing system designed for serverless and cloud applications. It features a custom OpenCL host library to enable seamless interaction with FPGAs 
and uses a \textit{time-sharing} approach to improve resource utilization and achieve multi-tenancy.
These efforts highlight the potential of FPGAs in serverless environments, but also the complexity of managing reconfigurable devices in multi-tenant contexts.

Regarding domain-specific accelerators, such as DPUs and NPUs, which target data-centric and ML/DL workloads, respectively, some research has been conducted, yet it is still limited.
%
DPUs are designed to offload data movement and networking tasks from CPUs, effectively improving communication efficiency in domains such as HPC, AI, and big data.
In this context, Liu et al.~\cite{liu_fuyao_2024} presented \textit{Fuyao}, a solution based on  \textit{Nightcore}~\cite{10.1145/3445814.3446701}, a serverless platform, aimed at improving intermediate data transfer in serverless environments. 
Fuyao’s architecture decouples control flow from data flow to alleviate CPU overhead. It offloads the intermediate data transfer controller for inter-node communication to DPUs, leveraging RDMA for fast inter-function communication and data transfer. This approach enables stateful, direct connections between functions, thereby bypassing traditional third-party forwarding methods usually found in existing serverless platforms. 
%
In contrast, Du et al.~\cite{du_serverless_2022} introduced \textit{Molecule}, a serverless runtime targeting heterogeneous machines. Unlike Fuyao, which employs DPUs solely for communication, Molecule supports accelerators as execution backends, leveraging DPUs to improve function density and FPGAs to boost performance in domain-specific applications
It introduces two abstractions to achieve this: \textit{XPU-Shim} and \textit{vectorized sandbox}.
The XPU-Shim serves as middleware, providing an indirection layer that offers a unified abstraction for managing and utilizing resources across different processing units (PUs). This layer is based on \textit{XPU calls}, a system call-like interface that, similarly to Fuyao, enables low-latency direct inter-function communication across different PUs, effectively supporting DAG-modeled serverless applications.
%
%
Further, the vectorized sandbox abstracts hardware heterogeneity and enables multiple serverless functions to be deployed and invoked concurrently on domain-specific accelerators. This is achieved by packaging multiple functions into a single FPGA image that can be flashed onto the device, enabling multi-tenancy.

Research on specialized accelerators for deep learning workloads, such as NPUs and TPUs, is even more limited.
For instance, Ma et al.~\cite{ma_widepipe_2021} proposed \textit{WidePipe}, a high-throughput serverless inference system designed for \textit{ML-as-a-Service (MLaaS)} workloads, which leverages NPU clusters to improve inference performance. While demonstrating the potential of NPUs for serverless ML workloads, current efforts remain largely confined to domain-specific proof-of-concept rather than general-purpose integration into FaaS platforms.


\subsubsection{Architectures}
Recent research has increasingly focused on architectural models that either adapt to or integrate with the serverless paradigm. 
These efforts often intersect with emerging trends such as resource disaggregation and heterogeneous computing, aiming to improve resource efficiency and enable effective use of accelerators within modern data center infrastructures. In parallel, domain-specific architectures have been proposed to tailor serverless computing to specific application domains or infrastructures. 
The following paragraphs summarize key contributions in these areas. Tables~\ref{tab:arc_disaggregated}, \ref{tab:arc_heterogeneous-dc}, and \ref{tab:arc_domain-specific-arch} provide an overview of the reviewed studies.

\begin{table*}[ht]
\footnotesize
\centering
\caption{Summary of studies on architectural approaches for resource disaggregation in serverless computing.}
\label{tab:arc_disaggregated}
\rowcolors{2}{white}{lightaccessblue}
\begin{tabularx}{\textwidth}{>{\raggedright\arraybackslash}l >{\raggedright\arraybackslash}X >{\raggedright\arraybackslash}p{3.25cm}}
\toprule
{\footnotesize\sffamily\bfseries\textcolor{accessblue}{Study}} & 
{\footnotesize\sffamily\bfseries\textcolor{accessblue}{Proposed Solution}} & 
{\footnotesize\sffamily\bfseries\textcolor{accessblue}{Target Object}} \\
\midrule
{\contentfont Malawski et al.~\cite{malawski_serverless_2022}} & 
{\contentfont Envision an architecture for scientific applications based on existing platforms and frameworks.} & 
{\contentfont Resource Disaggregation}\\

{\contentfont Smeliansky~\cite{smeliansky_network_2022}} & 
{\contentfont Propose an architecture that separates compute and network functions in disaggregated data centers.} & 
{\contentfont Resource Disaggregation}\\

{\contentfont Hu et al.~\cite{hu_skadi_2023}} & 
{\contentfont Propose an architecture designed for heterogeneous disaggregated resources.} & 
{\contentfont Resource Disaggregation} \\

{\contentfont Arjona et al.~\cite{arjona_transparent_2023}} & 
{\contentfont Extend serverless platforms to support legacy multi-threaded applications over disaggregated memory and compute resources.} & 
{\contentfont Memory Disaggregation} \\

{\contentfont Chuang et al.~\cite{chuang_aggregate_2023}} & 
{\contentfont Enable transparent aggregation of disaggregated compute and memory resources through distributed virtualization.} & 
{\contentfont Disaggregated Resource Virtualization} \\

{\contentfont Copik et al.~\cite{copik_software_2024}} & 
{\contentfont Leverage software-based disaggregation to improve resource utilization and flexibility in HPC environments.} & 
{\contentfont Software Resource Disaggregation} \\
\bottomrule
\end{tabularx}
\end{table*}

\paragraph{Resource Disaggregation}
Resource disaggregation has emerged as a transformative approach in modern data center architectures. Unlike traditional architectures, where compute, storage, and networking components are tightly coupled within individual servers, disaggregated data centers decouple these into independently managed and scalable resource pools. This separation allows each resource to scale independently, improving utilization and cost-efficiency by precisely matching resources to workload demands.
%
In this context, serverless computing acts as a key enabling paradigm for both traditional and disaggregated data center architectures. 
As described in Section~\ref{subsec:characteristics_of_serverless}, it abstracts the infrastructure, hiding from developers the complexities of resource provisioning and scaling. For HPC, AI, and big data applications, this abstraction facilitates the dynamic orchestration of disaggregated resources, enabling flexible allocation of compute, storage, and networking capabilities.

Recent research has investigated the intersection of serverless and disaggregated infrastructure. 
Malawski et al.~\cite{malawski_serverless_2022} envisioned a serverless architecture for scientific applications,  emphasizing how data center disaggregation, the convergence of HPC and cloud computing, and increasingly elastic resource management as factors that may ultimately lead to the widespread adoption of serverless computing in scientific domains.
Along similar lines, Smeliansky~\cite{smeliansky_network_2022} proposed  \textit{Network Powered by Computing}, a serverless architecture that tightly integrates with the network infrastructure and emphasizes resource disaggregation, enabling a clear separation between compute and network functions in disaggregated data centers.
Further, Hu et al.\cite{hu_skadi_2023} proposed \textit{Skadi}, a conceptual runtime and architecture for stateful serverless computing in disaggregated data centers. Applications are defined using domain-specific declarations, which are compiled into logical graphs resembling workflows and transformed into physically sharded execution graphs. Nodes, representing functions, are then scheduled across heterogeneous resources. To support deployment across diverse hardware, the authors proposed using \textit{MLIR}~\cite{9370308}, a framework for building domain-specific compilers. 
%
%
%
%
While these works highlight the potential of serverless computing to exploit disaggregated resources, they remain largely conceptual architectures with limited practical evaluation. Consequently, subsequent research have explored practical implementations.
%
%
For instance, Arjona et al.~\cite{arjona_transparent_2023} extended  \textit{Lithops}~\cite{9619947}, a serverless framework, to support the deployment of unmodified multi-threaded Python applications over disaggregated resources. Memory resource disaggregation was achieved through Redis, enabling existing applications to run in a serverless fashion. However, the evaluation conducted by the authors revealed that in serverless environments, data communication and shared memory across function instances remain significant bottlenecks, particularly for applications that heavily rely on data exchanges.

Resource fragmentation is a common issue in both HPC and cloud data centers and, if not addressed properly, could undermine the effectiveness of resource disaggregation. 
%
As discussed in Section~\ref{subsec:arch}, virtualization is a key enabler of serverless computing. Therefore, it also plays a crucial role in enabling resource disaggregation.
In this regard, Chuang et al.~\cite{chuang_aggregate_2023} proposed \textit{FragVisor}, a distributed VMM for disaggregated data centers. It aggregates resources from different machines and, through a distributed shared memory system, presents a unified view of the physical address space to guest virtual machines.
The authors evaluated FragVisor in the context of serverless computing by integrating it as an execution environment into the \textit{OpenLambda} platform~\cite{10.5555/3027041.3027047}. Compared to \textit{GiantVM}~\cite{10.1145/3381052.3381324}, a state-of-the-art distributed hypervisor, FragVisor demonstrated superior performance.

Resource disaggregation has also been explored in the context of traditional HPC clusters, which often experience low resource utilization due to over-provisioning and the inability of batch jobs to leverage temporarily idle resources, due to rigid allocation policies.  
Addressing this, Copik et al.~\cite{copik_software_2024} introduced  \textit{Software Resource Disaggregation}, an HPC-oriented FaaS architecture, and developed a proof-of-concept platform based on \textit{rFaaS}~\cite{copik_rfaas_2023}, aimed at improving resource utilization in HPC systems. The proposed approach leverages resource disaggregation to enable flexible and efficient allocation of compute and memory resources, thereby minimizing waste.
This is achieved by deploying function instances on idle or partially allocated nodes, effectively reclaiming underutilized memory and compute capacity. Experimental results showed that co-locating serverless functions with batch jobs incurs minimal overhead while significantly enhancing system efficiency compared to traditional exclusive node allocations.
%

That said, although serverless and resource disaggregation show strong potential for improving server consolidation across both cloud and HPC systems, challenges remain in efficiently managing communication, shared state, and heterogeneous resources, which future work must address to fully realize the benefits of disaggregated architectures.

\begin{table*}[ht]
\footnotesize
\centering
\caption{Summary of studies on architectural approaches for serverless computing in heterogeneous data centers.}
\label{tab:arc_heterogeneous-dc}
\rowcolors{2}{white}{lightaccessblue}
\begin{tabularx}{\textwidth}{l>{\raggedright\arraybackslash}Xl}
\toprule
{\footnotesize\sffamily\bfseries\textcolor{accessblue}{Study}} & 
{\footnotesize\sffamily\bfseries\textcolor{accessblue}{Proposed Solution}} & 
{\footnotesize\sffamily\bfseries\textcolor{accessblue}{Focus}} \\
\midrule
{\contentfont Werner et al.~\cite{werner_hardless_2022}} & 
{\contentfont Use pre-configured runtimes to abstract various accelerators from developers.} & 
{\contentfont Accelerator Abstraction} \\

{\contentfont Pfandzelter et al.~\cite{pfandzelter_kernel-as--service_2023}} & 
{\contentfont Propose a kernel-based programming model and architecture to simplify accelerator integration and resource sharing.} & 
{\contentfont Resource Sharing} \\

{\contentfont Vandebon et al.~\cite{vandebon_slate_2020}} & 
{\contentfont Present an architecture that maps functions to suitable accelerators.} & 
{\contentfont Accelerator-Aware Scheduling} \\

{\contentfont Bruel et al.~\cite{bruel_predicting_2024}} & 
{\contentfont Propose a serverless architecture to execute HPC and AI workflows in heterogeneous data centers.} & 
{\contentfont Heterogeneous Data Centers} \\
\bottomrule
\end{tabularx}
\end{table*}

\paragraph{Heterogeneous Data Centers}
With the growing adoption of domain-specific accelerators, modern data centers and consumer devices have become increasingly heterogeneous. In serverless computing, integrating these accelerators enables a complete disaggregation of computing resources and application-specific acceleration. Yet, supporting diverse hardware remains challenging, particularly in abstracting and exposing resources to developers.

While GPU support in serverless has been extensively studied, few works address other accelerator types. Werner et al.~\cite{werner_hardless_2022} introduced \textit{HARDLESS}, a serverless architecture that enables developers to use various accelerators via pre-configured runtimes tailored to domain-specific applications (e.g., Python 3 with PyTorch). 
These runtimes eliminate the need for developers to manage or even be aware of the underlying hardware. Instead, developers simply select the desired accelerated runtime, and the provider allocates the appropriate resources. 
Similarly, Pfandzelter et al.~\cite{pfandzelter_kernel-as--service_2023} introduced \textit{Kernel-as-a-Service (KaaS)}, both an architectural design and a programming model for hardware accelerators.
Developers register serverless kernels on a dedicated platform, which applications invoke through a \textit{KaaS server} that abstracts the underlying hardware, handles scheduling and resource provisioning, and supports resource sharing.
%
%
Both works illustrate how serverless abstractions can lower the barrier to heterogeneous accelerator usage. Yet, their reliance on predefined, often vendor-specific runtimes limits flexibility and generality. While suitable for well-defined workloads, these approaches may struggle with rapidly evolving accelerator ecosystems and increase the management burden for providers.

A key challenge in serverless and cloud computing is efficiently managing and provisioning heterogeneous accelerators, particularly for large-scale HPC and AI applications that require coordinated scheduling across diverse hardware.
Addressing this, Vandebon et al.~\cite{vandebon_slate_2020} introduced \textit{SLATE}, a serverless architecture to deploy HPC and AI applications in heterogeneous systems. It features a scheduler that maps function invocations to instance groups with specific environments and accelerators, and an auto-scaler that adjusts group size and composition on demand. Simulations on infrastructures combining CPU and FPGA resources demonstrated that SLATE improves cost efficiency for accelerator-bound functions, while maintaining comparable latency and execution time relative to homogeneous execution.


Beyond hardware-level management, another critical aspect is the co-design of architectures and programming models that enable applications to exploit heterogeneous resources effectively.
HPC and AI applications are increasingly organized as workflows, whose fine-grained, modular structure naturally lends itself to serverless execution and resource disaggregation, providing higher-level abstractions that enhance development, reusability, and deployment.
In this context, Bruel et al.~\cite{bruel_predicting_2024} identified nine principles for implementing HPC and AI workflows in heterogeneous serverless environments.  
The authors observed that decomposing applications into fine-grained functions aligns well with the granularity offered by accelerators, positioning workflows as the dominant programming model for next-generation HPC applications on serverless architectures. Furthermore, the authors proposed a conceptual architecture to deploy HPC and AI serverless workflows across heterogeneous infrastructures, aimed at improving server consolidation and reducing energy consumption and carbon footprint via geo-distributed scheduling.


\begin{table*}[ht]
\footnotesize
\centering
\caption{Summary of studies on domain-specific serverless architectures.}
\label{tab:arc_domain-specific-arch}
\rowcolors{2}{white}{lightaccessblue}
\begin{tabularx}{\textwidth}{>{\raggedright\arraybackslash}p{2.67cm} >{\raggedright\arraybackslash}X >{\raggedright\arraybackslash}l}
\toprule
{\footnotesize\sffamily\bfseries\textcolor{accessblue}{Study}} & 
{\footnotesize\sffamily\bfseries\textcolor{accessblue}{Proposed Solution}} & 
{\footnotesize\sffamily\bfseries\textcolor{accessblue}{Target Domain}} \\

\midrule
{\contentfont Prodan et al.~\cite{prodan_towards_2022} and Farahani et al.~\cite{farahani_towards_2023}} & 
{\contentfont Develop a domain-specific architecture tailored for serverless large-scale graph processing across the cloud continuum.} & 
{\contentfont Graph Processing} \\

{\contentfont Oakley et al.~\cite{oakley_fsd-inference_2024}} & 
{\contentfont Propose serverless communication strategies and inference algorithms for ML workloads.} & 
{\contentfont Machine Learning} \\

{\contentfont Mahapatra et al.~\cite{mahapatra_-storage_2024}} & 
{\contentfont Design an architecture that leverages storage-embedded domain-specific accelerators.} & 
{\contentfont In-Storage Computing} \\

{\contentfont Arevalo et al.~\cite{arevalo_enhancing_2024}} & 
{\contentfont Extend serverless to supercomputers using containers and HPC stack integration.} & 
{\contentfont HPC Infrastructures} \\

{\contentfont Hoefler et al.~\cite{hoefler_xaas_2024}} & 
{\contentfont Introduce a unified serverless architecture to support HPC/Cloud workloads.} & 
{\contentfont Hybrid HPC-Cloud} \\
\bottomrule
\end{tabularx}
\end{table*}

\paragraph{Domain-Specific Architectures}
The flexibility of serverless computing makes it attractive for a wide range of domain-specific applications, which require high scalability, low latency, and efficient resource utilization across cloud, HPC, and hybrid infrastructures. Domain-specific architectures can leverage the serverless model to optimize execution to workload characteristics, enabling efficient deployment of complex workloads, ranging from ML inference to large-scale graph processing on supercomputers.

One notable example where serverless shows significant potential is graph processing. Prodan et al.~\cite{prodan_towards_2022}, and subsequently Farahani et al.~\cite{farahani_towards_2023}, presented the \textit{Graph-Massivizer} project, a European Union-funded initiative aimed at developing a high-performance, scalable, and sustainable serverless platform for large-scale graph processing. The authors proposed a domain-specific architecture that enables efficient deployment and execution of such applications across the entire cloud continuum, from HPC data centers to edge devices.

Machine learning and deep learning represent another domain where serverless is promising. These workloads require fast networks and efficient inter-process communication mechanisms to enable rapid distributed computations, similar to those found in traditional HPC systems. Addressing this, Oakley et al.~\cite{oakley_fsd-inference_2024} introduced \textit{FSD-Inference}, a serverless architecture tailored for ML inference. The authors proposed two data communication approaches, one based on object storage and the other on a publish–subscribe queuing model, to achieve fast data transfers in distributed settings. In addition, the authors devised two fully serverless inference algorithms that leverage these communication strategies to perform distributed matrix-matrix and matrix-vector multiplications. 

The rise of domain-specific accelerators and disaggregated data centers has further created opportunities for researchers to devise specialized serverless architectures. 
Mahapatra et al.~\cite{mahapatra_-storage_2024} introduced \textit{DSCS-Serverless}, an architecture that utilizes accelerators embedded in storage devices to offload and execute computations directly on storage nodes. 
This design reduces communication overhead in disaggregated data centers by offloading functions, when feasible, onto accelerators located at storage nodes, allowing computation to be performed directly on data without consuming or interfering with the storage node’s resources and operations. 

%

Recent works have explored the intersection of cloud and HPC, investigating how cloud technologies can be used to deploy and orchestrate applications on supercomputers, and conversely, how cloud infrastructures can handle workloads traditionally associated with HPC systems.
In this context, Arevalo et al.~\cite{arevalo_enhancing_2024} proposed a serverless architecture for deploying highly parallel HPC applications on supercomputers using Lithops~\cite{9619947}. The authors extended Lithops to support Singularity and integrate with existing HPC software stacks such as SLURM, evaluating their solution on the MareNostrum~5 supercomputer at the Barcelona Supercomputing Center. Experimental evaluation showed that, while the serverless execution model improves scalability and performance, data communication and shared memory access remain major bottlenecks. Thus, further research on direct inter-function communication is needed to establish serverless computing as a practical alternative to traditional batch systems in HPC.
%
Orthogonally, Hoefler et al.~\cite{hoefler_xaas_2024} introduced \textit{Acceleration-as-a-Service (XaaS)}, a serverless architecture that unifies cloud and HPC resources to support resource-intensive, performance-sensitive workloads.
The architecture leverages containers, high-performance data communication (e.g., RDMA), and advanced resource management to efficiently schedule and execute heterogeneous tasks. By bridging the gap between HPC and cloud computing, XaaS aims to deliver both scalability and low-latency performance in applications such as real-time analytics, large-scale simulations, and ML~inference.


Overall, while these works demonstrate the potential of domain-specific serverless architectures, they are largely limited to conceptual designs with preliminary prototype evaluations. This highlights the potential of serverless in these contexts, but underscores the need for further research to move from conceptual results towards practical implementations.

\begin{table*}[htbp]
\footnotesize
\centering
\caption{Summary of studies on programming models and paradigms for serverless computing.}
\label{tab:serverless_programming_models}
\rowcolors{2}{white}{lightaccessblue}
\begin{tabularx}{\textwidth}{>{\raggedright\arraybackslash}p{2.60cm} >{\raggedright\arraybackslash}X >{\raggedright\arraybackslash}p{3.00cm} >{\raggedright\arraybackslash}l}
\toprule
{\footnotesize\sffamily\bfseries\textcolor{accessblue}{Study}} & 
{\footnotesize\sffamily\bfseries\textcolor{accessblue}{Proposed Solution}} & 
{\footnotesize\sffamily\bfseries\textcolor{accessblue}{Strategy}} & 
{\footnotesize\sffamily\bfseries\textcolor{accessblue}{Focus}} \\
\midrule
{\contentfont Pauloski et al.~\cite{pauloski_accelerating_2023}} & 
{\contentfont Propose a pass-by-reference model with object proxies for data access.} & 
{\contentfont Abstraction of storage and transfer backends} & 
{\contentfont Data Communication} \\

{\contentfont Copik et al.~\cite{copik_process-as--service_2024}} & 
{\contentfont Introduce Process-as-a-Service (PraaS) abstraction for persistent state and direct communication.} & 
{\contentfont State persistence and direct communication} & 
{\contentfont Data Communication} \\

{\contentfont Gupta et al.~\cite{gupta_serverless_2020}} & 
{\contentfont Use error-correcting codes to mitigate stragglers in matrix multiplication.} &  
{\contentfont Error-correcting codes for fault tolerance} & 
{\contentfont Data Communication} \\

{\contentfont Copik et al.~\cite{copik_fmi_2023} and Yuan et al.~\cite{yuan_smpi_2022}} & 
{\contentfont Provide MPI-like communication interface for direct inter-function communication in FaaS applications.} & 
{\contentfont MPI-like communication abstraction} & 
{\contentfont Message Passing Model} \\

{\contentfont Mathew et al.~\cite{mathew_pattern-based_2024}} & 
{\contentfont Design platform-agnostic patterns for modeling serverless workflows.} & 
{\contentfont Pattern-based modeling} & 
{\contentfont Data Processing Workflows} \\

{\contentfont Heus et al.~\cite{heus_transactions_2022}} & 
{\contentfont Develop a transactional workflow model with coordinator functions for stateful serverless workflows.} & 
{\contentfont 2PC and Saga coordination patterns} & 
{\contentfont Stateful Workflows} \\

{\contentfont Li et al.~\cite{li_rethinking_2023}} & 
{\contentfont Propose an “m-to-n” model for co-locating functions in shared sandboxes and mitigate cold start.} & 
{\contentfont Function co-location optimization} & 
{\contentfont Cold Start Mitigation} \\

{\contentfont Ejarque et al.~\cite{ejarque_enabling_2022}} & 
{\contentfont Propose a serverless service model optimized for HPC workflow execution on heterogeneous systems.} & 
{\contentfont HPC-oriented serverless paradigm} & 
{\contentfont Workflow Orchestration} \\

{\contentfont Wang et al.~\cite{wang_briskchain_2024}} & 
{\contentfont Offer a decentralized model for direct inter-function communication in serverless workflows.} & 
{\contentfont Worker-side orchestration approach} & 
{\contentfont Workflow Orchestration} \\
\bottomrule
\end{tabularx}
\end{table*}

\subsubsection{Paradigms and Programming Models}
Serverless computing is a promising paradigm for deploying large-scale parallel and distributed scientific and AI/ML applications. 
However, existing service and programming models are primarily designed for cloud-native workloads and often fail to meet the strict requirements of HPC, such as low-latency data movement and efficient orchestration. Although recent efforts mitigate some of these limitations, for instance, through faster interconnects or optimized communication protocols, architectural improvements alone are insufficient. New programming models and paradigms are needed to effectively capture parallelism, orchestrate workflows, and coordinate complex distributed computations in serverless environments. The following paragraphs review recent advances in this direction, with selected contributions summarized in Table~\ref{tab:serverless_programming_models}.

\paragraph{Data Communication}
While serverless computing has been proven to support highly parallel and distributed workloads, data and inter-function communication remain significant bottlenecks. To overcome these, several programming abstractions have been proposed to facilitate data movement in serverless applications, aiming at hiding from the developer the low-level details of underlying storage and communication layers.

As discussed in Section~\ref{subsec:service_models}, FaaS separates computation from storage. In practice, this means that storage access and inter-function communication are achieved via \textit{mediated} channels, often through cloud storage or other indirect mechanisms~\cite{copik_fmi_2023}. 
Pauloski et al.~\cite{pauloski_accelerating_2023} proposed a pass-by-reference programming model for serverless and federated applications, introducing the concept of \textit{object proxies}. These abstract the underlying storage or communication backend, exposing a simple reference interface to developers that is resolved lazily by functions when accessed.
This programming model was then implemented in \textit{ProxyStore}, a framework targeting Python applications that supports various storage and transfer systems (e.g., Kafka, UCX, and Redis), simplifying data movement in serverless applications.  
%
%
%
While ProxyStore provides a flexible abstraction across heterogeneous backends, it still relies on mediated channels, which can introduce overhead in workloads with frequent inter-function communication.

The separation of data and computation in FaaS is inefficient for certain workloads and cannot be fully addressed by composing stateless functions with remote cloud services.
To address this, Copik et al.~\cite{copik_process-as--service_2024} introduced \textit{Process-as-a-Service (PraaS)}, an abstraction defining long-lived cloud processes that expose persistent state and direct communication primitives to functions. 
While its persistence model recalls the \textit{virtual actors} in Orleans~\cite{bernstein2014orleans}, PraaS processes more closely resemble OS-level processes in a serverless environment. 
When resources are constrained, the system transparently swaps the process state, including user-defined objects and files, to disk or cloud storage. Data communication is done via \textit{direct} channels through a minimal message-passing interface (\texttt{send}/\texttt{recv}) that abstracts the underlying transport layer. 
%
%
%
In contrast to ProxyStore, which provides an abstraction over mediated channels, PraaS extends traditional FaaS by replacing ephemeral functions with persistent processes. This enables low-latency state access, fast invocations that bypass the control plane, and efficient, direct inter-process communication. However, PraaS is not fully FaaS-compliant and requires dedicated platform support, limiting portability across different serverless environments.

Recent research has also explored strategies for mitigating performance degradation due to stragglers, which can have a significant impact on the performance of HPC and ML applications. Gupta et al.~\cite{gupta_serverless_2020} proposed a model for straggler mitigation and fault tolerance targeting matrix multiplication. This approach, based on error-correcting codes, employs parallel encoding and decoding over cloud-stored data using serverless workers, thereby reducing computation, communication, and storage bottlenecks by reconstructing outputs from stragglers or faulty serverless workers.
While effective, this solution is highly specialized for matrix operations, highlighting the need for more general strategies to mitigate stragglers in broader classes of serverless HPC and ML workloads.

\paragraph{Message Passing Models}
Recent research has investigated how message-passing paradigms, traditionally used in HPC, can be adapted to serverless environments to support high-performance workloads. Several efforts have focused on designing MPI-inspired programming models to general-purpose FaaS platforms.

Yuan et al.~\cite{yuan_smpi_2022} proposed a two-level parallelism scheme to support MPI programming in serverless. It divides parallelism at function and worker levels,  where the function level reflects the number of function instance replicas (named \textit{parallel replicas}), and the worker level denotes the degree of intra-function parallelism within each instance.
The authors implemented this scheme in a framework built atop OpenFaaS, named \textit{SMPI}. It enables MPI programs to run serverlessly by launching MPI processes based on the number of workers within each function instance. Intra-function communication is efficiently managed by the MPI runtime, but inter-function coordination is mediated by OpenFaaS’s gateway and scheduler. This tight integration, coupled with the lack of communication optimizations, limits performance scalability, particularly in multi cluster scenarios, and portability across FaaS platforms.

Targeting commercial FaaS platforms, Copik et al.~\cite{copik_fmi_2023} introduced \textit{FaaS Message Interface (FMI)}, a message-passing library built atop AWS Lambda that provides MPI-like point-to-point and collective communication operations.
%
FMI defines two types of  channels: mediated channels, implemented through object storage or indirect methods (e.g., AWS S3, Redis, and AWS DynamoDB) like ProxyStore~\cite{pauloski_accelerating_2023}, and direct channels, enabled via a TCP hole-punching mechanism.
Direct channels enable developers to bypass mediated paths when low-latency or high-throughput communication is required.
The authors evaluated FMI's collective operations against OpenMPI's equivalents, achieving comparable performance, particularly when relying on direct TCP communication. 

Taken together, these efforts show the promise of adapting MPI-like abstractions to FaaS. Both SMPI and FMI exploit direct communication to approach the low-latency and high-throughput demands of HPC and AI applications. Yet, achieving this requires tight coupling to specific runtimes or platforms, which may limits their generality and portability across different serverless infrastructures.

\paragraph{Workflows}
The current literature has shown an increasing interest in programming models for workflow-based applications. Many applications naturally decompose into modular, reusable, and independently scalable stages, making them particularly well-suited to workflow representation. This decomposition seamlessly aligns with the serverless execution model, making such applications a natural fit for this paradigm.

In HPC, scientific pipelines, such as those for digital twins, climate modeling, and urgent computing, typically consist of multiple stages, including simulation, data analysis, and AI training or inference. Consequently, these applications are increasingly modeled as workflows.
In this context, Ejarque et al.~\cite{ejarque_enabling_2022} introduced \textit{HPC Workflow-as-a-Service (HPCWaaS)}, a serverless service model targeting complex workflows on heterogeneous HPC systems. HPCWaaS follows FaaS principles but specifically targets workflows in federated HPC environments rather than general-purpose applications. By abstracting infrastructure management, it extends serverless principles to scientific computing, allowing users to trigger complex workflows via simple REST API calls.
To support HPCWaaS, the authors developed eFlows4HPC, a dedicated software stack that integrates tools for workflow development, data logistics, and dynamic execution. By automating deployment through orchestration standards (e.g., TOSCA) and runtime systems (e.g., PyCOMPSs), HPCWaaS provides flexible execution while addressing portability, reusability, and reproducibility in scientific workflows across federated infrastructures.


Despite the active development of dedicated platforms for deploying serverless applications on HPC infrastructures, general-purpose FaaS platforms remain appealing for workloads that do not require specialized hardware or dedicated resources, as they reduce both operational complexity and execution costs.
However, the heterogeneity of programming models across FaaS platforms significantly hinders application portability. Each platform employs distinct paradigms for composing functions and orchestrating workflows, complicating migration and interoperability and ultimately leading to vendor lock-in. To address these challenges, Matthew et al.~\cite{mathew_pattern-based_2024} proposed a platform-agnostic workflow modeling framework based on \textit{Enterprise Integration Patterns (EIPs)}. The framework defines 20 patterns for serverless data processing pipelines, targeting big data, real-time analytics, eScience, and data science applications. The proposed framework delineates the essential components for data pipeline construction, mapping them onto the underlying FaaS implementation to provide a solid methodological foundation that mitigates vendor lock-in and promotes interoperability across serverless~platforms.

Another open concern is support for transactional workflows in stateful applications. FaaS platforms such as \textit{Azure Durable Functions} and frameworks like \textit{Beldi} support stateful computations, however their capabilities for transactional workflows remain limited. Heus et al.~\cite{heus_transactions_2022} proposed a programming model for orchestrating transactional workflows across stateful serverless functions, targeting serializable and eventual consistency guarantees. It introduces \textit{coordinator functions}, entities capable of managing transactions through either the two-phase commit protocol or the Saga pattern. These abstractions enable distributed transactions without burdening developers with low-level error handling or consistency concerns.
The authors implemented the model on \textit{Apache Flink StateFun}, a stateful dataflow system with exactly-once semantics. Experimental evaluations showed that their approach achieves higher throughput than CockroachDB and Beldi, albeit with increased latency due to checkpointing for exactly-once processing, demonstrating the potential of serverless for stateful stream processing applications.

In addition to state management, performance overheads from isolation motivate new deployment models.
Typically, serverless platforms enforce strong isolation by instantiating each function in a dedicated environment, improving security and fault tolerance but introducing overhead from cold starts, inter-function communication, and runtime management.
To mitigate these inefficiencies, Li et al.~\cite{li_rethinking_2023} proposed the \textit{``m-to-n''} deployment model. In contrast to the more conventional \textit{``one-to-one''} model (as adopted by platforms like AWS Step Functions), and the \textit{``many-to-one''} model, 
which co-locates multiple functions in a single sandbox, 
the ``m-to-n'' model maps $m$ functions to $n$ execution units. 
This mapping reduces overhead by strategically partitioning workflows and co-locating functions, thereby lowering startup latency and communication costs.
%
%
To support the ``m-to-n'' deployment model, the authors introduced \textit{wraps}, an abstraction that encapsulates a subset of workflow functions and defines the granularity of deployment, and evaluated it through \textit{Chiron}, a deployment manager built on OpenFaaS. Experiments against platforms such as AWS Step Functions, OpenFaaS, SAND, and Faastlane demonstrated performance improvements in latency, throughput, and memory usage, demonstrating the model’s effectiveness. Nonetheless, identifying optimal function partitioning and execution modes in serverless workflows remains an open challenge for future research.

Finally, orchestration itself has become a central bottleneck. Most serverless workflow engines adopt a master-side orchestration model, where a centralized controller schedules functions and manages state propagation. While simplifying coordination, this design prevents direct inter-function communication and introduces overhead from repeated interactions with the master, creating bottlenecks in high-performance or data-intensive workflows.  
Hence, achieving optimal function composition remains challenging—a problem commonly referred to as the serverless trilemma~\cite{10.1145/3133850.3133855}.
To overcome these limitations, Wang et al.~\cite{wang_briskchain_2024} proposed a decentralized \textit{worker-side} orchestration pattern  
for serverless workflows. In this model, each function transmit its output directly to the subsequent function, eliminating the need for mediated channels or master coordination. 
The authors implemented this approach in \textit{BriskChain}, a serverless workflow engine based on OpenWhisk. Experimental evaluations against OpenWhisk and Apache Composer demonstrated a significant reduction in communication overhead and latency, highlighting the potential of decentralized orchestration for serverless workflows.

\begin{table*}[ht]
\footnotesize
\centering
\caption{Summary of studies on platforms for high-performance serverless computing.}
\label{tab:platform}
\rowcolors{2}{white}{lightaccessblue}
\begin{tabularx}{\textwidth}{>{\raggedright\arraybackslash}p{2.60cm} >{\raggedright\arraybackslash}X >{\raggedright\arraybackslash}l}
\toprule
{\footnotesize\sffamily\bfseries\textcolor{accessblue}{Study}} & 
{\footnotesize\sffamily\bfseries\textcolor{accessblue}{Proposed Solution}} & 
{\footnotesize\sffamily\bfseries\textcolor{accessblue}{Focus}} \\
\midrule
{\contentfont Shillaker et al.~\cite{shillaker_faasm_2020}} & 
{\contentfont Develop a lightweight isolation abstraction that supports stateful functions through direct memory sharing.} & 
{\contentfont Stateful Computations} \\

{\contentfont Chen et al.~\cite{chen_yuanrong_2024}} & 
{\contentfont Introduce a general-purpose serverless platform targeting a vast array of computing-demanding workloads.} & 
{\contentfont High-performance Workloads} \\

{\contentfont Chard et al.~\cite{chard_funcx_2020} and Li et al.~\cite{li_funcx_2022}} & 
{\contentfont Develop a distributed serverless platform for heterogeneous, federated infrastructures targeting scientific applications.} & 
{\contentfont Federated Resources} \\

{\contentfont Copik et al.~\cite{copik_rfaas_2023,copik_software_2024}} & 
{\contentfont Use RDMA for fast function invocation and task offloading in HPC applications.} & 
{\contentfont Low-latency Invocation} \\

{\contentfont Petrosyan et al.~\cite{petrosyan_serverless_2022, petrosyan_optimizing_2023}} & 
{\contentfont Propose a serverless platform based on Kubernetes and containerization to support compute-intensive workloads on HPC clouds.} & 
{\contentfont HPC Cloud Deployment} \\

{\contentfont Nanos et al.~\cite{nanos_hardware-accelerated_2023}} & 
{\contentfont Present a hardware-accelerated platform supporting serverless execution on heterogeneous edge-cloud resources using vAccel and microVMs.} & 
{\contentfont Accelerator Integration} \\

{\contentfont Nguyen et al.~\cite{nguyen_qfaas_2024}} & 
{\contentfont Introduce a serverless framework for deploying quantum functions with multi-SDK and quantum backend support.} & 
{\contentfont Quantum-as-a-Service (QaaS)} \\
\bottomrule
\end{tabularx}
\end{table*}

\subsubsection{Platforms and Frameworks}
While the current serverless landscape includes a broad range of commercial and open-source platforms, most target general-purpose cloud workloads and fall short for HPC, AI, or big data applications. These domains require platforms that deliver high-performance through efficient parallel task orchestration and low-latency communication, while exploiting the architectural features of HPC systems—capabilities largely absent in existing cloud-oriented solutions. Additionally, big data analytics, AI/ML, and scientific computing demand specialized frameworks to address their specific execution requirements. Thus, recent research has focused on serverless platforms and frameworks tailored to high-performance workloads.
The following paragraphs review these efforts, while Tables~\ref{tab:platform}, \ref{tab:general-purpose-frameworks}, and \ref{tab:domain-specific} summarize the corresponding studies.

\newcommand{\mt}[2]{\makecell[c]{#1 \\ #2}}
\newcommand{\mtt}[3]{\makecell[c]{#1 \\ #2 \\ #3}}
\newcommand{\ms}[1]{\makecell[c]{#1}}

\paragraph{Platforms}
%
Serverless computing is emerging as a promising execution model for big data, AI, and HPC applications across cloud, HPC, and hybrid infrastructures. 
Nevertheless, effectively supporting parallel compute-intensive workloads presents two main challenges: data and communication overheads, and the runtime footprint of execution environments. While lightweight sandboxing technologies mitigate the latter, efficient data access and inter-function communication remain critical bottlenecks. Furthermore, most existing platforms target FaaS, emphasizing stateless, event-driven execution. Yet many applications require stateful execution and, particularly in workflow-based ones, low-latency communication.

One approach to support large-scale parallelism and low-latency data access is to enable direct memory sharing between serverless functions.
In this context, Shillaker et al.~\cite{shillaker_faasm_2020} introduced \textit{Faasm}, a distributed serverless runtime targeting data-intensive workloads that leverages WebAssembly. 
Although WebAssembly has been shown to be a suitable lightweight execution environment for serverless computing~\cite{9214403, kjorveziroski_webassembly_2023,gackstatter_pushing_2022,hall_execution_2019}, it lacks native mechanisms for direct memory sharing. 
To address this, Faasm introduces \textit{Faaslets}, a lightweight isolation abstraction that allows stateful functions to share memory regions within the same address space. The authors also extended WASI to ease porting of OpenMP and MPI applications~\cite{shillaker_faabric_2023,305991}.
%
To reduce cold start latency, Faasm employs a snapshot-based technique, similar to \textit{SEUSS}~\cite{10.1145/3342195.3392698}, based on \textit{Proto-Faaslets}, snapshots of arbitrary execution states that can be restored into a new Faaslet on any host across the cluster. 
%
%
Evaluation on ML workloads against Knative, a Kubernetes-based serverless platform, showed improved execution times and lower memory footprint. Comparisons with Docker containers demonstrated that Faaslets and Proto-Faaslets greatly reduce cold start latency.
Despite its promise, WebAssembly still offers limited support for key features, including core libraries (e.g., BLAS and LAPACK)~\cite{shillaker_faasm_2020}, accelerator integration~\cite{305991}, and multithreading\footnote{Despite ongoing efforts in formalizing threads in WebAssembly specifications. See \url{https://github.com/WebAssembly/shared-everything-threads}.}, which limits Faasm’s applicability to diverse compute-intensive workloads.

%
Alternatively, other efforts aim for platforms with broad workload support, addressing both general-purpose and high-performance workloads.
For instance, Chen et al.~\cite{chen_yuanrong_2024} presented \textit{YuanRong}, a general-purpose serverless platform adopted by Huawei. YuanRong supports a broad range of workloads, including big data analytics, ML/DL training and inference, and HPC. 
It exposes a unified programming interface, the \textit{Open Serverless Interface (OSI)}, which provides system call-like abstractions for function management, interaction between stateless and stateful functions, and data access via objects and streams by abstracting underlying storage systems and BaaS platforms. Functions communicate through an actor-like abstraction that enables RPC-based invocation with data passed by reference, reducing transfer overheads.
%
%
%
Nonetheless, YuanRong’s proprietary nature and the lack of disclosed architectural details hinder its comparison with other platforms, highlighting the research gap in developing open, high-performance serverless platforms.

Recent research has also focused on serverless platforms for scientific applications, enabling scientists and researchers to leverage the benefits of serverless computing in terms of scalability, flexibility, and abstraction from resource management.
Chard et al.~\cite{chard_funcx_2020} introduced \textit{funcX} (now \textit{Globus Compute}), a federated serverless platform designed to support the deployment of scientific applications across heterogeneous infrastructures, including, but not limited to, HPC systems. 
Its architecture addresses scientific computing requirements by prioritizing low-latency execution, integration with institutional authentication, and interoperability across diverse computational resources.
To manage federated resources, funcX decouples the control and execution planes. The control plane consists of a cloud-hosted management service that exposes a REST API for function registration, invocation, and monitoring, while the execution plane relies on funcX endpoints that deploy functions locally and independently manage resource allocation and load balancing.
Function execution is container-based, with support for Docker, Singularity, and Shifter.
Further, Li et al.~\cite{li_funcx_2022} extended funcX and evaluated its performance and scalability over 130,000 concurrent workers, achieving performance comparable to commercial FaaS platforms and confirming its suitability for large-scale scientific workloads.
Nevertheless, its reliance on Redis and shared file systems for data transfer may introduce overheads for applications with frequent inter-function communication.

One of the main barriers to serverless adoption in HPC is the lack of support for high-speed communication typical of modern HPC clusters.
Addressing this, Copik et al.~\cite{copik_rfaas_2023} proposed \textit{rFaaS}, an RDMA-accelerated FaaS platform targeting compute-intensive applications. It employs RDMA to reduce communication latency during function invocation, enabling direct memory operations between clients and serverless function executors, thus circumventing TCP/IP networking overheads.
rFaaS makes use of \texttt{ibverbs}, a user-level RDMA communication library, for managing high-throughput, low-latency network operations, and employs Docker as an execution environment.
To facilitate its use in HPC contexts, rFaaS provides a programming interface inspired by the C++23 \textit{executors}\footnote{Further information about C++23 executors is available at \url{https://www.open-std.org/jtc1/sc22/wg21/docs/papers/2018/p0443r9.html}.}.
This enables applications to offload tasks to remote serverless function executors, enabling a dynamic increase in computing resources when required. 
%
%
In follow-up work, Copik et al.~\cite{copik_software_2024} extended rFaaS with \texttt{libfabric} to support Cray systems using
\textit{uGNI}, and added support for Singularity and Sarus as container-based execution environments. Both offer native support to HPC systems, enabling access to compute and I/O devices, integration with job schedulers such as Slurm, and out-of-the-box support for MPI applications.
Compared to state-of-the-art, rFaaS achieves single-digit microsecond invocation latency, far outperforming platforms such as OpenWhisk, which operate in the hundreds of milliseconds. Leveraging RDMA is a promising direction for future work to enable low-latency, direct inter-function communication required by tightly coupled HPC applications.

In recent years, the use of \textit{HPC cloud} has increased, combining HPC capabilities with the scalability and flexibility of traditional clouds~\cite{10.1145/3150224}. 
Rather than replacing on-premises systems, HPC clouds are typically used in hybrid deployments, where private clusters are extended with public cloud resources for \textit{cloud bursting}. This model enables dynamic provisioning during peak loads, yet challenges in workload orchestration and resource management remain. The serverless paradigm has emerged as a promising approach to address these issues.
%
For instance, Petrosyan et al.~\cite{petrosyan_serverless_2022, petrosyan_optimizing_2023} introduced \textit{Shoc}, an under-development serverless platform designed for compute-intensive workloads on HPC clouds. By leveraging Kubernetes and utilizing Docker and Singularity as execution environments, Shoc enables serverless execution of compute- and data-intensive workloads over cloud, abstracting the complexities of deploying HPC infrastructures.
Its architecture includes a CLI-based frontend, a container registry, an application builder that automates container image creation for HPC applications (e.g., MPI, Spark), and a Kubernetes-integrated executor that dynamically assigns workloads to suitable nodes.
Although still in its early phases, Shoc demonstrates the potential of serverless as a unifying paradigm for both cloud and HPC, facilitating seamless use of cloud infrastructures to run HPC workloads. Yet, its immaturity and limited evaluation currently prevent meaningful comparison with existing platforms.

Furthermore, recent literature has investigated the integration of hardware accelerators into serverless platforms. In this context, Nanos et al.~\cite{nanos_hardware-accelerated_2023} presented \textit{SERRANO}, a platform that supports the deployment of hardware-accelerated serverless functions across heterogeneous resources, 
combining edge devices, located near end users, with HPC infrastructures capable of supporting compute-intensive workloads. 
To enable seamless offloading of tasks to accelerators, the authors extended the \textit{vAccel} framework~\footnote{vAccel is available at \url{https://github.com/nubificus/vaccel}.}, which decouples applications from hardware-specific code by exposing pre-defined, hardware-accelerated functions. Deployment of vAccel-enabled applications is supported through the use of microVMs, specifically via
\textit{Kata Containers}~\footnote{Kata Containers is available at \url{https://github.com/kata-containers/kata-containers}.}, a lightweight VMM implemented in Rust and built on top of \texttt{rustvmm}, offering strong isolation and security while preserving the simplicity of traditional containerization.
Despite the promises of accelerator-aware serverless platforms, aspects such as scheduling and resource sharing remain largely unresolved.

Finally, the rapid evolution of quantum computing is driving the need for specialized cloud solutions tailored to its requirements. The emergence of \textit{Quantum-as-a-Service (QaaS)}~\cite{AHMAD2024102094}, exemplified by \textit{IBM Quantum}~\cite{ibm_quantum} and \textit{AWS Braket}~\cite{aws_braket}, allows users to access quantum resources via third-party providers, abstracting infrastructure management. However, it introduces challenges such as vendor lock-in, hindering the portability of applications across platforms and SDKs.
%
%
To address these limitations, Nguyen et al.~\cite{nguyen_qfaas_2024} proposed \textit{QFaaS}, a serverless platform for deploying quantum applications. QFaaS leverages the serverless model to consolidate quantum resources in multi-tenant environments and enables hybrid workflows combining classical and quantum functions, relevant for quantum machine learning.
%
QFaaS enables deployment of quantum functions as cloud-accessible services. It supports multiple quantum SDKs, ensuring compatibility with existing tools and libraries, and allows function execution on both self-hosted quantum simulators and third-party providers backed by physical quantum hardware.
%
%
To mitigate cold starts, QFaaS enforces a keep-alive timeout on active function instances and leverages cache-based mechanisms. Specifically, it caches the QASM of previously transpiled quantum circuits to avoid recompilation upon repeated invocations of the same quantum function. 
While this work, the first of its kind to our knowledge, shows promise for developing quantum and hybrid quantum-cloud serverless applications, research in this field remains in its infancy, and further efforts are needed to make this emerging model production-ready.

\begin{table}[ht]
\scriptsize
\centering
\caption{Feature comparison of serverless platforms.}
\label{tab:platform_comparison}
\rowcolors{2}{lightaccessblue}{white}
\begin{tabularx}{\columnwidth}{>{\raggedright\arraybackslash}p{1.95cm} c c >{\raggedright\arraybackslash}c c }
\toprule
{\scriptsize\sffamily\bfseries\textcolor{accessblue}{Platform}} & 
{\scriptsize\sffamily\bfseries\textcolor{accessblue}{\mt{Stateful}{Support\textsuperscript{†}}}} & 
{\scriptsize\sffamily\bfseries\textcolor{accessblue}{\mt{Low-Lat.}{Comm.}}} & 
{\scriptsize\sffamily\bfseries\textcolor{accessblue}{\mt{Exec.}{Env.}}} & 
{\scriptsize\sffamily\bfseries\textcolor{accessblue}{Accel.}} \\
\midrule
{\contentfont Faasm~\cite{shillaker_faasm_2020}} & 
{\contentfont \cmark} & 
{\contentfont \cmark} & 
{\contentfont Faaslets} & 
{\contentfont \xmark} \\

{\contentfont YuanRong~\cite{chen_yuanrong_2024}} & 
{\contentfont \cmark} & 
{\contentfont \mt{\cmark}{(RPC by ref.)}} & 
{\contentfont \mt{Baremetal,}{Containers}} & 
{\contentfont \xmark} \\

{\contentfont funcX~\cite{chard_funcx_2020,li_funcx_2022}} & 
{\contentfont \xmark} & 
{\contentfont \xmark} & 
{\contentfont \mtt{Singularity,}{Docker,}{Shifter}} & 
{\contentfont GPUs} \\

{\contentfont rFaaS~\cite{copik_rfaas_2023,copik_software_2024}} & 
{\contentfont \xmark} & 
{\contentfont \mt{\cmark}{(RDMA)}} & 
{\contentfont \mtt{Singularity,}{Docker,}{Sarus}} & 
{\contentfont \xmark} \\

{\contentfont Shoc~\cite{petrosyan_serverless_2022, petrosyan_optimizing_2023}} & 
{\contentfont \xmark} & 
{\contentfont n/s} & 
{\contentfont \mt{Singularity,}{Docker}} & 
{\contentfont GPUs} \\

{\contentfont SERRANO~\cite{nanos_hardware-accelerated_2023}} & 
{\contentfont \xmark} & 
{\contentfont \xmark} & 
{\contentfont \mt{Kata}{Containers}} & 
{\contentfont \mt{GPUs}{FPGAs}} \\

{\contentfont QFaaS~\cite{nguyen_qfaas_2024}} & 
{\contentfont \xmark} & 
{\contentfont \xmark} & 
{\contentfont Docker} & 
{\contentfont \xmark} \\
\bottomrule
\end{tabularx}
\smallskip

\hspace{0.15cm}
\footnotesize{\contentfont n/s: not specified; \cmark: yes; \xmark: no; $\dagger$: provided by the platform.}
\end{table}

Overall, while these platforms address different aspects of performance, usability, and in some cases specific domains, their capabilities remain fragmented, with no single solution meeting the full requirements of large-scale, compute-intensive applications. Features such as low-latency  invocation and communication, native stateful function support, and accelerator integration are not consistently provided. Table~\ref{tab:platform_comparison} summarizes the main characteristics of the surveyed platforms, underscoring the gaps that motivate further research.

\begin{table*}[ht]
\footnotesize
\centering
\caption{Summary of studies on serverless frameworks for general-purpose HPC and scientific applications.}
\label{tab:general-purpose-frameworks}
\rowcolors{2}{white}{lightaccessblue}
\begin{tabularx}{\textwidth}{>{\raggedright\arraybackslash}p{3.50cm} >{\raggedright\arraybackslash}X >{\raggedright\arraybackslash}l}
\toprule
{\footnotesize\sffamily\bfseries\textcolor{accessblue}{Study}} & 
{\footnotesize\sffamily\bfseries\textcolor{accessblue}{Proposed Solution}} & 
{\footnotesize\sffamily\bfseries\textcolor{accessblue}{Target Object}} \\
\midrule
{\contentfont Pérez et al.~\cite{perez_serverless_2018}} & 
{\contentfont Propose a container-aware framework for the deployment of parallel event-driven scientific applications.} & 
{\contentfont Scientific Applications} \\

{\contentfont Arif et al.~\cite{arif_canary_2022}} & 
{\contentfont Devise a checkpoint-restore fault-tolerance framework for stateful functions.} & 
{\contentfont Fault Tolerance} \\

{\contentfont Mohanty et al.~\cite{mohanty_faastloop_2024}} & 
{\contentfont Leverage loop-level optimizations and adaptive code deployment in serverless HPC workloads.} & 
{\contentfont Performance Optimization} \\

{\contentfont Jana et al.~\cite{jana_dagit_2023}} & 
{\contentfont Develop a framework for workflow lifecycle management on FaaS platforms.} & 
{\contentfont Serverless Workflows} \\

{\contentfont Carver et al.~\cite{carver_wukong_2020}} & 
{\contentfont Present a framework for the development of workflow-modeled applications on public cloud platforms.} & 
{\contentfont Serverless Workflows} \\

{\contentfont Li et al.~\cite{li_unifaas_2024}} & 
{\contentfont Introduce a framework for distributed scientific workflows with dynamic scheduling and profiling.} & 
{\contentfont Scientific Workflows} \\

{\contentfont Related Studies~\cite{malawski_serverless_2020,andrei_da_silva_enabling_2025,thurimella_serverless_2025}} & 
{\contentfont Integrate existing workflow engines with serverless platforms to enable scalable task-parallel scientific workflows.} & 
{\contentfont Scientific Workflows} \\

{\contentfont Roy et al.~\cite{roy_daydream_2022, basu_roy_starship_2024,roy_mashup_2022}} & 
{\contentfont Address bottlenecks in serverless workflows via cold start mitigation, function scheduling, and resource provisioning.} & 
{\contentfont Performance Optimization} \\
\bottomrule
\end{tabularx}
\end{table*}

\paragraph{General Purpose Frameworks}
High-performance applications often impose strict requirements on execution models, performance, and programmability. General-purpose serverless platforms tend to fall short of these demands due to limits on available programming languages, libraries, and execution environments. This increases the complexity of porting HPC, AI, and big data applications to serverless infrastructure. 
To bridge this gap, recent research has introduced frameworks that extend these platforms to adapt them to the needs of such large-scale workloads.
%

Despite the promise of serverless for highly parallel scientific workloads, commercial FaaS platforms are often not tailored to such applications. Pérez et al.~\cite{perez_serverless_2018} introduced \textit{SCAR}, a framework designed to simplify the deployment of highly parallel, event-driven applications on AWS Lambda by leveraging custom Docker-based runtimes. 
SCAR provides support to custom execution environments, enabling the use of programming languages not natively supported by the serverless platform. It achieves that by introducing \textit{SCAR Supervisor}, a lightweight agent embedded within the function that retrieves Docker images into the function’s ephemeral storage and orchestrates container instantiation. 
%
To overcome the lack of native Docker support in AWS Lambda, SCAR integrates \texttt{udocker}\footnote{udocker is available at \url{https://github.com/indigo-dc/udocker}.}, enabling container execution in user space without requiring root privileges. 
Further, SCAR introduces a specialized programming model for \textit{High Throughput Computing} workloads, enabling highly parallel applications, such as scientific computing, video transcoding, and image processing, to run efficiently on general-purpose FaaS platforms, thus extending the applicability of serverless computing to domains traditionally reliant on IaaS or HPC infrastructures.

Another problem in current platforms is the lack of robust fault-tolerance mechanisms to handle hardware and software failures. While retry-based approaches offer simple-to-implement strategies for failure recovery, they may not be ideal for long-running tasks such as those found in ML/DL and scientific workflows. To address this, Arif et al.~\cite{arif_canary_2022} proposed \textit{Canary}, 
a framework designed to improve fault tolerance in general-purpose FaaS platforms by mitigating the impact of stateful function failures. 
It employs a checkpoint-restore mechanism that tracks the state of each function during its execution and, in case of failure, loads the latest checkpoint into a new function instance. Evaluation of Canary on OpenWhisk using the \textit{SeBS} benchmark~\cite{10.1145/3464298.3476133} demonstrated faster recovery times compared to default retry-based strategies.

In the context of HPC workloads, applications often consist of large portions of code structured as loops, typically referred to as kernels. In traditional non-serverless environments, this structure enables a wide range of code optimizations. Depending on data dependencies, loop iterations can be executed in parallel across multiple processes to reduce execution times. Additionally, iterations may be grouped into chunks (or blocks) based on the granularity of computation. Mohanty et al.~\cite{mohanty_faastloop_2024} investigated the feasibility of applying such optimizations in serverless environments by introducing \textit{FAAStloop}, a framework designed to improve the performance of HPC applications through loop-level optimizations and dynamic application profiling.
Developers annotate loops in their code using pragmas, and
FAAStloop automatically generates and selects optimized code variants based on user goals and runtime profiling.
Experiments on AWS Lambda showed up to $3.3\times$ speedup and $2.1\times$ cost reduction compared to baseline approaches, confirming the feasibility of such optimizations in serverless HPC contexts.

While serverless computing offers a promising execution model for deploying scientific workflows, effectively orchestrating them remains a challenge.  Consequently, recent works have addressed this by extending existing workflow engines originally designed for non-serverless environments.
One of the earliest contributions in this regard was presented by Malawski et al.~\cite{malawski_serverless_2020}, who integrated the \textit{HyperFlow}~\cite{BALIS2016147} workflow engine with AWS Lambda and Google Cloud Run functions to support task-level parallelism through a direct executor model, where tasks are dispatched as HTTP requests to cloud functions. Experimental evaluation demonstrated the feasibility and scalability of serverless execution for fine-grained, high-throughput workflows, while also highlighting limitations related to execution time constraints imposed by general-purpose serverless platforms and the need for hybrid models that integrate traditional IaaS resources to handle specific workloads. Similarly, da Silva et al.~\cite{andrei_da_silva_enabling_2025} proposed a Knative-based serverless backend integrated with the \textit{WfCommons} suite~\cite{wfcommons}. It introduces a Knative translator and deploys \textit{WfBench} as a service to generate and execute synthetic HPC workflows. Results showed that not all types of workflows benefit equally from serverless execution and that not all the steps of a workflow are necessarily well-suited for this model, emphasizing the need for hybrid models that combine serverless and traditional execution environments.
Thurimella et al.~\cite{thurimella_serverless_2025} explored a similar direction by integrating Knative with Pegasus. Experimental results showed that Knative execution achieves performance close to native execution while maintaining container-level isolation, making it a viable solution for executing highly parallel scientific workflows in shared HPC environments.

In contrast, other studies have proposed standalone solutions specifically designed for serverless environments.
Jana et al.~\cite{jana_dagit_2023} presented \textit{DAGit}, a serverless framework built on top of Apache OpenWhisk to support the end-to-end lifecycle of general-purpose DAG-based workflows. DAGit uses JSON-based specifications to register functions, workflows, and triggers, and its orchestrator supports a comprehensive set of workflow primitives.
Furthermore, Li et al.~\cite{li_unifaas_2024} proposed \textit{UniFaaS}, a serverless framework developed on top of the Globus Compute platform to support the deployment of scientific workflows across distributed cyberinfrastructures. UniFaaS extends Globus with a Python-based programming interface that supports task parallelism and DAG-modeled workflows, enabling tasks to be expressed as Python functions or as calls to external backends. The interface employs Python futures to represent ongoing function invocations and to support asynchronous programming. This mechanism is also employed to dynamically construct task graphs at runtime. Additionally, UniFaaS incorporates an execution profiler and a scheduler that support custom task mapping across available endpoints, offering multiple policies to optimize placement on heterogeneous and dynamic resources. 
%

Finally, various research efforts have focused on providing frameworks for deploying scientific workflows on public serverless platforms~\cite{roy_daydream_2022, basu_roy_starship_2024, carver_wukong_2020} or have explored hybrid execution models that map workflows on both FaaS and IaaS platforms~\cite{roy_mashup_2022}.
In this context, Carver et al.~\cite{carver_wukong_2020} presented \textit{Wukong}, a serverless framework designed to orchestrate DAG-modeled workflows on AWS Lambda. Wukong adopts a Python-based programming interface inspired by \textit{Dask} and reuses its DAG generator to simplify workflow expression and enable automatic translation into task graphs for parallel execution.
%
%
To support efficient data handling, Wukong manages intermediate and final data through a distributed storage cluster of Redis instances deployed on AWS Fargate, offering low-latency access and high throughput for both large and small objects, outperforming traditional solutions such as AWS S3.
%

\begin{table*}[ht]
\footnotesize
\centering
\caption{Summary of studies on serverless frameworks for domain-specific applications.}
\label{tab:domain-specific}
\rowcolors{2}{white}{lightaccessblue}
\begin{tabularx}{\textwidth}{>{\raggedright\arraybackslash}p{3.90cm} >{\raggedright\arraybackslash}X >{\raggedright\arraybackslash}l}
\toprule
{\footnotesize\sffamily\bfseries\textcolor{accessblue}{Study}} & 
{\footnotesize\sffamily\bfseries\textcolor{accessblue}{Proposed Solution}} & 
{\footnotesize\sffamily\bfseries\textcolor{accessblue}{Domain}} \\
\midrule
{\contentfont Giménez-Alventosa et al.~\cite{gimenez-alventosa_framework_2019}} & 
{\contentfont Develop a framework based on the MapReduce parallel computing pattern for handling distributed workloads.} & 
{\contentfont Big Data Analytics} \\

{\contentfont Müller et al.~\cite{muller_lambada_2020}} & 
{\contentfont Design a fully serverless framework for orchestrating distributed queries using cloud-native services.} & 
{\contentfont Big Data Analytics} \\

{\contentfont Skluzacek et al.~\cite{skluzacek_serverless_2021}} & 
{\contentfont Build a scalable and automated approach for extracting metadata from large, distributed repositories.} & 
{\contentfont Big Data Analytics} \\

{\contentfont Related Studies~\cite{sanchez-artigas_seer_2022,eizaguirre_seer_2024}} & 
{\contentfont Introduce a shuffle management system to optimize inter-function data exchange during analytical operations.} & 
{\contentfont Big Data Analytics} \\

{\contentfont Related Studies~\cite{zhang_mark_2019,zhang_enabling_2022,ali_batch_2020}} & 
{\contentfont Design a general-purpose serverless inference system for latency-sensitive ML tasks with batching optimizations.} & 
{\contentfont ML Inference} \\

{\contentfont Yang et al.~\cite{yang_infless_2022}} & 
{\contentfont Co-design a serverless framework and inference engine to natively support GPU-accelerated ML workloads.} & 
{\contentfont ML Inference} \\

{\contentfont Toader et al.~\cite{toader_graphless_2019}} & 
{\contentfont Implement a serverless graph processing system leveraging external storage and coordination services.} & 
{\contentfont Graph Processing} \\

{\contentfont Liu et al.~\cite{liu_faasgraph_2024}} & 
{\contentfont Propose a serverless-native architecture for efficient distributed graph processing with reduced communication overhead.} & 
{\contentfont Graph Processing} \\

{\contentfont Luckow et al.~\cite{luckow_performance_2019}} & 
{\contentfont Extend existing resource management systems to support stream processing workloads under the serverless paradigm.} & 
{\contentfont Stream Processing} \\

{\contentfont Rotchford et al.~\cite{rotchford_laminar_2024}} & 
{\contentfont Propose a modular serverless system for composing and executing streaming data workflows with low-latency communication.} & 
{\contentfont Stream Processing} \\

{\contentfont Kica et al.~\cite{kica_serverless_2023}} & 
{\contentfont Introduce serverless support for executing large-scale VVUQ workflows in scientific computing domains.} & 
{\contentfont Scientific Computing} \\
\bottomrule
\end{tabularx}
\end{table*}

\paragraph{Domain-specific Frameworks}
Serverless computing has proven to be a versatile paradigm capable of supporting a broad spectrum of workloads. Yet, certain domains, such as big data analytics, ML/DL inference and training, and others, have requirements that are not fully addressed by current general-purpose platforms and frameworks. Hence, recent research has introduced domain-specific frameworks tailored to the needs of these workloads.

One of the most promising use cases for serverless is big data analytics. As discussed in Section~\ref{subsec:characteristics_of_serverless}, its scalability and pay-as-you-go model make it particularly well suited for processing large and variable volumes of data.
%
Consequently, 
research has increasingly focused on designing serverless frameworks tailored for such workloads. One of the first efforts in this direction is \textit{MapReduce on Lambda (MARLA)}~\cite{gimenez-alventosa_framework_2019}, a framework targeting Python-based workloads that provides a serverless implementation of the well-known MapReduce parallel pattern~\cite{10.1145/1327452.1327492}. 
MARLA is built on top of AWS Lambda and uses S3 for data storage.
Similarly, Müller et al.~\cite{muller_lambada_2020} presented \textit{Lambada}, a distributed data-processing framework for serverless environments. It employs a local driver to orchestrate query execution via AWS Lambda workers, and leverages AWS S3 for bulk storage, DynamoDB for metadata, and SQS for messaging. To mitigate cold start and storage latency, it implements a tree-based worker invocation scheme for rapid scale-out, a scan operator that balances S3 throughput and cost, and a fully serverless exchange operator for data shuffling.
Additionally, other works have addressed complementary aspects of data analytics. Skluzacek et al.~\cite{skluzacek_serverless_2021} proposed \textit{Xtract}, an automated and scalable framework for bulk metadata extraction from large distributed data repositories. Built on top of Globus Compute, it leverages metadata extractors that are executed on remote and heterogeneous computing endpoints.
Further, Sánchez-Artigas~\cite{sanchez-artigas_seer_2022} and T. Eizaguirre~\cite{eizaguirre_seer_2024} introduced and later extended \textit{SEER}, a shuffle manager designed for serverless data analytics. SEER addresses the challenges of inter-function communication in the context of data analytics operations such as \texttt{join} and \texttt{groupBy}. It leverages cloud object storage for managing intermediate data and uses a runtime mechanism that dynamically selects between two shuffle implementations based on the volume of data observed during execution. Initially built on top of \textit{PyWren-IBM}~\cite{10.1145/3284028.3284029}, SEER has since been migrated to Lithops~\cite{9619947}.

Recent research has increasingly explored the use of the serverless paradigm to support ML workloads, especially for the inference phase, where strict latency requirements must be met to enable real-time execution.
A common optimization to improve throughput in ML inference is \textit{request batching}, which aggregates multiple inference requests for simultaneous execution, increasing GPU utilization and amortizing invocation overheads. In the serverless context, this technique has been adopted in early works such as \textit{MArk}~\cite{zhang_mark_2019} and \textit{BATCH}~\cite{ali_batch_2020}. MArk is a general-purpose inference serving system that hides provisioning latency in IaaS by leveraging serverless functions to handle occasional load spikes, targeting ML workloads that demand low response time and elastic scaling. It was later extended with architectural refinements and improved batching strategies~\cite{zhang_enabling_2022}.
BATCH, in contrast, is a fully serverless inference framework. Built on AWS Lambda, it improves request batching in FaaS by integrating a profiling strategy and an analytical model to optimize key parameters, such as memory and batch size, thereby balancing performance with cost efficiency under the serverless pay-per-use model. 
Compared to AWS SageMaker, a state-of-the-art ML cloud platform, both MArk and BATCH reduce costs more effectively by exploiting the bursty nature of inference workloads and leveraging serverless to minimize or eliminate overprovisioning. SageMaker’s feedback-based scaling instead depends on manually configured autoscaling rules, which often require overprovisioning to avoid SLO violations, resulting in higher costs.\footnote{Since 2021, AWS offers SageMaker Serverless Inference for workloads with intermittent or infrequent traffic.}
More recently, Yang et al.~\cite{yang_infless_2022} introduced \textit{INFless}, a serverless framework for ML inference. Unlike MArk and BATCH, which rely on external batching layers on top of commercial serverless platforms not optimized for ML, INFless adopts a native design that tightly couples the inference engine with the underlying serverless infrastructure. This co-designed architecture significantly improves throughput and reduces latency. Built on OpenFaaS, INFless integrates NVIDIA-Docker and NVIDIA MPS to enable GPU acceleration, incorporates an auto-scaling engine, and implements a tailored cold start mitigation policy. Compared against both OpenFaaS with GPU support and BATCH, INFless achieved significantly higher throughput and lower latency, underscoring the benefits of serverless-native co-designed architectures for ML workloads.

Another area of application for serverless is graph processing. 
The latter has a wide range of applications, spanning social network analysis, scientific simulation, graph-based ML, big data, and more.
Such applications typically run in highly distributed or HPC environments and require substantial data communication, which poses challenges in bringing these applications to serverless environments.
Toader et al.~\cite{toader_graphless_2019} proposed \textit{Graphless}, a graph processing framework built on top of AWS Lambda. It leverages AWS S3 for storage and uses Redis for inter-function communication. 
Experimental evaluation against Apache Giraph and GraphMat 
showed lower performance, due to data communication overhead.
This highlights the importance of devising specific optimizations that target the characteristics of graph processing applications to make them viable for deployment on serverless infrastructure.
Building upon these limitations, Liu et al.~\cite{liu_faasgraph_2024} proposed \textit{FaaSGraph}, a graph processing framework co-designed with the underlying serverless infrastructure. 
Preliminary analysis showed that directly deploying existing graph processing frameworks on serverless platforms led to severe performance degradation due to communication overhead and cold start latency, and excessive pre-processing costs. This highlighted the need for co-designed frameworks tightly integrated with the underlying serverless infrastructure, similar to the approach used in INFless and other domain-specific serverless frameworks.
%
%
FaaSGraph addresses the data communication problem by adopting a subgraph-centric model and employing the \textit{BSP (Bulk Synchronous Parallel)} model to process graphs iteratively. This enables coarse-grained computation and, compared to a fully vertex-centric approach, reduces communication overhead. To optimize inter-function communication, FaaSGraph introduces locality-aware resource fusion, enabling shared memory communication and resource sharing among co-located functions.
Evaluation showed that FaaSGraph, while competitive for light to medium workloads, was outperformed by traditional non-serverless frameworks in high-load scenarios due to communication and state-management constraints in current serverless platforms.

Stream processing has become essential for a wide range of applications that rely on real-time processing of data generated by multiple sources, including monitoring systems, autonomous vehicles, financial services, and scientific simulations.
Depending on the specific scenario, computation may be placed at different points along the cloud continuum. Yet, stream processing tasks that are both computationally and data intensive are typically executed on HPC infrastructures. Adopting a serverless execution model for such applications poses several challenges due to the complexity of managing stateful computations across distributed, ephemeral functions, and communication bottlenecks that can increase end-to-end latency. As a result, few works in the current serverless literature specifically target stream processing.
Luckow et al.~\cite{luckow_performance_2019} presented an extension of \textit{Pilot-Streaming}~\cite{8588652}, a framework for deploying and managing streaming frameworks across HPC infrastructures. The extension adds support for deploying stream processing applications on serverless platforms. In addition, the authors developed \textit{StreamInsight}, a framework for analyzing and predicting the performance of streaming systems and applications, which leverages \textit{Universal Scalability Model (USL)}-based performance modeling. StreamInsight was evaluated across serverless, HPC, and cloud environments, and on a variety of applications. Results confirmed the feasibility of serverless for stream processing workloads, while highlighting trade-offs in performance and cost that depend on the application and platform characteristics.
Further, Rotchford et al.~\cite{rotchford_laminar_2024} presented \textit{Laminar 2.0}, an improved version of the serverless stream processing framework \textit{Laminar}~\cite{10.1145/3624062.3624280}, built on top of the \textit{dispel4py} Python library~\cite{7079021}. 
It introduces several improvements compared to its previous version, including dynamic workflow execution, enhanced dependency and resource management, HTTP/2-based true streaming for low-latency data transfer between client and execution engine, and advanced code search to assist developers during workflow composition.  

Finally, digital twins are increasingly employed across domains from healthcare to aerospace and environmental engineering. In that regard, techniques such as \textit{Verification, Validation, and Uncertainty Quantification (VVUQ)} have proven essential to ensure the reliability of digital twins in fields such as healthcare and many others~\cite{sel_survey_2025}. VVUQ tasks, such as sensitivity analysis and uncertainty propagation, can involve tens of thousands of independent model executions and are typically executed on HPC clusters. However, many VVUQ tasks consist of embarrassingly parallel simulations with short execution times, making them excellent candidates for serverless execution.
Kica et al.~\cite{kica_serverless_2023} presented \textit{CloudVVUQ}, an extension of \textit{EasyVVUQ} supporting serverless execution on several commercial platforms. CloudVVUQ was evaluated on real-world applications from computational medicine. Results demonstrated the suitability of serverless to support such workloads, showing competitive performance and scalability across varying configurations, and confirming its potential as a viable alternative to traditional HPC for VVUQ tasks.


\begin{table*}[htbp]
\footnotesize
\centering
\caption{Summary of studies on cold start mitigation techniques.}
\label{tab:cold_start_mitigation}
\rowcolors{2}{white}{lightaccessblue}
\begin{tabularx}{\textwidth}{>{\raggedright\arraybackslash}p{3.10cm} >{\raggedright\arraybackslash}X >{\raggedright\arraybackslash}l}
\toprule
{\footnotesize\sffamily\bfseries\textcolor{accessblue}{Study}} & 
{\footnotesize\sffamily\bfseries\textcolor{accessblue}{Proposed Solution}} & 
{\footnotesize\sffamily\bfseries\textcolor{accessblue}{Strategy}} \\
\midrule
{\contentfont Yang et al.~\cite{yang_infless_2022}} & 
{\contentfont Design a policy that combines short- and long-term historical patterns to drive adaptive pre-warming decisions.} & 
{\contentfont Invocation Prediction} \\

{\contentfont Lu et al.~\cite{lu_smiless_2024}} & 
{\contentfont Construct a co-optimization framework that predicts future invocations and allocates resources for DAG-based applications.} & 
{\contentfont Invocation Prediction} \\

{\contentfont Sankaranarayanan et al.~\cite{sankaranarayanan_pulse_2025}} & 
{\contentfont Develop a dynamic strategy to adjust keep-alive durations and select between model variants based on runtime behavior.} & 
{\contentfont ML Model Selection} \\

{\contentfont Hong et al.~\cite{hong_optimus_2024}} & 
{\contentfont Use inter-function model transformation and meta-operators to re-purpose preloaded models within warm, but idle, containers.} & 
{\contentfont ML Model Reuse} \\

{\contentfont Roy et al.~\cite{roy_daydream_2022}} & 
{\contentfont Propose a speculative execution model with delayed specialization to reduce cold start latency in DAG-modeled scientific applications.} & 
{\contentfont Pre-warming} \\

{\contentfont Fuerst et al.~\cite{fuerst_iluvatar_2023}} & 
{\contentfont Optimize control plane through pre-warmed container and network namespace pools.} & 
{\contentfont Pre-warming} \\

{\contentfont Bai et al.~\cite{bai_faast_2024}} & 
{\contentfont Employ VM snapshots and improve networking setup to accelerate function initiation.} & 
{\contentfont Snapshotting} \\
\bottomrule
\end{tabularx}
\end{table*}

\subsubsection{Scheduling and Resource Management}
Serverless computing introduces new challenges in scheduling and resource management due to its on-demand nature and highly dynamic execution model. Resources must be allocated within milliseconds while accommodating diverse workloads across heterogeneous infrastructures. These constraints complicate scheduling and provisioning, particularly in the presence of cold start latency, heterogeneous accelerators, and data-intensive applications.
Recent research has addressed these issues through cold start mitigation, adaptive scaling, accelerator-aware placement, and hybrid execution models. Storage and memory access have also emerged as critical factors, strongly influencing the performance of serverless applications in big data and HPC scenarios. The following paragraphs review key contributions in these areas, while Tables~\ref{tab:function_scheduling}, ~\ref{tab:scaling_scheduling}, ~\ref{tab:accelerator_scheduling}, ~\ref{tab:hybrid_execution}, and ~\ref{tab:storage_optimization} summarize the surveyed~studies.

\paragraph{Cold Start Reduction}
Since the introduction of serverless computing, a wide range of solutions for cold start mitigation have been proposed. A commonly employed approach involves the use of pre-warming techniques, wherein a fixed pool of warm containers is maintained a priori or kept alive for a fixed time after function execution. Yet, these strategies suffer from the drawback of keeping part of the available resources idle, which leads to suboptimal resource utilization. Dynamic scheduling techniques, on the other hand, attempt to mitigate cold starts more efficiently~\cite{10.1007/978-3-030-13342-9_15, 8975850, 254430, 10.1145/3503222.3507750}, often relying on ML to analyze historical invocation patterns. While promising, these approaches require time to collect data about function invocations and analyze time-series patterns, making them unsuitable for immediate deployment. Moreover, they often struggle to scale, as they depend on maintaining invocation histories for each function or workflow individually.
These limitations are particularly problematic for HPC applications, which, unlike typical FaaS workloads, may not consist of short-lived functions with predictable or recurring patterns. To overcome this, the literature has also explored other strategies for reducing cold start latency, including lightweight execution environments~\cite{246288, hall_execution_2019, 9214403, shillaker_faasm_2020}, cache- and snapshot-based techniques~\cite{10.1145/3342195.3392698, 216031}, and reuse of execution environments across multiple functions~\cite{9213020, 215935}.

Depending on the application domain, existing cold start mitigation techniques may be ineffective, as they often lack awareness of workload characteristics.
In the context of ML workloads, Yang et al.~\cite{yang_infless_2022} proposed \textit{LSTH}, a cold start mitigation approach based on the \textit{Hybrid Histogram Policy (HHP)} originally introduced by Shahrad et al.~\cite{254430}, and integrated it into \textit{INFLESS}, a serverless framework for ML inference. LSTH addresses the limitations of static histogram-based policies by combining short-term and long-term idle time distributions to dynamically adjust both pre-warming and keep-alive windows, better capturing periodic workload patterns while remaining responsive to short-term fluctuations, thereby reducing cold start latency.
Complementary to this, Sankaranarayanan et al.~\cite{sankaranarayanan_pulse_2025} introduced \textit{PULSE}, a dynamic 10-minute keep-alive mechanism tailored for serverless ML inference. PULSE extends prior function prediction techniques by selecting the most appropriate model variant and determining how long it should be maintained warm within a 10-minute window, minimizing keep-alive costs while preserving service quality. It leverages historical invocation traces to balance 
model accuracy, service time, and memory consumption. 
To reduce memory pressure during high-load periods, PULSE selectively downgrades specific models to lower-accuracy variants, minimizing resource consumption without significantly compromising performance.
%
In contrast to LSTH and PULSE, Hong et al.~\cite{hong_optimus_2024} addressed the problem from a model loading perspective. They proposed \textit{Optimus}, a low-latency serverless inference system that mitigates cold start by accelerating model loading through inter-function model transformation. 
Since model structure loading dominates cold start latency, Optimus repurposes containers to serve other ML functions by reusing the structured state of already-loaded models.

With the growing adoption of serverless computing for ML inference, modeling such applications as DAGs has become increasingly common. This approach improves scalability of individual stages and enables efficient use of hardware heterogeneity in modern clusters. However, cold start mitigation becomes more complex in DAG-based applications due to cascading effects across dependent functions. Addressing this, Lu et al.~\cite{lu_smiless_2024} introduced \textit{SMIless}, a serverless ML inference platform built on OpenFaaS. SMIless co-optimizes resource provisioning and cold start mitigation by employing adaptive pre-warming windows, dynamically adjusted based 
on each function’s position in the DAG and predicted invocation patterns. It combines offline profiling, which measures initialization and inference times under diverse configurations, with a dual LSTM-based online predictor to forecast the number of invocations and inter-arrival times. 
Through these predictions, SMIless selects the most effective initialization and execution strategies. Compared with state-of-the-art approaches, it reduces costs while meeting SLA requirements.

Scientific applications represent another class of workloads that benefit from DAG-based modeling. However, they are often dynamic and irregular in terms of invocation patterns and execution times, requiring specialized cold start mitigation strategies.
%
In this context, Roy et al.~\cite{roy_daydream_2022} introduced \textit{DayDream}, a framework designed to mitigate cold starts in dynamic scientific workflows. DayDream introduces the concept of \textit{hot starts}, where execution environments are booted and maintained ahead of time but only specialized to a function upon invocation. This decouples the runtime environment from the function code, partially amortizing cold start overhead and keep-alive cost. 
Rather than relying on time-series predictions, DayDream dynamically predicts the number of required instances by modeling the distribution of phase concurrency using statistical distributions (e.g., Weibull).
Since different components in a DAG may require different resource configurations, DayDream supports heterogeneous function instances by offering two execution tiers, high-end and low-end.
Evaluated against solutions such as \textit{Pegasus}, a traditional workflow manager, and \textit{Serverless in the Wild}~\cite{254430}, an ARIMA-based predictor for serverless workloads, DayDream significantly outperformed both in service time and cost, confirming its effectiveness for scientific workflows.

While most research on serverless computing has focused on reducing cold start latency through invocation prediction and pre-warming, less attention has been paid to the control plane, whose overheads also contribute significantly to cold start latency and overall execution time.
Fuerst et al.~\cite{fuerst_iluvatar_2023} addressed this gap by introducing \textit{Ilúvatar}, an extensible serverless research platform featuring a fast, modular control plane. Ilúvatar adopts a worker-centric architecture in which workers are decoupled from a centralized orchestrator, managing load balancing and performance monitoring locally. To reduce function execution latency and mitigate cold starts, Ilúvatar minimizes overheads along the function’s critical path.
Each worker maintains a pool of pre-warmed containers managed through \textit{Least Recently Used (LRU)} and \textit{Greedy-Dual-Size-Frequency}~\cite{10.1145/3445814.3446757} policies, and reuses pre-created virtual network namespaces to avoid costly setup operations during container initialization.
%
Further optimizations target communication between the worker and the container agent. Ilúvatar reduces overheads by caching an HTTP client per container and enabling connection pooling. Each worker also maintains an invocation queue for incoming requests, while short-lived functions may bypass the queue to further reduce latency.
Orthogonally, Bai et al.~\cite{bai_faast_2024} proposed \textit{Faast}, a solution that mitigates cold start latency by leveraging virtual machine snapshots. 
Built on Firecracker, Faast introduces several optimizations to improve snapshot restoration time. It employs a preparation phase to generate a lightweight working set that excludes stale memory pages, thereby reducing disk I/O during startup. 
%
To further lower control-plane overhead, Faast integrates NAT within the \texttt{virtio-net} device model, bypassing network namespace setup and enabling faster connectivity.
Both Ilúvatar and Faast highlight complementary strategies to reduce cold start latency, showing that effective mitigation must also address the invocation path.

\begin{table*}[htbp]
\footnotesize
\centering
\caption{Summary of studies on function scheduling strategies in serverless computing.}
\label{tab:function_scheduling}
\rowcolors{2}{white}{lightaccessblue}
\begin{tabularx}{\textwidth}{>{\raggedright\arraybackslash}p{2.8cm} >{\raggedright\arraybackslash}X >{\raggedright\arraybackslash}l}
\toprule
{\footnotesize\sffamily\bfseries\textcolor{accessblue}{Study}} & 
{\footnotesize\sffamily\bfseries\textcolor{accessblue}{Proposed Solution}} & 
{\footnotesize\sffamily\bfseries\textcolor{accessblue}{Strategy}} \\
\midrule
{\contentfont Denninnart et al.~\cite{denninnart_improving_2019}} 
& {\contentfont Implement probabilistic pruning to defer or drop tasks unlikely to meet deadlines.}
& {\contentfont Probabilistic Task Pruning} \\

{\contentfont Schirmer et al.~\cite{schirmer_profaastinate_2023}} 
& {\contentfont Develop a deferred execution system that queues non-critical function calls based on deadlines and resource availability.} 
& {\contentfont Deferred Execution} \\

{\contentfont Przybylski et al.~\cite{przybylski_using_2022}} 
& {\contentfont Integrate SLURM and Singularity to deploy functions on idle HPC resources.} 
& {\contentfont Opportunistic Scheduling} \\

{\contentfont Da Costa Marques et al.~\cite{da_costa_marques_preliminary_2023}} 
& {\contentfont Leverage preemptible cloud instances to host serverless workloads on spot VMs for cost-efficient execution.} 
& {\contentfont Opportunistic Scheduling} \\

{\contentfont Carver et al.~\cite{carver_wukong_2020}} 
& {\contentfont Partition HPC workflow DAGs into subgraphs and assign each to a Lambda executor for localized task scheduling.} 
& {\contentfont Decentralized Scheduling} \\

{\contentfont Chen et al.~\cite{chen_switchflow_2024}} 
& {\contentfont Predict function invocations and deploy them on the most suitable platform.} 
& {\contentfont Predictive Platform Selection} \\

{\contentfont Li et al.~\cite{li_unifaas_2024}} 
& {\contentfont Implement a scheduler that profiles resources and function characteristics for heterogeneity-aware task placement.}
& {\contentfont Profile-driven Scheduling} \\
\bottomrule
\end{tabularx}
\end{table*}

\paragraph{Function Scheduling}
Efficient function scheduling is critical for ensuring the responsiveness, scalability, and robustness of serverless platforms. 
This challenge becomes particularly significant in heterogeneous and resource-constrained environments, where systems must dynamically allocate resources to ephemeral, stateless functions under highly variable workloads. This dynamic nature introduces challenges, particularly in avoiding SLA violations, resource under-utilization, and inefficient task placement across heterogeneous infrastructures. Recent research has explored a variety of scheduling strategies, from probabilistic pruning and deferred execution to platform-aware orchestration and opportunistic resource usage, 
with the goal of improving scheduling in both HPC and cloud contexts.

Denninnart et al.~\cite{denninnart_improving_2019} introduced a probabilistic task pruning mechanism for heterogeneous serverless systems. The proposed approach improves system robustness in oversubscribed scenarios by proactively deferring or discarding tasks unlikely to meet their deadlines. This mechanism relies on probabilistic execution-time models to estimate task completion likelihood. 
Evaluated on the \textit{Queen Bee 2} HPC cluster, part of the \textit{Louisiana Optical Network Infrastructure (LONI)}, it significantly increased the percentage of tasks completed on time, especially under high workload intensities.
Similarly, Schirmer et al.~\cite{schirmer_profaastinate_2023} presented \textit{ProFaaStinate}, a system built on top of Nuclio that enables deferred execution of non-time-sensitive functions during peak load, when resources are scarce. ProFaaStinate places asynchronous invocations into a priority queue and schedules them based on developer-defined deadlines and current resource availability, improving platform responsiveness and reducing overload during high-demand phases.
Yet, developers must manually configure allowable delays for each function, making the solution suitable for simple applications but less practical for complex workflows.

In HPC infrastructures, resource under-utilization and node idleness remains a non-trivial challenge. 
To address this, research has explored co-locating FaaS workloads on idle HPC resources to improve server consolidation~\cite{copik_software_2024,przybylski_using_2022}. Przybylski et al.~\cite{przybylski_using_2022} proposed \textit{HPC-Whisk}, an extension of Apache OpenWhisk that integrates with SLURM and uses Singularity containers to deploy serverless functions on idle nodes without disrupting running jobs. 
Experiments on a production cluster showed that HPC-Whisk covered up to $90\%$ of idle time slots without affecting the execution of primary workloads.
Addressing the same problem from a cloud perspective, Da Costa Marques et al.~\cite{da_costa_marques_preliminary_2023} evaluated the use of spot cloud instances to support HPC serverless workloads by leveraging AWS spot instances. The authors deployed \textit{funcX} endpoints on preemptible virtual machines and conducted a preliminary evaluation comparing execution time and cost against on-demand instances. Results showed that, despite the lack of availability guarantees, spot instances can significantly reduce cost for short-lived, fault-tolerant workloads. Collectively, these studies demonstrate that both HPC and cloud environments can benefit from opportunistic serverless execution models to improve overall resource utilization.

Targeting HPC workflows, Carver et al.~\cite{carver_wukong_2020} proposed \textit{Wukong}, a decentralized and locality-aware serverless parallel framework designed to scale DAG-modeled applications efficiently. Instead of relying on a centralized scheduler, Wukong partitions a DAG into static subgraphs, assigning each to a Lambda-based executor responsible for dynamically scheduling and executing tasks.
Nonetheless, HPC tasks typically exhibit high and diverse resource demands, and existing scheduling approaches often overlook the heterogeneity of modern serverless platforms. In this context, Chen et al.~\cite{chen_switchflow_2024} proposed \textit{SwitchFlow}, a system that optimizes HPC workflow execution by selectively deploying workflow stages on the most suitable serverless platform. 
The scheduler collaborates with serverless-side controllers for node selection, while a behavior model applies time-series analysis to predict future function invocations and select the appropriate platform, thereby reducing cold starts. The prediction is implemented using a \textit{Long Short-Term Memory (LSTM)} model. Additionally, SwitchFlow employs a \textit{Deep Q-Network (DQN)} to handle bursty workloads by switching between frameworks in real time based on performance metrics. Experimental results demonstrated significant improvements in execution time, latency, and service availability compared to standalone deployments on either Fission or OpenWhisk.
Similarly, UniFaaS~\cite{li_unifaas_2024} incorporates an execution profiler and a heterogeneity-aware scheduler that facilitates the custom mapping of workflow tasks to available endpoints, including HPC clusters, cloud nodes, and edge systems. UniFaaS adopts an observe-predict-decide approach to optimize task placement by profiling both resource states and function characteristics and supports multiple scheduling strategies, including a delay mechanism and re-scheduling, to adapt to resource dynamics and minimize makespan.

\begin{table*}[htbp]
\footnotesize
\centering
\caption{Summary of studies on resource scaling and scheduling in serverless computing.}
\label{tab:scaling_scheduling}
\rowcolors{2}{white}{lightaccessblue}
\begin{tabularx}{\textwidth}{>{\raggedright\arraybackslash}p{3.6cm} >{\raggedright\arraybackslash}X >{\raggedright\arraybackslash}l}
\toprule
{\footnotesize\sffamily\bfseries\textcolor{accessblue}{Study}} & 
{\footnotesize\sffamily\bfseries\textcolor{accessblue}{Proposed Solution}} & 
{\footnotesize\sffamily\bfseries\textcolor{accessblue}{Strategy}} \\
\midrule
{\contentfont Related Works~\cite{enes_real-time_2020,castellanos-rodriguez_serverless-like_2024}} 
& {\contentfont Develop a container-level scaling framework based on user-defined policies and post-execution profiling.} 
& {\contentfont Policy-Driven Scaling} \\

{\contentfont Cheng et al.~\cite{cheng_autonomous_2018}} 
& {\contentfont Develop a framework that autonomously schedules resources for stream processing using workload-aware scaling.} 
& {\contentfont Workload-Aware Scheduling} \\
\bottomrule
\end{tabularx}
\end{table*}

\paragraph{Resource Scaling and Scheduling}
An important aspect in serverless computing is how efficiently resources are scaled and scheduled based on application characteristics. While FaaS platforms typically handle scaling automatically, certain applications may benefit from custom scaling and scheduling policies. Stream processing exemplifies this scenario, as individual stages mapped to serverless functions can be replicated dynamically according to input rate fluctuations. For instance, Enes et al.~\cite{enes_real-time_2020} presented a platform for automatic, real-time container resource scaling targeting big data applications. The proposed solution operates in two phases. First, it performs resource scaling at runtime, driven by user-defined policies. Second, after execution, it collects detailed logs that provide fine-grained insights into resource usage throughout the execution. In a follow-up study, the same authors extended this solution to support YARN clusters~\cite{castellanos-rodriguez_serverless-like_2024}.
Addressing scheduling in serverless stream processing applications, Cheng et al.~\cite{cheng_autonomous_2018} proposed \textit{ARS(FaaS)}. It dynamically allocates resources based on stream workload characteristics, enabling elastic and efficient execution of concurrent streaming applications. ARS(FaaS) employs Apache Storm to ingest data from mobile devices, processes the streams using OpenWhisk serverless functions, and stores the results in a NoSQL database for low-latency access.
%
While these studies explore adaptive scaling and scheduling mechanisms tailored to application-specific workloads, this research area remains largely unexplored, leaving fine-grained resource control an open challenge in serverless computing.

\begin{table*}[htbp]
\footnotesize
\centering
\caption{Summary of studies on accelerator scheduling strategies in serverless computing.}
\label{tab:accelerator_scheduling}
\rowcolors{2}{white}{lightaccessblue}
\begin{tabularx}{\textwidth}{>{\raggedright\arraybackslash}p{2.4cm} >{\raggedright\arraybackslash}X >{\raggedright\arraybackslash}l}
\toprule
{\footnotesize\sffamily\bfseries\textcolor{accessblue}{Study}} & 
{\footnotesize\sffamily\bfseries\textcolor{accessblue}{Proposed Solution}} & 
{\footnotesize\sffamily\bfseries\textcolor{accessblue}{Strategy}} \\
\midrule

{\contentfont Zhao et al.~\cite{zhao_gpu-enabled_2023}} 
& {\contentfont Propose a scheduling policy that prioritizes requests directed to models already cached in GPU memory.} 
& {\contentfont Locality-aware Scheduling} \\

{\contentfont Bhasi et al.~\cite{bhasi_paldia_2024}} 
& {\contentfont Implement a hybrid GPU sharing mechanism guided by request rate prediction and workload profiling.} 
& {\contentfont Spatio-temporal GPU Sharing} \\

{\contentfont Hui et al.~\cite{hui_esg_2024}} 
& {\contentfont Design a two-phase scheduler that combines A*-based configuration search with resource- and locality-aware function placement.} 
& {\contentfont Spatial GPU Sharing} \\

{\contentfont Prakash et al.~\cite{prakash_optimizing_2021}}
& {\contentfont Propose a dynamic kernel slicing strategy and EDF-based scheduler to improve GPU utilization on virtualized hardware.} 
& {\contentfont Kernel Slicing} \\

{\contentfont Garg et al.~\cite{garg_faaster_2021}} 
& {\contentfont Develop a scheduling heuristic to determine optimal slicing and GPU allocation based on runtime profiling.} 
& {\contentfont Kernel Slicing} \\

{\contentfont Ma et al.~\cite{ma_widepipe_2021}} 
& {\contentfont Apply Q-learning to jointly choose batch size and schedule jobs on NPUs.} 
& {\contentfont RL-based Scheduling} \\

{\contentfont Bernard et al.~\cite{lannurien_herofake_2023}} 
& {\contentfont Develop a two-stage scheduler that minimizes energy consumption using offline profiling and EDF-based policy.} 
& {\contentfont Energy-aware Scheduling}  \\

{\contentfont Rattihalli et al.~\cite{rattihalli_fine-grained_2023}} 
& {\contentfont Extend Kubernetes with a fine-grained, energy-aware scheduler for heterogeneous devices based on energy profiling.} 
& {\contentfont Energy-aware Scheduling} \\
\bottomrule
\end{tabularx}
\end{table*}

\paragraph{Accelerator Scheduling}

With serverless computing increasingly used for compute-intensive workloads, accelerator scheduling has become a critical research area. Consequently, recent studies have focused on improving accelerator utilization while meeting SLOs and addressing sustainability and energy efficiency concerns.

Machine learning and deep learning inference are among the main use cases for accelerators in serverless platforms, with GPUs representing the predominant type employed in such applications. 
In this context, Zhao et al.~\cite{zhao_gpu-enabled_2023} proposed a GPU-enabled serverless ML inference framework, built on OpenFaaS, that employs a locality-aware, load-balancing policy to prioritize requests targeting GPUs with already cached models, reducing latency and cache misses. The framework integrates a cache manager that maintains frequently used models in GPU memory using an LRU policy, while task scheduling is managed through a global queue and per-GPU queues. When a GPU becomes idle, the scheduler selects tasks from the global queue based on arrival time and cache locality, dispatching them to the appropriate GPU. 
Other approaches focus on reducing costs and improving resource utilization through spatio-temporal multiplexing. Bhasi et al.~\cite{bhasi_paldia_2024} proposed \textit{PALDIA}, a serverless framework that enables SLO-compliant and cost-effective execution of ML inference workloads. PALDIA integrates a hardware selection model that selects cost-efficient CPU or GPU nodes based on predicted request rates and workload profiles. To improve GPU utilization, it leverages NVIDIA MPS and implements a hybrid spatio-temporal GPU sharing mechanism that dynamically balances job interference and queuing delays.
Similarly, focusing on DNN-based workflows, Hui et al.~\cite{hui_esg_2024} proposed \textit{ESG}, a scheduling algorithm that exploits shareable GPUs to improve resource efficiency. ESG employs NVIDIA MIG for GPU spatial sharing and adopts \textit{application-function-wise (AFW)} job queues to co-locate function invocations from the same application on the same machine, reducing inter-function communication overhead. Its scheduling process follows a two-step design: first, ESG computes optimal resource configurations using an A*-search algorithm with dual-blade pruning to narrow the search space based on estimated execution time and cost; second, it maps function invocations to invokers considering current resource availability and data locality. 
When evaluated against INFless and other platforms, ESG achieved the highest SLO hit rates while significantly reducing costs, showing that separating configuration planning from placement reduces scheduling complexity and improves scalability.
As discussed in Section~\ref{par:gpu-support}, kernel slicing has also been explored as an approach to improve GPU utilization.
However, the dynamic nature of serverless makes it challenging to quickly determine the number of slices per kernel and assign them to the appropriate GPUs.
Prakash et al.~\cite{prakash_optimizing_2021} proposed a dynamic slicing strategy for vGPUs that adjusts slice sizes to current GPU load and uses an EDF-based scheduler to avoid dispatching tasks unlikely to meet deadlines, minimizing wasted GPU time.
Analogously, Garg et al.~\cite{garg_faaster_2021} proposed a heuristic that determines the optimal number and size of slices for each function invocation and maps them to available GPUs based on runtime profiling, maximizing resource utilization.
While both solutions have proven effective under low arrival rates, they tend to struggle under high load conditions, highlighting that dynamically determining optimal slice sizes at runtime remains a major challenge under peak workloads.

Although most existing work on accelerator scheduling in the serverless literature has focused on GPUs, more attention is being given to other types, such as NPUs, TPUs, and FPGAs. 
Ma et al.~\cite{ma_widepipe_2021} proposed \textit{WidePipe}, a reinforcement learning-based scheduling strategy targeting NPU clusters. WidePipe uses Q-learning to co-adapt resource allocation and request batch size based on device status, thereby improving resource utilization and inference throughput.
Sustainability and energy efficiency have also emerged as key concerns. Bernard et al.~\cite{lannurien_herofake_2023} introduced \textit{HeROfake}, a framework for deploying short-lived, interactive deepfake detection tasks on heterogeneous serverless platforms leveraging both GPUs and FPGAs. HeROfake adopts a two-phase scheduling approach. In the offline phase, each function is profiled to collect metadata such as memory usage, execution time, cold start latency, and energy consumption. 
This metadata is then used in the online phase, where an EDF-based policy and queue length analysis guide task placement to satisfy QoS constraints while minimizing energy usage.
%
Similarly, Rattihalli et al.~\cite{rattihalli_fine-grained_2023} extended Kubernetes with a 
fine-grained, energy-aware scheduler supporting both FPGAs and GPUs. The proposed solution integrates an energy profiler that captures CPU energy consumption measurements at various load levels. Further, the scheduler takes CPU and memory requirements from the user and estimates the energy cost of running the task on the available nodes based on each load, selecting the node expected to consume the least energy.
Despite these efforts, research on energy-aware scheduling for heterogeneous accelerators in serverless platforms remains in its early stages, with limited studies addressing high-load, real-world scenarios.

\begin{table*}[htbp]
\footnotesize
\centering
\caption{Summary of studies on hybrid execution strategies.}
\label{tab:hybrid_execution}
\rowcolors{2}{white}{lightaccessblue}
\begin{tabularx}{\textwidth}{>{\raggedright\arraybackslash}p{2.3cm} >{\raggedright\arraybackslash}X >{\raggedright\arraybackslash}l}
\toprule
{\footnotesize\sffamily\bfseries\textcolor{accessblue}{Study}} & 
{\footnotesize\sffamily\bfseries\textcolor{accessblue}{Proposed Solution}} & 
{\footnotesize\sffamily\bfseries\textcolor{accessblue}{Strategy}} \\
\midrule
{\contentfont Finol et al.~\cite{finol_exploiting_2024}} 
& {\contentfont Propose a hybrid executor that distributes Java threads across local resources and serverless functions for irregular parallel workloads.} 
& {\contentfont Local–Serverless Co-execution} \\

{\contentfont Ray et al.~\cite{roy_mashup_2022}} 
& {\contentfont Design a workflow manager that integrates profiling and a decision controller to assign workflow stages to VMs or serverless functions.} 
& {\contentfont Hybrid Workflow Orchestration} \\

{\contentfont Chahal et al.~\cite{chahal_isesa_2022}} 
& {\contentfont Propose an ML-based framework to offload suitable HPC and AI workloads to AWS Lambda when on-premises resources are saturated.} 
& {\contentfont Hybrid Task Offloading} \\

{\contentfont Aubin et al.~\cite{aubin_helastic_2021}} 
& {\contentfont Introduce a dual-layer model that offloads long-running tasks from FaaS to containers based on timeout thresholds.} 
& {\contentfont Hybrid Task Offloading} \\
\bottomrule
\end{tabularx}
\end{table*}

\paragraph{Hybrid Execution}
Serverless computing shows promise for highly dynamic parallel applications with unbalanced and irregular workloads, which are difficult to run efficiently on statically provisioned clusters due to fluctuating resource demands that can cause over- or under-provisioning.
It addresses these challenges by enabling elastic scaling for peak loads, akin to cloud bursting, without the overcommitment of persistent resources. Additionally, hybrid deployments are possible, where short-lived workloads run serverlessly while compute-intensive tasks use traditional resources.
Hence, hybrid approaches combining serverless and on-premises infrastructures have been explored to improve resource utilization, scalability, and cost-efficiency.

Several works illustrate this hybrid execution paradigm. Finol et al.~\cite{finol_exploiting_2024} proposed a serverless hybrid executor that enables the instantiation of Java threads across both serverless functions and local resources. Evaluated across different highly parallel applications, this approach outperformed execution on static Spark clusters and large local machines, showing how hybrid execution models can overcome the rigidity of static provisioning by dynamically adapting to fluctuating workloads.
Further, Ray et al.~\cite{roy_mashup_2022} presented \textit{Mashup}, a proof-of-concept workflow manager for orchestrating HPC workflows across traditional VMs and serverless platforms. 
Mashup supports two execution strategies. The first, used as a baseline, offloads stages with a parallelism degree exceeding the available resources on the local HPC cluster to external serverless platforms. The second integrates a \textit{Placement Decision Controller (PDC)} that profiles each workflow stage and estimates execution times on both VMs and serverless platforms. 
Based on profiling data, the controller selects the most suitable execution site. Evaluation on real-world workflows showed that Mashup significantly reduced execution time and cost compared to traditional workflow managers, highlighting the potential of hybrid execution for scientific workflows.

By combining serverless platforms with on-premises infrastructure, hybrid execution models offer a compelling solution for efficiently addressing peak loads, when local resources are saturated.
In this regard, Chahal et al.~\cite{chahal_isesa_2022} introduced \textit{iSeSA}, a framework that automatically offloads HPC and AI workloads to AWS Lambda once on-premises capacity is reached.
Building on previous work~\cite{10.1145/3452413.3464789}, the authors extended their methodology to support the migration of both HPC and AI applications to the cloud. 
iSeSA integrates an ML-based decision framework that assesses application suitability for serverless execution, automatically deploys the workload, and tunes instance configurations. 
In contrast, Aubin et al.\cite{aubin_helastic_2021} proposed \textit{Helastic}, a dual-layer elastic execution model tailored for the bioinformatics application jModelTest. Unlike iSeSA, which offloads tasks from on-premises infrastructure to FaaS platforms, Helastic defaults to the FaaS layer for short-lived tasks and offloads long-running ones to a container-based backend upon exceeding a predefined timeout. This layered approach combines the scalability and cost-efficiency of FaaS with the stability and flexibility of container-based execution, effectively addressing applications characterized by heterogeneous execution patterns.

\begin{table*}[htbp]
\footnotesize
\centering
\caption{Summary of studies on storage optimization strategies in serverless computing.}
\label{tab:storage_optimization}
\rowcolors{2}{white}{lightaccessblue}
\begin{tabularx}{\textwidth}{>{\raggedright\arraybackslash}p{3.9cm} >{\raggedright\arraybackslash}X >{\raggedright\arraybackslash}l}
\toprule
{\footnotesize\sffamily\bfseries\textcolor{accessblue}{Study}} & 
{\footnotesize\sffamily\bfseries\textcolor{accessblue}{Proposed Solution}} & 
{\footnotesize\sffamily\bfseries\textcolor{accessblue}{Strategy}} \\
\midrule
{\contentfont HoseinyFarahabady et al.~\cite{hoseinyfarahabady_data-intensive_2021}} 
& {\contentfont Develop a consolidation model and a resource controller to reduce I/O interference and ensure response time for data-intensive workloads.} 
& {\contentfont Workload Consolidation} \\

{\contentfont Kang et al.~\cite{kang_towards_2024}} 
& {\contentfont Propose runtime-aware job scheduling and online data merging to mitigate I/O bottlenecks in serverless deep learning training.} 
& {\contentfont Locality-aware Scheduling} \\

{\contentfont Sly-Delgado et al.~\cite{sly-delgado_taskvine_2023}} 
& {\contentfont Introduce a scheduler that co-optimizes data movement and task placement by tracking input file locations across cluster nodes.} 
& {\contentfont Locality-aware Scheduling} \\

{\contentfont Elshamy et al.~\cite{elshamy_study_2023}} 
& {\contentfont Analyze orchestration models and propose prefetching and placement techniques to reduce remote data access overheads in workflows.} 
& {\contentfont Prefetching} \\

{\contentfont Zahir et al.~\cite{zahir_sas_2025}} 
& {\contentfont Present a scheduler that uses Bayesian inference on execution logs to prefetch data speculatively on edge machines.} 
& {\contentfont Speculative Prefetching} \\

{\contentfont Roy et al.~\cite{basu_roy_starship_2024}} 
& {\contentfont Optimize storage-function tier selection in HPC workflows using Levenberg–Marquardt-based configuration tuning.} 
& {\contentfont Storage-Tier Selection} \\
\bottomrule
\end{tabularx}
\end{table*}

\paragraph{Storage}
A key bottleneck in serverless computing, especially for data-intensive and scientific applications, is the latency introduced by memory and storage operations. When multiple functions are instantiated concurrently, the overhead incurred from accessing and reading data can become a significant performance constraint. 
To mitigate these limitations, several studies have proposed techniques that optimize memory and storage access.

Data-intensive workloads are particularly sensitive to I/O overheads, which can cause severe performance degradation under high consolidation. HoseinyFarahabady et al.~\cite{hoseinyfarahabady_data-intensive_2021} analyzed these challenges in the context of serverless and showed that shared resource contention (e.g., last-level cache, disk buffers) among workloads can lead to significant throughput degradation. 
To mitigate this, the authors developed 
a model-predictive controller that dynamically adjusts resource allocation across multiple physical hosts. 
The controller includes a system model estimating performance degradation from consolidation, an ARIMA-based event rate predictor, and an optimizer that selects co-location strategies to balance I/O throughput and QoS.
Evaluated against OpenWhisk, this approach reduced the overall QoS violation rate by $90\%$ and increased the throughput usage of storage devices by $39\%$.

For deep learning workloads, Kang et al.~\cite{kang_towards_2024} proposed \textit{CNDLSys}, a serverless platform designed to support training tasks. It adopts cooperative job scheduling and dataset placement to exploit data locality by co-locating subsequent jobs that share datasets.
%
In addition, it integrates a runtime-aware online data merging and loading technique aimed at mitigating I/O bottlenecks and improving data loading. Experimental results showed that CNDLSys significantly increased local data hit rates and reduced data loading times, thereby improving overall training performance.

Efficient data movement and communication are also crucial in large-scale scientific workflows and DAG-modeled parallel applications.
\textit{TaskVine}~\cite{sly-delgado_taskvine_2023} addresses this challenge in HPC clusters exploiting aggregate local storage and network bandwidth to accelerate the execution of serverless workflows.
It achieves this by tightly coupling data movement with task scheduling, allowing 
to co-schedule compute and data by tracking the availability of required input files across workers. When possible, TaskVine co-locates tasks on nodes that already hold the necessary data or are expected to receive it, minimizing contention on shared filesystems and reducing transfer overhead.
%
Elshamy et al.~\cite{elshamy_study_2023} compared serverful and serverless orchestration of scientific workflows. Specifically, the authors evaluated two serverless orchestration approaches: centralized, and decentralized. Using Montage~\cite{10.1504/IJCSE.2009.026999} as a case study, the authors found that remote data access over distributed file systems introduces significant overheads. To mitigate this, the authors proposed two optimization techniques: prefetching file privileges and container placement. The former involves proactively acquiring file privileges from previous nodes (or stages) while waiting for input data, thereby overlapping privilege acquisition with computation. The latter instead co-locates containers that access shared data on the same physical node, improving data locality. Both strategies were shown to significantly reduce data access overhead in serverless workflow execution.
Similarly, Zahir et al.~\cite{zahir_sas_2025} proposed a scheduling technique that exploits data locality and speculative prefetching to reduce I/O overhead in scientific workflows. The authors introduced a \textit{Speculation Aware Scheduler (SAS)} that leverages historic execution logs to anticipate upcoming tasks using a Bayesian inference model. By preloading required data into memory before task execution on edge machines, the system improves data locality and significantly reduces data access latency.
Finally, Roy et al.~\cite{basu_roy_starship_2024} presented \textit{StarShip}, a framework for executing HPC workflows in serverless environments. 
The authors focused on the storage and compute heterogeneity offered by platforms such as AWS Lambda, which provides multiple tiers of serverless functions, varying in computational resources, and different storage options. 
StarShip adopts the Levenberg–Marquardt optimization method, a combination of gradient descent and the Gauss–Newton method, to select the most suitable storage-function pair for each stage in the workflow. Experimental results showed significant improvements in both service time and cost reduction compared to existing methods.

Overall, these studies show that serverless performance hinges on efficient memory and storage management. Techniques such as data-locality scheduling, prefetching, and adaptive storage allocation mitigate I/O overhead, but scaling them across heterogeneous platforms and complex workflows remains challenging. Further research is essential to fully exploit serverless for data-intensive applications.

\begin{table*}[htbp]
\footnotesize
\centering
\caption{Summary of studies on evaluation and testing techniques for serverless platforms and applications.}
\label{tab:evaluation_testing}
\rowcolors{2}{white}{lightaccessblue}
\begin{tabularx}{\textwidth}{>{\raggedright\arraybackslash}p{3.10cm} >{\raggedright\arraybackslash}X >{\raggedright\arraybackslash}l}
\toprule
{\footnotesize\sffamily\bfseries\textcolor{accessblue}{Study}} & 
{\footnotesize\sffamily\bfseries\textcolor{accessblue}{Proposed Solution}} & 
{\footnotesize\sffamily\bfseries\textcolor{accessblue}{Technique}} \\
\midrule

{\contentfont Spillner et al.~\cite{spillner_faaster_2018}} 
& {\contentfont Compare the performance of the FaaS model with that of conventional monolithic algorithm execution for scientific workloads.} 
& {\contentfont Benchmarking} \\

{\contentfont Malla et al.~\cite{malla_hpc_2020}} 
& {\contentfont Compare cost and performance of FaaS and IaaS platforms for HPC applications.} 
& {\contentfont Benchmarking} \\

{\contentfont Decker et al.~\cite{decker_performance_2022}} 
& {\contentfont Assess the performance of two serverless frameworks for HPC applications.} 
& {\contentfont Benchmarking} \\

{\contentfont Mileski et al.~\cite{mileski_high-performance_2023}} 
& {\contentfont Design a stress-testing tool for evaluating serverless and cloud services.} 
& {\contentfont Load Generation} \\

{\contentfont Mastenbroek et al.~\cite{mastenbroek_opendc_2021}} 
& {\contentfont Provide a simulation environment for modeling serverless platforms.} 
& {\contentfont Simulation} \\

{\contentfont Talluri et al.~\cite{talluri_exde_2024}} 
& {\contentfont Provide a framework to evaluate scheduling mechanisms using real traces.} 
& {\contentfont Simulation} \\

{\contentfont Bauer et al.~\cite{bauer_globus_2024}} 
& {\contentfont Collect and analyze real-world execution traces from a FaaS platform.} 
& {\contentfont Data Collection} \\

\bottomrule
\end{tabularx}
\end{table*}

\subsubsection{Evaluation and Testing}
Evaluating serverless platforms is essential to understand their performance characteristics, scalability, and suitability for diverse workloads.
To this end, recent research has explored several directions, including comparisons with traditional execution models, benchmarking of existing platforms, and the development of dedicated tools and datasets for systematic evaluation.
%
%
The remainder presents an overview of representative studies, while Table~\ref{tab:evaluation_testing} provides a summary.

Numerous studies have compared serverless computing against traditional execution models for compute-intensive workloads.
Spillner et al.~\cite{spillner_faaster_2018} compared FaaS-based approaches against conventional monolithic execution across various scientific and HPC tasks, including $\pi$ calculation, face detection, password cracking, and precipitation forecasting. Results demonstrated that serverless, when supported by proper parallelization strategies and platform-aware optimizations, can serve as viable alternatives for scientific computing.
%
Similarly, Malla et al.~\cite{malla_hpc_2020} analyzed the performance–cost trade-offs between FaaS and IaaS for embarrassingly parallel HPC workloads. The authors deployed a human protein similarity comparison application on Google Cloud Run functions and Google Compute Engine.
Although IaaS achieved $1.65\times$ faster execution, FaaS reduced costs by $14\%$–$40\%$, confirming its suitability for cost-efficient parallel workloads despite moderate performance penalties.
%
%
Expanding this perspective, Decker et al.~\cite{decker_performance_2022} compared two open-source serverless frameworks, OpenFaaS and Nuclio, for HPC-oriented data processing workloads such as image manipulation and byte reversal. Experiments indicated that Nuclio, tailored for data science applications, outperformed OpenFaaS by a factor of $1.5\times$ in terms of data throughput, showing the potential of platforms tailored for scientific workloads.


Beyond comparative studies, serverless computing has also been leveraged as an enabler for performance benchmarking itself. Mileski et al.~\cite{mileski_high-performance_2023} proposed \textit{HPSRG}, a high-performance serverless request generator designed to evaluate the scalability of serverless and cloud services. Capable of generating up to $100{,}000$ requests per second, it was developed within the CardioHPC project to support the evaluation of \textit{ECG Streaming}, a serverless solution for real-time ECG analysis.


Complementary to experimental benchmarking, simulation frameworks provide controlled environments to analyze serverless behavior under configurable and reproducible workloads.
Mastenbroek et al.~\cite{mastenbroek_opendc_2021} presented \textit{OpenDC}, a platform for the simulation of data center operations. It supports a wide range of emerging cloud-datacenter technologies and applications, including serverless computing and ML. The authors validated OpenDC through a series of use cases, encompassing serverless computing with a FaaS platform deployment, machine learning, and \textit{HPC-as-a-Service (HPCaaS)} infrastructures, among numerous others.
%
Building upon this foundation, Talluri et al.~\cite{talluri_exde_2024} extended OpenDC with \textit{ExDe}, a trace-driven discrete-event simulation framework aimed at evaluating scheduler architectures and mechanisms for serverless data-processing systems. ExDe models components such as placers, brokers, and data managers within a conceptual abstraction called the \textit{scheduler frame}, enabling systematic exploration of architectural and coordination strategies. 

Finally, datasets derived from real-world serverless executions provide a complementary foundation for empirical evaluation.
Bauer et al.~\cite{bauer_globus_2024} collected extensive traces from the Globus Compute platform, spanning 31 weeks of activity from 252 users and $277{,}386$ registered functions. This dataset offers valuable insight into workload dynamics and function invocation patterns, supporting further studies on scheduling, cold-start mitigation, and workload characterization.

\begin{table*}[htbp]
\footnotesize
\centering
\caption{Summary of studies on serverless-based applications in scientific and analytical domains.}
\label{tab:serverless_applications}
\rowcolors{2}{white}{lightaccessblue}
\begin{tabularx}{\textwidth}{>{\raggedright\arraybackslash}p{3.50cm} >{\raggedright\arraybackslash}X >{\raggedright\arraybackslash}l}
\toprule
{\footnotesize\sffamily\bfseries\textcolor{accessblue}{Study}} & 
{\footnotesize\sffamily\bfseries\textcolor{accessblue}{Use Case}} & 
{\footnotesize\sffamily\bfseries\textcolor{accessblue}{Domain}} \\
\midrule

{\contentfont Ishakian et al.~\cite{ishakian_serving_2018}} 
& {\contentfont Deep learning inference for image classification using serverless functions.} 
& {\contentfont ML Inference} \\

{\contentfont Jambi~\cite{jambi_serverless_2022}} 
& {\contentfont Crisis prediction via ML inference deployed on Google Cloud Run functions.} 
& {\contentfont ML Inference} \\

{\contentfont Sisniega et al.~\cite{sisniega_efficient_2024}} 
& {\contentfont ML inference pipeline with batch covariate drift detection using OSCAR.} 
& {\contentfont ML Inference} \\

{\contentfont Aytekin et al.~\cite{aytekin_exploiting_2019}} 
& {\contentfont Parallel optimization for regularized logistic regression using AWS Lambda.} 
& {\contentfont Numerical Computing} \\

{\contentfont Gupta et al.~\cite{gupta_oversketch_2018}} 
& {\contentfont Approximate distributed matrix multiplication optimized for AWS Lambda.} 
& {\contentfont Numerical Computing} \\

{\contentfont Witte et al.~\cite{witte_event-driven_2019,witte_event-driven_2020}} 
& {\contentfont Seismic imaging and inversion workflow on AWS Lambda.} 
& {\contentfont Scientific Computing} \\

{\contentfont Ward et al.~\cite{ward_cloud_2023}} 
& {\contentfont AI-guided simulation workflows using Globus Compute.} 
& {\contentfont AI-Driven Simulation} \\

{\contentfont Related Studies~\cite{gusev_cardiohpc_2022,ristov_serverless_2023}} 
& {\contentfont Real-time ECG stream processing using CaaS- and FaaS-based offerings.}  
& {\contentfont Health Monitoring} \\

\bottomrule
\end{tabularx}
\end{table*}

\subsubsection{Serverless-based Applications}
Since the advent of serverless computing, research has increasingly explored porting large-scale, compute-intensive applications designed for HPC clusters to serverless architectures. 
The remainder outlines representative efforts across AI, scientific, and domain-specific applications, with Table~\ref{tab:serverless_applications} summarizing selected works by domain and use case.

There is growing interest in using serverless to handle AI training and inference tasks.
Ishakian et al.~\cite{ishakian_serving_2018} evaluated the suitability of serverless computing  for deep learning inference, developing an image classification application deployed on a FaaS platform. Results indicated that warm starts incur acceptable latency, whereas cold starts may introduce significant overhead.
Similarly, Jambi~\cite{jambi_serverless_2022} implemented a serverless ML inference application to predict real-world crises using social media data, deployed using cloud and serverless services provided by Google Cloud. Results demonstrated that serverless functions can handle thousands of concurrent requests; however, the authors suggested using batching to further reduce latency and cost in ML applications.
Building on this trend, Sisniega et al.~\cite{sisniega_efficient_2024} employed the \textit{OSCAR}~\cite{8814513} framework to implement a batch covariate drift detection system integrated with a ML inference pipeline. 
Evaluation showed that the use of the serverless paradigm enables the decoupling of drift detection from ML tasks without interference, allowing complex inference pipelines to be deployed with minimal overhead and improved resource efficiency.

Beyond AI, several studies examined serverless computing for numerical and simulation-driven workloads.
Aytekin et al.~\cite{aytekin_exploiting_2019} developed a parallel optimization algorithm for solving a regularized logistic regression problem on AWS Lambda. While it scaled well, the authors noticed some limitations, as the stateless nature and limited execution time of FaaS platforms pose challenges for optimization tasks, requiring explicit state management and fault tolerance for long-running algorithms.
%
Differently, Ward et al.~\cite{ward_cloud_2023} implemented two AI-guided simulation workflows on heterogeneous resources leveraging the Globus Compute platform. Evaluation confirmed the feasibility of the approach, achieving performance matching with traditional HPC workflow systems.
Yet, certain HPC applications cannot be efficiently ported in serverless environments as they are, primarily due to platform constraints and the lack of direct inter-function communication. These limitations highlight the need for tailored solutions. For example, Gupta et al.~\cite{gupta_oversketch_2018} proposed \textit{OverSketch}, an approximate distributed matrix multiplication algorithm optimized for serverless execution. Implemented on AWS Lambda, the approach demonstrates how serverless-specific algorithms can improve numerical computing performance, particularly by reducing compute time.
Further, Witte et al.~\cite{witte_event-driven_2019,witte_event-driven_2020} developed a seismic imaging and inversion workflow 
that replaces traditional HPC clusters with a FaaS architecture on AWS Lambda.
Experimental results showed benefits in performance, scalability, and reduced idle times. 
%
Although the event-driven nature of FaaS introduces overhead from repeated resource initialization, it substantially reduces operational costs by eliminating idle resources.
Consequently, serverless deployment emerges as a practical and cost-effective alternative to on-premise clusters for complex scientific workloads.  

Finally, the serverless paradigm has also been explored in the healthcare domain, particularly within the CardioHPC project~\cite{gusev_cardiohpc_2022, ristov_serverless_2023}.
Gusev et al.~\cite{gusev_cardiohpc_2022} presented an application for monitoring electrocardiogram (ECG) data collected from wearable sensors. The authors assessed the scalability of serverless platforms by comparing two distinct implementations: a container-based approach using Google Cloud Run and a FaaS-based approach using AWS Lambda. Experimental results compared the two approaches, providing insights into their respective performance and suitability. Building on this, Ristov et al.~\cite{ristov_serverless_2023} proposed a conceptual architecture for real-time ECG stream processing. The solution integrates both CaaS and FaaS, through Google Cloud Run and AWS Lambda, respectively, assigning them to different components of the workflow to exploit their complementary strengths.

\begin{table*}[htbp]
\footnotesize
\centering
\caption{Summary of studies on modeling techniques for serverless applications.}
\label{tab:app_modeling_serverless}
\rowcolors{2}{white}{lightaccessblue}
\begin{tabularx}{\textwidth}{l>{\raggedright\arraybackslash}Xl}
\toprule
{\footnotesize\sffamily\bfseries\textcolor{accessblue}{Study}} & 
{\footnotesize\sffamily\bfseries\textcolor{accessblue}{Proposed Approach}} & 
{\footnotesize\sffamily\bfseries\textcolor{accessblue}{Focus}} \\
\midrule
{\contentfont Spataru et al.~\cite{spataru_tufa_2022}} 
& {\contentfont Extend TOSCA to integrate support for accelerator-aware services.} 
& {\contentfont Accelerator-aware Modeling} \\

{\contentfont Tom-Ata et al.~\cite{tom-ata_polymorphic_2023}} 
& {\contentfont Analyze application source code to derive alternative architectural forms and recommend suitable configurations using ML/DL techniques.} 
& {\contentfont Architecture Polymorphism} \\
\bottomrule
\end{tabularx}
\end{table*}

\subsubsection{Application Modeling}
With the growing adoption of the serverless paradigm, existing application modeling approaches designed for traditional cloud environments must be adapted to capture its unique deployment characteristics. Such frameworks help structure applications to improve efficiency and reduce operational costs. The remainder outlines research efforts in this direction, while Table~\ref{tab:app_modeling_serverless} summarizes the selected studies.

TOSCA is one of the most widely used languages for describing relationships and dependencies between services and components in cloud applications.
While well established for traditional cloud deployments, it requires extensions to address the specific characteristics of serverless applications.
%
Spataru et al.~\cite{spataru_tufa_2022} took an initial step in this direction with \textit{TUFA}, an extension of TOSCA that incorporates accelerator-aware services. TUFA enables the specification of applications that include GPUs, FPGAs, and MICs in cloud-continuum deployments. Although it does not yet address the serverless paradigm, future extensions are planned as part of the EU-funded SERRANO project, which aims to support FaaS-based operational workflows.
%
%
Complementarily, Tom-Ata et al.~\cite{tom-ata_polymorphic_2023} proposed a static analysis approach to extract insights from source code and identify alternative polymorphic architectural forms to improve performance. These forms span a range of execution models, from container-based to serverless functions. 
The authors developed a pipeline that integrates machine learning and deep learning techniques to recommend architectural variants and hardware configurations that reflect the polymorphic capabilities of the application.

Despite these efforts, research on modeling techniques explicitly designed for serverless remains in its early stages, with limited works addressing the paradigm’s dynamic and fine-grained nature, thus leaving room for further investigation.

\subsection{RQ3: Use Cases}
\label{subsec:use_cases}
\begin{figure}[ht]
    \centering
    \includegraphics[width=0.95\columnwidth]{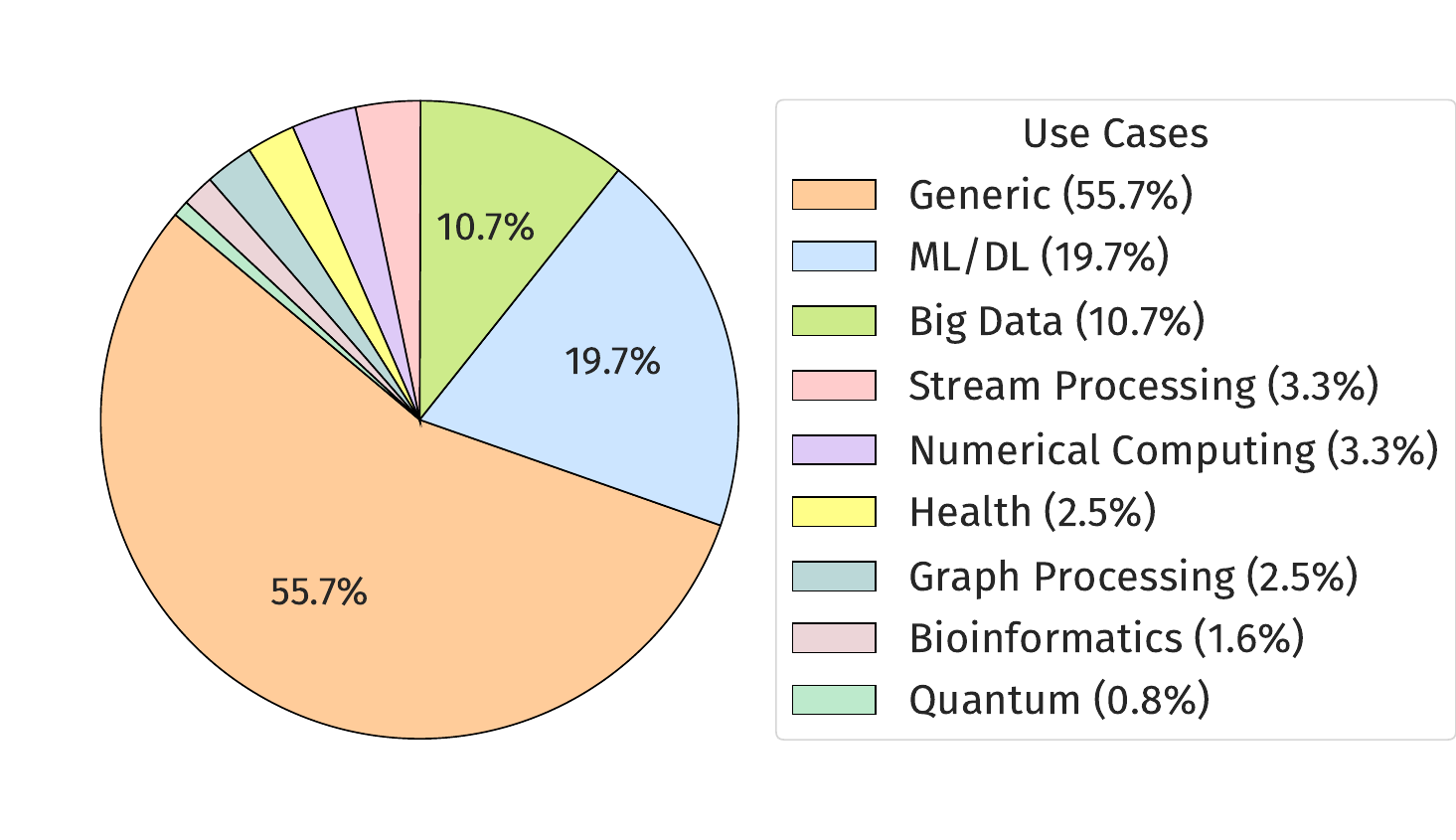}
    \caption{Distribution of use cases addressed in the selected studies.}
    \label{fig:usecasedist}
\end{figure}

To answer RQ3, we built a taxonomy of use cases following the methodology previously presented in Section~\ref{subsec:taxonomy}. For each research article taken into consideration, we focused on the specific domain addressed and labeled it accordingly. Nevertheless, several works do not focus on specific domains but rather on generic high-performance computing use cases. Hence, we labeled the latter as \textit{“Generic”}. Figure~\ref{fig:usecasedist} provides the distribution of use cases across the selected articles.

The majority of the surveyed research articles, circa $55.7\%$, target generic parallel and compute-intensive workloads, providing contributions across multiple aspects, including platforms~\cite{shillaker_faasm_2020, li_funcx_2022, chard_funcx_2020, copik_rfaas_2023}, frameworks~\cite{perez_serverless_2018, carver_wukong_2020, roy_daydream_2022}, and exploratory studies evaluating the suitability and performance of serverless for scientific applications~\cite{spillner_faaster_2018}. While the serverless landscape includes a wide array of platforms, most are designed for generic cloud and event-driven workloads. As a result, recent research has focused on developing specific solutions for parallel and compute-intensive scenarios~\cite{shillaker_faasm_2020, li_funcx_2022, chard_funcx_2020, copik_rfaas_2023, perez_serverless_2018}, with a particular emphasis on data access and inter-function communication~\cite{pauloski_accelerating_2023, copik_process-as--service_2024, copik_fmi_2023} optimizations, as well as workflow orchestration~\cite{wang_briskchain_2024, roy_mashup_2022}. This highlights the growing need for specialized serverless solutions that can meet the demands and characteristics of high-performance computing use cases.
The second most represented use case is machine learning and deep learning, accounting for $19.7\%$ of the surveyed works. In this context, the literature presents preliminary studies on the use of serverless for ML and DL applications~\cite{ishakian_serving_2018}, investigations into enabling accelerator support to boost performance~\cite{naranjo_accelerated_2020, satzke_efficient_2021, dhakal_fine-grained_2023}, and the design of domain-specific frameworks~\cite{ali_batch_2020, yang_infless_2022, zhang_mark_2019} targeting ML/DL inference. In addition, several studies also focus on resource scheduling and management techniques~\cite{hong_optimus_2024, lannurien_herofake_2023} tailored for ML/DL workloads. From this, we observe a strong interest in optimizing serverless architectures to support ML/DL workloads both at the infrastructure and framework levels, in certain cases also co-designing both~\cite{yang_infless_2022} to achieve better performance and integration.
Further, big data represents another relevant use case, targeted by $10.7\%$ of the surveyed works. Contributions in this area primarily focus on the development of tailored frameworks~\cite{muller_lambada_2020, gimenez-alventosa_framework_2019, sanchez-artigas_seer_2022, eizaguirre_seer_2024} and programming models based on patterns~\cite{mathew_pattern-based_2024}. Due to its inherent scalability and elasticity, serverless computing presents a promising model for big data processing pipelines. This motivates the growing research interest in this domain.
Moreover, stream processing is addressed by $3.3\%$ of the works, with studies focusing on real-time and continuous data workflows~\cite{rotchford_laminar_2024, luckow_performance_2019, cheng_autonomous_2018, yang_exploring_2023}. Numerical computing is another area of interest, comprising $3.3\%$ of the works, and includes studies proposing serverless approaches for distributed and event-driven computation~\cite{gupta_oversketch_2018, witte_event-driven_2020, kica_serverless_2023}.
Health-related applications also emerge as a relevant use case, particularly through works associated with the CardioHPC project~\cite{ristov_serverless_2023, mileski_high-performance_2023, gusev_cardiohpc_2022}, which account for approximately $2.5\%$ of the reviewed studies. These studies demonstrate the potential of serverless to support real-time data analytics in healthcare environments and highlight the flexibility and multifaceted nature of this paradigm.
Graph processing is another addressed use case, comprising $2.5\%$ of the surveyed works. Here, the majority of contributions focus on adapting the serverless paradigm for large-scale graph analytics~\cite{toader_graphless_2019, prodan_towards_2022, farahani_towards_2023}.
Finally, less represented but notable use cases include bioinformatics~\cite{aubin_helastic_2021, da_costa_marques_preliminary_2023}, comprising $1.6\%$ of the works, and quantum computing, represented by a single preliminary study~\cite{nguyen_qfaas_2024}, highlighting the versatility of the serverless paradigm across a wide range of domains.

\subsection{RQ4: Publication Venues and Trends}
\label{subsec:trends}
%
To answer RQ4, we conducted an in-depth analysis of publication trends over the selected period and examined the venues in which these studies appeared. We also analyzed the most influential works by citation count, highlighting those that have most significantly shaped research on high-performance serverless computing. The following summarizes our findings.

\subsubsection{Trends}
\begin{figure}[ht]
    \centering
    \includegraphics[width=0.85\columnwidth]{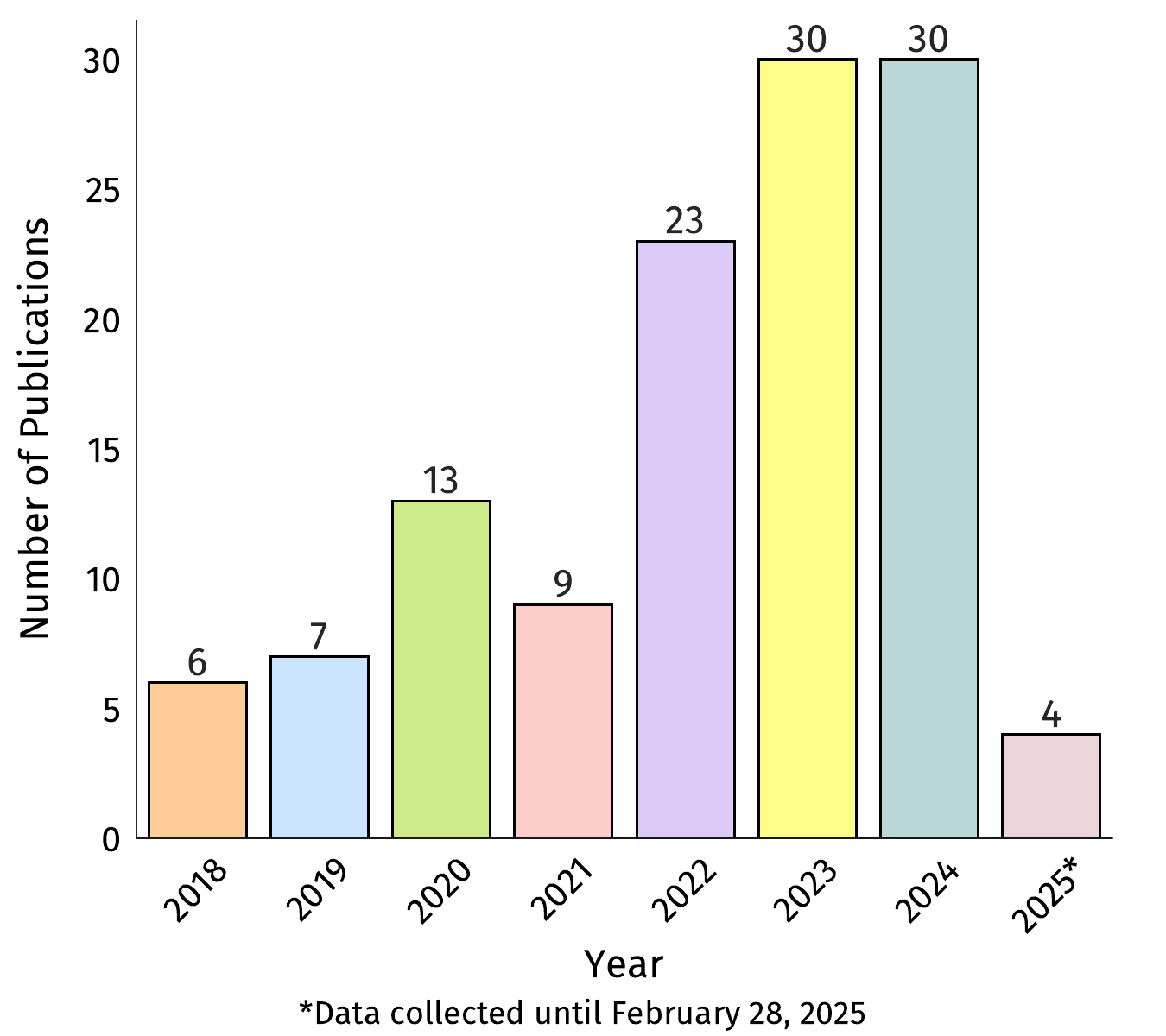}
    \caption{Publication trends over the years.}
    \label{fig:histoyears}
\end{figure}

\begin{figure}[ht]
    \centering
    \includegraphics[width=0.80\columnwidth]{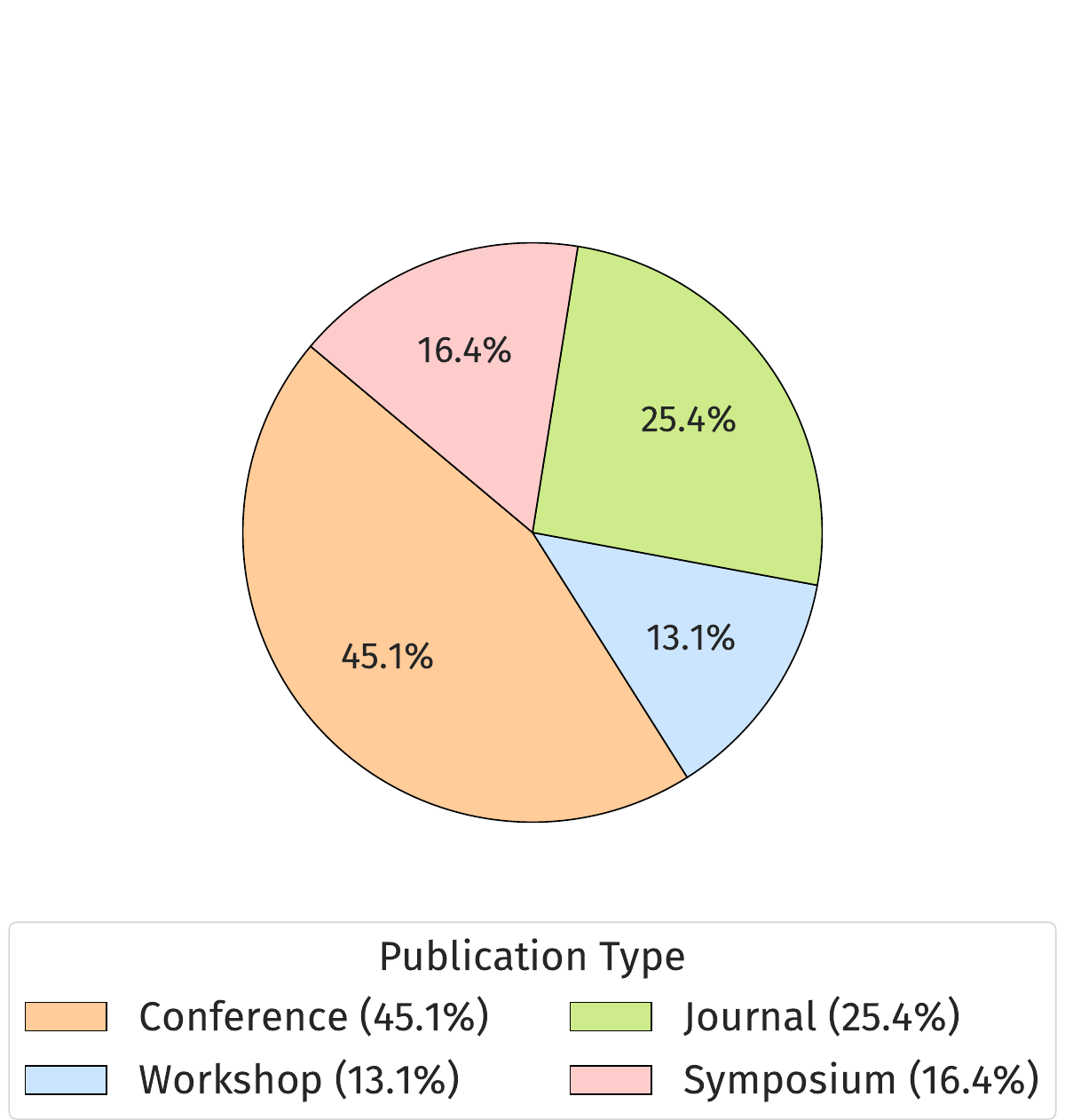}
    \caption{Distribution of selected works across publication venues.}
    \label{fig:piepubs}
\end{figure}

We examined publication trends from 2018 onward, when early studies began to explore the potential of serverless computing for accelerating HPC, AI, and big data workloads. Initial contributions focused on leveraging this execution model for parallel scientific computations~\cite{gupta_oversketch_2018, spillner_faaster_2018}, highlighting how the elasticity and event-driven nature of serverless platforms can improve performance and resource utilization. These characteristics align well with scientific applications composed of independent, parallelizable tasks.
Since then, as shown in Figure~\ref{fig:histoyears}, interest in the field has steadily grown, with notable contributions applying this execution model to machine learning and graph processing~\cite{zhang_mark_2019, toader_graphless_2019, ali_batch_2020}. Although there was a downward trend in publication counts during 2020 and 2021—consistent with the global COVID-19 pandemic—a constant increase in publications has been observed since 2021. 
Finally, recent works increasingly focus on designing more complex serverless frameworks and platforms aimed at high-performance workloads~\cite{copik_rfaas_2023, chen_switchflow_2024, ejarque_enabling_2022, li_unifaas_2024}, reflecting a shift from feasibility studies to architecture design and performance optimization, hence indicating that this research field has reached a higher level of maturity.

\subsubsection{Influential Research Works}
\label{subsubsec:works}
\begin{table*}[t]
\footnotesize
\centering
\caption{Top-cited studies within the selected works as of February 28, 2025.}
\label{tab:mostcitedpapers}
\rowcolors{2}{white}{lightaccessblue}
\begin{tabularx}{\textwidth}{
  >{\raggedright\arraybackslash}p{2.7cm}  
  >{\raggedright\arraybackslash}X         
  >{\raggedright\arraybackslash}p{3.9cm}  
  >{\centering\arraybackslash}p{1.0cm}    
}
\toprule
{\footnotesize\sffamily\bfseries\textcolor{accessblue}{Study}} &
{\footnotesize\sffamily\bfseries\textcolor{accessblue}{Research Direction}} &
{\footnotesize\sffamily\bfseries\textcolor{accessblue}{Sub-category}} &
{\footnotesize\sffamily\bfseries\textcolor{accessblue}{Citations}} \\
\midrule

{\contentfont Shillaker et al.~\cite{shillaker_faasm_2020}} &
{\contentfont Platforms and Frameworks} &
{\contentfont Platforms} &
{\contentfont 239} \\

{\contentfont Chard et al.~\cite{chard_funcx_2020}} &
{\contentfont Platforms and Frameworks} &
{\contentfont Platforms} &
{\contentfont 146} \\

{\contentfont Malawski et al.~\cite{malawski_serverless_2020}} &
{\contentfont Platforms and Frameworks} &
{\contentfont General Purpose Framework} &
{\contentfont 108} \\

{\contentfont Perez et al.~\cite{perez_serverless_2018}} &
{\contentfont Platforms and Frameworks} &
{\contentfont General Purpose Framework} &
{\contentfont 100} \\

{\contentfont Ishakian et al.~\cite{ishakian_serving_2018}} &
{\contentfont Platforms and Frameworks} &
{\contentfont General Purpose Framework} &
{\contentfont 95} \\

{\contentfont Müller et al.~\cite{muller_lambada_2020}} &
{\contentfont Platforms and Frameworks} &
{\contentfont Domain-Specific Framework} &
{\contentfont 87} \\

{\contentfont Spillner et al.~\cite{spillner_faaster_2018}} &
{\contentfont Evaluation and Testing} &
{\contentfont Performance Comparison} &
{\contentfont 86} \\

{\contentfont Carver et al.~\cite{carver_wukong_2020}} &
{\contentfont Platforms and Frameworks, Scheduling and Resource Management} &
{\contentfont General Purpose Framework} &
{\contentfont 79} \\

{\contentfont Ali et al.~\cite{ali_batch_2020}} &
{\contentfont Platforms and Frameworks} &
{\contentfont Domain-Specific Framework} &
{\contentfont 75} \\

{\contentfont Gupta et al.~\cite{gupta_oversketch_2018}} &
{\contentfont Serverless-based Applications} &
{\contentfont Numerical Computing} &
{\contentfont 65} \\

\bottomrule
\end{tabularx}
\end{table*}

We analyzed citation counts for all 122 selected studies as of February 28, 2025, to assess their impact and contributions within the field of high-performance serverless computing. Table~\ref{tab:mostcitedpapers} presents the top 10 most cited papers, along their research directions, sub-categories, and citation counts.

The most cited work is by Shillaker et al.~\cite{shillaker_faasm_2020}, which introduced Faasm, a distributed serverless runtime for data-intensive workloads that enables stateful functions to share memory.
With 239 citations, this study has attracted considerable attention for its exploration of efficient inter-function communication and shared-memory mechanisms, addressing a fundamental limitation of traditional FaaS platforms. 

The second most influential work, by Chard et al.~\cite{chard_funcx_2020}, presented funcX (now Globus Compute), a federated serverless platform for scientific computing across heterogeneous infrastructures. 
With 146 citations, this platform addresses the specific needs of scientific workflows, including low-latency execution, seamless interoperability between cloud and HPC resources, and integration with institutional authentication.

Among the top 10 most cited papers, the majority fall within the \textit{Platforms and Frameworks} research direction. Specifically, four works propose general-purpose frameworks~\cite{malawski_serverless_2020, perez_serverless_2018, ishakian_serving_2018, carver_wukong_2020}, two introduce domain-specific frameworks~\cite{muller_lambada_2020, ali_batch_2020}, and two present complete serverless platforms~\cite{shillaker_faasm_2020, chard_funcx_2020}. This distribution shows the strong interest in the research community in specialized platforms and frameworks to support high-performance workloads, representing a primary focus area within high-performance serverless computing research.
In particular, Carver et al.~\cite{carver_wukong_2020}, with 79 citations, spans multiple directions, including both \textit{Platforms and Frameworks} and \textit{Scheduling and Resource Management}. The proposed framework addresses resource scheduling and management for serverless workflows in high-performance contexts, showing the close relationship between platform design and resource orchestration in this domain.

On the other hand, Spillner et al.~\cite{spillner_faaster_2018} is the only study among the top ten classified under \textit{Evaluation and Testing}. With 86 citations, this early work provides key insights into the performance characteristics and feasibility of serverless for scientific applications.
Similarly, Gupta et al.~\cite{gupta_oversketch_2018} is the only contribution from the \textit{Serverless-based Applications} direction among the top ten, with 65 citations. This early exploratory work focused on applying serverless computing to numerical applications, assessing the feasibility and potential benefits of the paradigm for compute-intensive scientific workloads.

Collectively, these works reveal that platform and framework development, combined with efficient data communication and resource management strategies, represent the most influential research contributions in high-performance serverless computing. The prominence of works on inter-function communication, memory sharing, and execution across heterogeneous resources reflects the recognition from the research community of these aspects as key enablers for adopting serverless in HPC, AI, and big data domains.

\begin{table*}[htbp]
\centering
\footnotesize
\caption{Full names and corresponding acronyms of publication venues.}
\label{tab:venue_names}
\rowcolors{2}{white}{lightaccessblue}
\begin{tabularx}{\textwidth}{l>{\raggedright\arraybackslash}X}
\toprule
{\footnotesize\sffamily\bfseries\textcolor{accessblue}{Acronym}} & 
{\footnotesize\sffamily\bfseries\textcolor{accessblue}{Full Name}} \\
\midrule
\contentfont\textbf{SC} & \contentfont\textit{International Conference for High Performance Computing, Networking, Storage, and Analysis} \\
\contentfont\textbf{ASPLOS} & \contentfont\textit{ACM International Conference on Architectural Support for Programming Languages and Operating Systems} \\
\contentfont\textbf{CLOUD} & \contentfont\textit{IEEE International Conference on Cloud Computing} \\
\contentfont\textbf{IC2E} & \contentfont\textit{IEEE International Conference on Cloud Engineering} \\
\contentfont\textbf{EuroSys} & \contentfont\textit{European Conference on Computer Systems} \\
\contentfont\textbf{FGCS} & \contentfont\textit{Future Generation Computer Systems} \\
\contentfont\textbf{JPDC} & \contentfont\textit{Journal of Parallel and Distributed Computing} \\
\contentfont\textbf{Computer} & \contentfont\textit{IEEE Computer} \\
\contentfont\textbf{TPDS} & \contentfont\textit{IEEE Transactions on Parallel and Distributed Systems} \\
\contentfont\textbf{IPDPS} & \contentfont\textit{IEEE International Parallel and Distributed Processing Symposium} \\
\contentfont\textbf{HPDC} & \contentfont\textit{International Symposium on High-Performance Parallel and Distributed Computing} \\
\contentfont\textbf{CCGrid} & \contentfont\textit{IEEE/ACM International Symposium on Cluster, Cloud and Internet Computing} \\
\contentfont\textbf{SoCC} & \contentfont\textit{ACM Symposium on Cloud Computing} \\
\bottomrule
\end{tabularx}
\end{table*}


\begin{table}[htbp]
\small
\centering
\caption{Most frequent publication venues by category.}
\label{tab:top_venues}

\begin{minipage}{0.49\textwidth}
\centering
\rowcolors{2}{white}{lightaccessblue}
\begin{tabularx}{\textwidth}{>{\raggedright\arraybackslash}X|r|r}
\toprule
{\footnotesize\sffamily\bfseries\textcolor{accessblue}{Conference}} & 
{\footnotesize\sffamily\bfseries\textcolor{accessblue}{Works}} & 
{\footnotesize\sffamily\bfseries\textcolor{accessblue}{\%}} \\
\midrule
\contentfont SC & \contentfont 7 & \contentfont 12.7\% \\
\contentfont ASPLOS & \contentfont 5 & \contentfont 9.1\% \\
\contentfont CLOUD & \contentfont 3 & \contentfont 5.5\% \\
\contentfont IC2E & \contentfont 3 & \contentfont 5.5\% \\
\contentfont \textbf{Others ($\leq$ 2)} & \contentfont 37 & \contentfont 67.3\% \\
\bottomrule
\end{tabularx}
\end{minipage}
\hfill
\vspace{1em}
\begin{minipage}{0.49\textwidth}
\centering
\rowcolors{2}{white}{lightaccessblue}
\begin{tabularx}{\textwidth}{>{\raggedright\arraybackslash}X|r|r}
\toprule
{\footnotesize\sffamily\bfseries\textcolor{accessblue}{Workshop}} & 
{\footnotesize\sffamily\bfseries\textcolor{accessblue}{Works}} & 
{\footnotesize\sffamily\bfseries\textcolor{accessblue}{\%}} \\
\midrule
\contentfont Workshop of SC & \contentfont 6 & \contentfont 37.5\% \\
\contentfont Workshop of EuroSys & \contentfont 2 & \contentfont 12.5\% \\
\contentfont Workshop of IPDPS & \contentfont 2 & \contentfont 12.5\% \\
\contentfont Workshop of HPDC & \contentfont 1 & \contentfont 6.2\% \\
\contentfont \textbf{Others ($\leq$ 1)} & \contentfont 5 & \contentfont 31.2\% \\
\bottomrule
\end{tabularx}
\end{minipage}

\vspace{1em}

\begin{minipage}{0.49\textwidth}
\centering
\rowcolors{2}{white}{lightaccessblue}
\begin{tabularx}{\textwidth}{>{\raggedright\arraybackslash}X|r|r}
\toprule
{\footnotesize\sffamily\bfseries\textcolor{accessblue}{Journal}} & 
{\footnotesize\sffamily\bfseries\textcolor{accessblue}{Works}} & 
{\footnotesize\sffamily\bfseries\textcolor{accessblue}{\%}} \\
\midrule
\contentfont FGCS & \contentfont 13 & \contentfont 41.9\% \\
\contentfont JPDC & \contentfont 3 & \contentfont 9.7\% \\
\contentfont Computer & \contentfont 2 & \contentfont 6.5\% \\
\contentfont TPDS & \contentfont 2 & \contentfont 6.5\% \\
\contentfont \textbf{Others ($\leq$ 1)} & \contentfont 11 & \contentfont 35.5\% \\
\bottomrule
\end{tabularx}
\end{minipage}
\hfill
\vspace{1em}
\begin{minipage}{0.49\textwidth}
\centering
\rowcolors{2}{white}{lightaccessblue}
\begin{tabularx}{\textwidth}{>{\raggedright\arraybackslash}X|r|r}
\toprule
{\footnotesize\sffamily\bfseries\textcolor{accessblue}{Symposium}} & 
{\footnotesize\sffamily\bfseries\textcolor{accessblue}{Works}} & 
{\footnotesize\sffamily\bfseries\textcolor{accessblue}{\%}} \\
\midrule
\contentfont IPDPS & \contentfont 6 & \contentfont 30.0\% \\
\contentfont HPDC & \contentfont 5 & \contentfont 25.0\% \\
\contentfont CCGrid & \contentfont 3 & \contentfont 15.0\% \\
\contentfont SoCC & \contentfont 3 & \contentfont 15.0\% \\
\contentfont \textbf{Others ($\leq$ 1)} & \contentfont 3 & \contentfont 15.0\% \\
\bottomrule
\end{tabularx}
\end{minipage}
\end{table}

\subsubsection{Publication Venues}
We conducted an analysis of publication venues to better understand how the serverless paradigm has been adopted in the scientific community over time. As shown in Figure~\ref{fig:piepubs}, $45.1\%$ of the papers were published in conferences, $25.4\%$ in journals, $16.4\%$ in symposiums, and $13.1\%$ in workshops, which typically represent early-stage or exploratory research.
Table~\ref{tab:top_venues} summarize the top four venues for each category. Venues with one or two contributions are aggregated under the \textit{"Others"} label.
Moreover, to provide a clearer overview, we report in Table~\ref{tab:venue_names} the full names of the most frequently encountered venues and the relative acronyms. 

A noticeable part of the publication, $45.1\%$, appeared in conferences. The \textit{International Conference for High Performance Computing, Networking, Storage, and Analysis (SC)} accounts for the largest number of contributions. Given its strong focus on high-performance computing, its prominent role confirms the relevance of serverless computing for HPC-related research. 
One quarter of the works were published in journals. These contributions often present mature results, with detailed methodologies and comprehensive evaluations. Among journal venues, the majority of articles appeared in \textit{Future Generation Computer Systems (FGCS)}, which aligns with its scope focused on distributed and high-performance systems, cloud infrastructures, and large-scale data processing. This further highlights the applicability of the serverless paradigm in domains such as HPC and big data analytics.  
Roughly $16.4\%$ of the works were published in symposiums. The most represented among them is the \textit{International Parallel and Distributed Processing Symposium (IPDPS)}, which covers a broad range of topics in parallel and distributed computing. This aligns with the observation that serverless is increasingly explored as an execution model suitable for highly parallel and scalable workloads. Notable contributions published at IPDPS include studies on resource disaggregation~\cite{copik_software_2024} and deep learning inference~\cite{ward_cloud_2023}.
Finally, $13.1\%$ of the contributions appeared in workshops. These typically present early-stage or exploratory work, often aiming to assess feasibility or introduce preliminary designs. Examples include initial studies exploring the use of serverless platforms for scientific computing and real-time analytics~\cite{chahal_isesa_2022, gusev_cardiohpc_2022, sly-delgado_taskvine_2023, rotchford_laminar_2024}.

\subsection{RQ5: Author Influence and Collaborations}
\label{subsec:bibliometric}
To provide a comprehensive overview of the current research landscape of serverless computing, we conducted a bibliometric analysis of the 122 selected papers. 
First, we performed a co-authorship network analysis to map the relationships among researchers, identify central authors, and reveal patterns of collaboration across institutions. 
Second, we examined the most-cited authors of the surveyed papers to identify the main contributors.
The following section presents our findings.

\subsubsection{Co-Authorship Network and Collaboration Patterns}
We constructed a network graph starting from the 122 papers, focusing on authors with three or more publications to emphasize sustained contributions.
Out of the 496 identified authors, only 31 ($6.25\%$ of the total) met the inclusion threshold, resulting in a network comprising 66 edges.

\begin{figure*}[ht]
    \centering
    \includegraphics[width=0.90\textwidth]{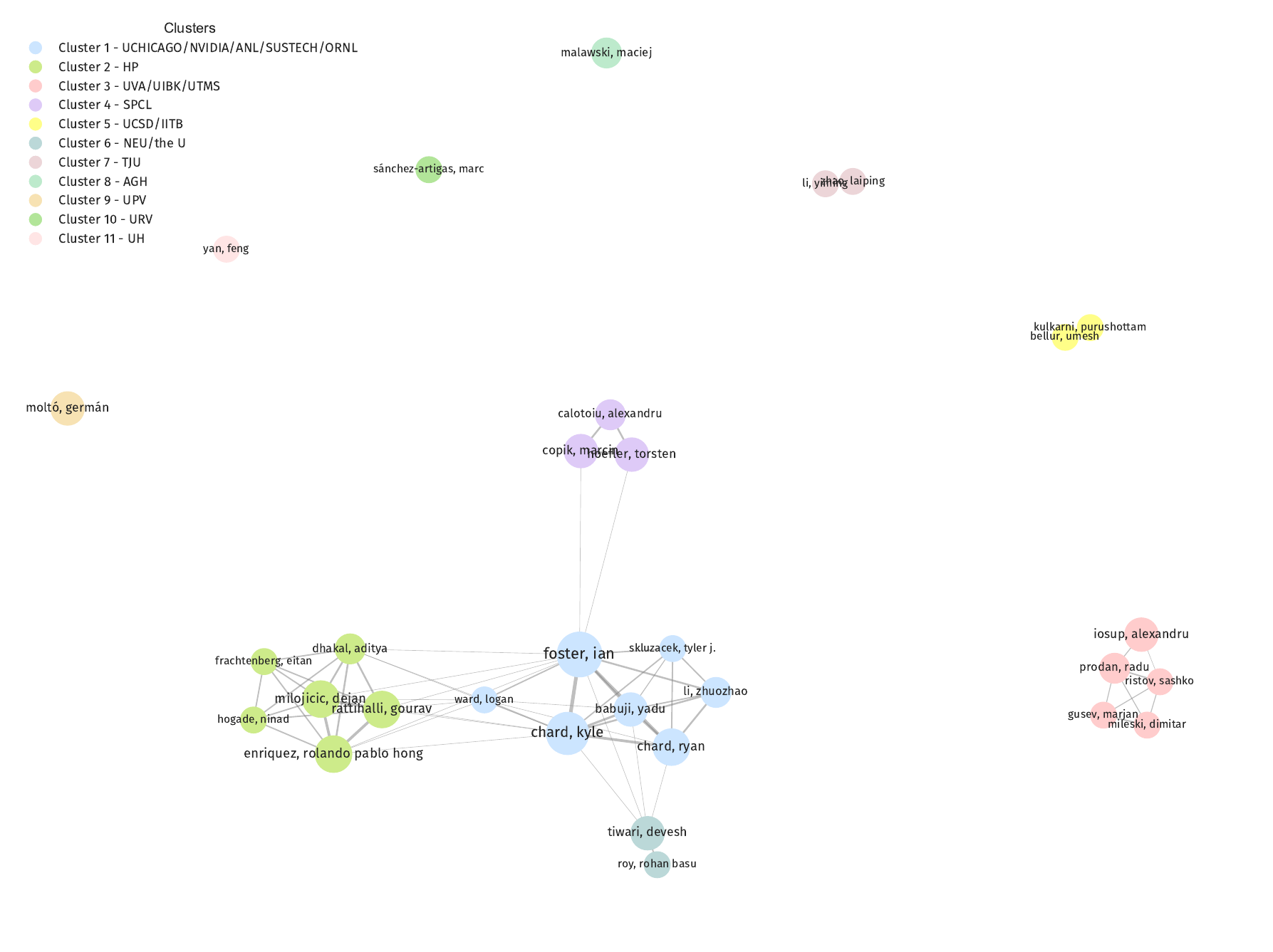}
    \caption{Co-authorship network of authors with three or more published papers.}
    \label{fig:coauthgraph}
\end{figure*}

Figure~\ref{fig:coauthgraph} depicts the resulting co-authorship network, where nodes represent authors and the edges indicate co-authorship. 
The size of each node represents the number of publications, its degree indicates how many authors the node collaborated with at least once, and the edge strength reflects the number of co-authored works between two authors.

We applied the VOS clustering method~\cite{WALTMAN2010629}, which detected 11 clusters. 
The distance between nodes and clusters represents an approximate measure of the similarity within the set.
The following describes each cluster, with percentages indicating its relative size within the set of 31 authors:
\begin{enumerate}
    \item \textit{Cluster 1}: Formed by 7 nodes ($22.6\%$). It includes researchers from the University of Chicago, Nvidia, Argonne National Laboratory, Southern University of Science and Technology, and Oak Ridge National Laboratory. The focus of the works within this cluster is on federated serverless for scientific computing~\cite{chard_funcx_2020, li_funcx_2022, li_unifaas_2024, pauloski_accelerating_2023, ward_cloud_2023, dhakal_fine-grained_2023, skluzacek_serverless_2021}, hybrid HPC-Cloud serverless architectures~\cite{hoefler_xaas_2024}, and related studies~\cite{bauer_globus_2024}.
    \item \textit{Cluster 2}: Formed by 6 nodes ($19.4\%$). It comprises researchers from Hewlett Packard and works on serverless for machine learning~\cite{dhakal_fine-grained_2023, rattihalli_fine-grained_2023, bruel_predicting_2024}, for scientific workflows~\cite{andrei_da_silva_enabling_2025, thurimella_serverless_2025}, and accelerator support~\cite{pfandzelter_kernel-as--service_2023}.
    \item \textit{Cluster 3}: Formed by 5 nodes ($16.1\%$). It contains researchers from the University of Skopje, University of Innsbruck, and  University of Amsterdam. Among the 8 contributions, three of them focus on serverless graph processing~\cite{toader_graphless_2019, prodan_towards_2022, farahani_towards_2023}, two on healthcare~\cite{gusev_cardiohpc_2022, ristov_serverless_2023}, and three on scheduling techniques for data processing~\cite{talluri_exde_2024} and other aspects~\cite{mileski_high-performance_2023, mastenbroek_opendc_2021}.
    \item \textit{Cluster 4}: Formed by 3 nodes ($9.7\%$). It includes researchers from the ETH Z\"urich with contributions on aspects such as serverless programming models~\cite{copik_fmi_2023}, serverless execution on HPC clusters~\cite{copik_rfaas_2023}, and paradigms and abstractions~\cite{copik_process-as--service_2024, copik_software_2024, hoefler_xaas_2024}.
    \item \textit{Cluster 5}: Formed by 2 nodes ($6.5\%$). It comprises researchers from the University of California San Diego and Indian Institute of Technology Bombay. It contains contributions on accelerator support~\cite{garg_faaster_2021, prakash_optimizing_2021} and frameworks for serverless workflows~\cite{jana_dagit_2023}.
    \item \textit{Cluster 6}: Formed by 2 nodes ($6.5\%$). It consists of researchers from the University of Utah and Northeastern University. Contributions are mainly focused on serverless support for scientific workflows~\cite{roy_daydream_2022, roy_mashup_2022, basu_roy_starship_2024}.
    \item \textit{Cluster 7}: Formed by 2 nodes ($6.5\%$). It includes researchers from the Tianjin University.  Their contributions revolve around the usage of DPUs in serverless environments~\cite{liu_fuyao_2024}, low-latency, high-throughput inference~\cite{yang_infless_2022}, and deployment models~\cite{li_rethinking_2023}.
    \item \textit{Cluster 8}: Formed by Maciej Malawski from the AGH University of Krakow. His research focuses on  the deployment of scientific applications and workflows on serverless environments~\cite{malawski_serverless_2020, malawski_serverless_2022, kica_serverless_2023} and scheduling approaches for HPC contexts~\cite{przybylski_using_2022}.
    \item \textit{Cluster 9}: Formed by Germán Moltó from the Polytechnic University of Valencia. He addresses topics such as accelerator support~\cite{naranjo_accelerated_2020, naranjo_delgado_acceleration_2023},  platforms and frameworks for high-throughput workloads~\cite{perez_serverless_2018,gimenez-alventosa_framework_2019}, and serverless ML inference~\cite{sisniega_efficient_2024}.
    \item \textit{Cluster 10}: Formed by Marc Sánchez-Artigas from the Universitat Rovira i Virgili. His work focuses on storage optimization~\cite{eizaguirre_seer_2024, sanchez-artigas_seer_2022}, and hybrid execution approaches for unbalanced and irregular workloads~\cite{finol_exploiting_2024}.
    \item \textit{Cluster 11}: Formed by Feng Yan from the University of Houston. His research focuses on techniques for serverless ML inference~\cite{ali_batch_2020, zhang_enabling_2022, zhang_mark_2019}.
\end{enumerate}

Out of the 11 clusters, four (Clusters 1, 2, 4, and 6) together form the \textit{Giant Connected Component (GCC)}, also called the \textit{Largest Connected Component (LCC)}, of the network. The GCC highlights strong interconnections among prominent research groups, including the Globus Lab led by Ian Foster (Cluster 1), the SPCL at ETH Z\"urich led by Torsten Hoefler (Cluster 4), the Goodwill Computing Lab led by Devesh Tiwari (Cluster 6), and affiliated researchers at Hewlett Packard (Cluster 2). Cluster 1 is the largest and most central within the GCC, connecting to Clusters 4, 6, and 2 through three bridging publications~\cite{hoefler_xaas_2024, bauer_globus_2024, dhakal_fine-grained_2023}, which correspond to the links between Clusters 1 and 4, 1 and 6, and 1 and 2, respectively.
Within the cluster 1, the most prolific authors are Ian Foster, Kyle Chard and Ryan Chard, with 9, 8 and 6 total papers published respectively.
In the cluster 2, Rolando Pablo Hong Enriquez, Dejan Milojicic, and Gourav Rattihalli stand out with 6 published papers.
In the cluster 4, Torsten Hoefler and Marcin Copik are the most active researchers in their group, with 5 publications in total.
Finally, in the cluster 6, the smallest in the GCC, Devesh Tiwari is the most active researcher, with 5 contributions.
%
%
%
With a diameter of 4, the GCC shows moderate separation between authors, meaning collaboration paths involve up to four intermediaries. This indicates that active authors frequently collaborate with each~another.

%
The average node degree across the entire graph is $4.26$, this indicates that authors collaborated with roughly four others on average. Within the GCC, this increases to $6.22$, reflecting denser interconnections. Ian Foster, the most connected node, serves as the main hub of the GCC, with a degree of $13$, directly linking to all other GCC groups through his research team. Two additional hub researchers that facilitate cross-cluster collaborations are Dejan Milojicic, with a degree of $8$, and Kyle Chard, with a degree of $11$.
The average link strength across the entire network is $2.82$, rising slightly to $2.91$ within the GCC, indicating marginally stronger repeated collaborations.
Taking into consideration the whole graph, the top five strongest edges, in terms of recurrent collaborations, all reside within the GCC:
\begin{itemize}
    \item Ian Foster and Kyle Chard (8 co-authored works)~\cite{skluzacek_serverless_2021, pauloski_accelerating_2023, ward_cloud_2023, dhakal_fine-grained_2023, li_funcx_2022, chard_funcx_2020, bauer_globus_2024, li_unifaas_2024}.
    \item Kyle Chard and Ryan Chard (6 co-authored works)~\cite{skluzacek_serverless_2021, ward_cloud_2023, li_funcx_2022, chard_funcx_2020, bauer_globus_2024, li_unifaas_2024}.
    \item Ryan Chard and Ian Foster (6 co-authored works)~\cite{skluzacek_serverless_2021, ward_cloud_2023, chard_funcx_2020, li_funcx_2022, bauer_globus_2024, li_unifaas_2024}.
    \item Enriquez Rolando Pablo Hong and Dejan Milojicic (6 co-authored works)~\cite{andrei_da_silva_enabling_2025, dhakal_fine-grained_2023, rattihalli_fine-grained_2023, pfandzelter_kernel-as--service_2023, bruel_predicting_2024, thurimella_serverless_2025}.
    \item Enriquez Rolando Pablo Hong and Gourav Rattihalli (6 co-authored works)~\cite{andrei_da_silva_enabling_2025, dhakal_fine-grained_2023, rattihalli_fine-grained_2023, pfandzelter_kernel-as--service_2023, bruel_predicting_2024, thurimella_serverless_2025}.
\end{itemize}
Overall, these links highlight frequent and impactful collaborations within the serverless computing research community.

The \textit{Total Link Strength (TLS)}, the sum of all edge strengths for a given node, has an average value of $12.0$ across the full graph, increasing to $18.11$ within the GCC. The nodes with the highest TLS values are Ian Foster (36), Kyle Chard (34), and Dejan Milojicic (25). Logan Ward acts as a key connector between Clusters 1 and 2, highlighting strong ties between these research groups. High TLS values indicate frequent and repeated collaborations, reinforcing local network density.

Overall, our co-authorship network analysis highlights a distinct yet growing high-performance serverless computing research community with a structured collaboration pattern. However, the $\ge 3$ publication threshold may exclude less prolific authors, potentially under-representing them. For instance, Shillaker et al.~\cite{shillaker_faasm_2020}, as discussed in Section~\ref{subsubsec:works}, is the most cited work among the surveyed papers, yet none of its authors appear in the co-authorship network we analyzed, indicating that highly impactful contributors may be overlooked.

\subsubsection{Influential Authors}
\begin{figure}[ht]
    \centering    \includegraphics[width=0.85\columnwidth]{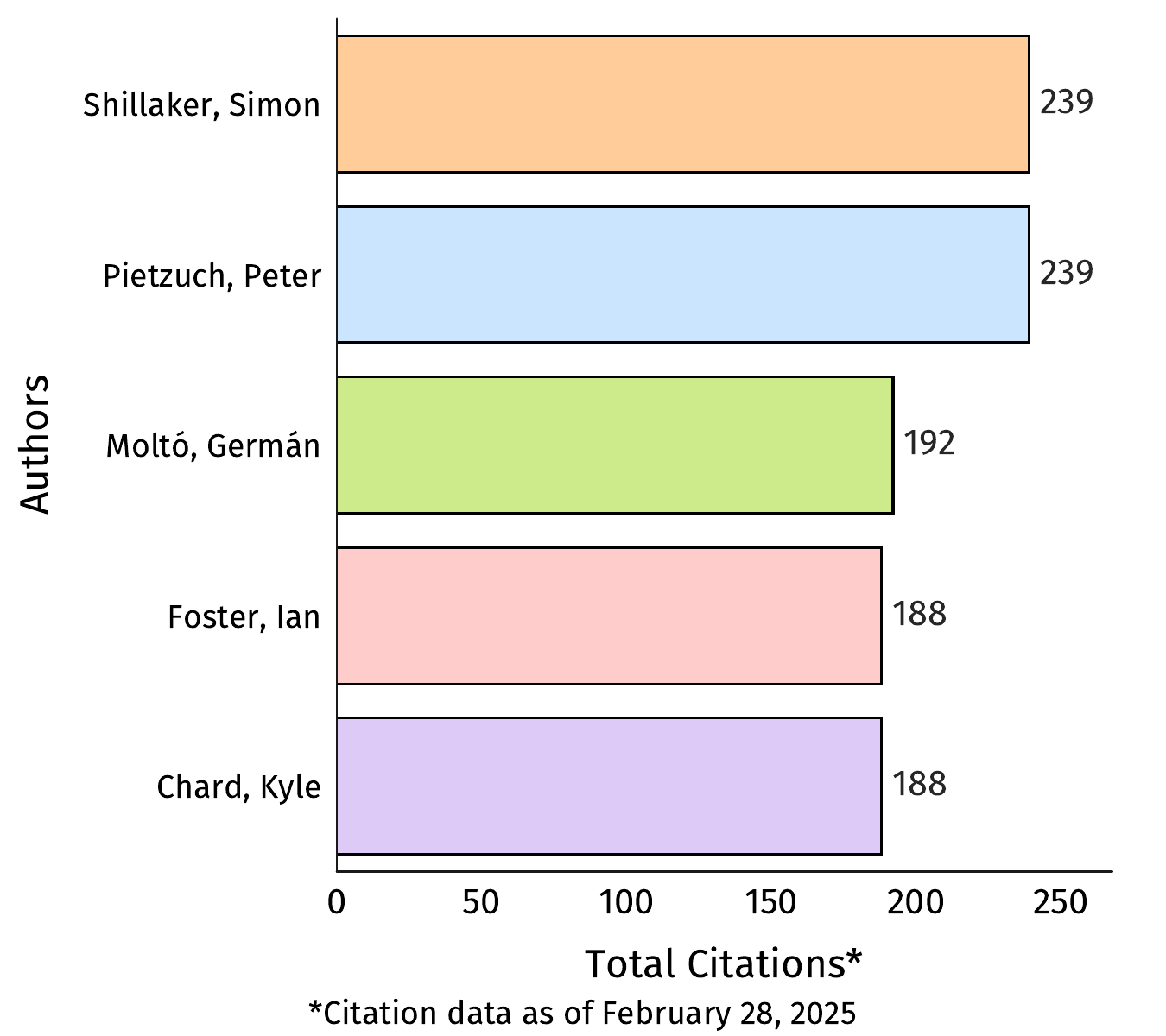}
    \caption{Top five authors by citation count.}
    \label{fig:mostcites}
\end{figure}

We quantified author influence by aggregating total citation counts across all publications, selecting the top five most cited researchers in our dataset. 
As shown in Figure~\ref{fig:mostcites}, the most impactful authors are Simon Shillaker and Peter Pietzuch, both authors of Faasm~\cite{shillaker_faasm_2020}. This work, which proposes direct memory sharing between serverless functions to create a distributed serverless runtime for data-intensive workloads, has achieved 239 citations. As discussed in Section~\ref{subsec:trends}, it is the most cited paper among the surveyed publications, reflecting strong interest in platforms tailored for high-performance serverless computing.
The third most cited author is Germán Moltó~\cite{gimenez-alventosa_framework_2019, naranjo_accelerated_2020, sisniega_efficient_2024, naranjo_delgado_acceleration_2023, perez_serverless_2018}, whose research focuses on serverless and cloud computing, with particular emphasis on hardware accelerator support.
Ian Foster and Kyle Chard are also among the most active researchers in the dataset, collectively authoring eight publications~\cite{li_funcx_2022, chard_funcx_2020, bauer_globus_2024, li_unifaas_2024, dhakal_fine-grained_2023, ward_cloud_2023, pauloski_accelerating_2023, skluzacek_serverless_2021} and achieving a total of 188 citations. Most of their work focuses on the development of the Globus platform, with Chard et al.~\cite{chard_funcx_2020} being the second most cited paper in our dataset.

Our citation analysis reveals that research in high-performance serverless computing is driven by two complementary research areas: platform innovations addressing core architectural challenges and frameworks enabling deployment across heterogeneous infrastructures. Moreover, we noticed sustained contributions in accelerator integration, exemplified by the work of Germán Moltó, underscoring the importance of heterogeneous computing support for AI and HPC workloads. Overall, the presence of highly cited authors indicates that the field is increasingly maturing toward a structured research area while also revealing opportunities for broader community engagement and research collaboration.

\section{Discussion}
\label{sec:discussion}
The evolving landscape of high-performance serverless computing, as revealed by our systematic literature review of 122 papers, reflects both substantial progress and persistent challenges. This section examines the implications of our findings, assesses the maturity of the research field, and identifies key gaps and opportunities for future work that must be addressed to unlock the full potential of serverless computing for HPC, AI, and big data workloads.

\subsection{Synthesis of Key Findings}
Our systematic literature review of 122 papers, published between 2018 and early 2025, reveals that high-performance serverless computing has evolved from isolated research efforts into a more cohesive research area. The eight research directions identified in our taxonomy do not exist in silos. Rather, they represent different parts of a single challenge: making serverless practical for compute-intensive workloads.

As discussed in Section~\ref{subsec:taxonomy}, \textit{Platform and Framework Development} ($29.5\%$) and \textit{Scheduling and Resource Management} ($28.6\%$) dominate the surveyed literature.
This distribution suggests that the field is still grappling with fundamental infrastructure challenges related to resource provisioning, function orchestration, and platform design for HPC, AI, and big data workloads.
The dominance of these directions reflects the recognition of the research community that existing general-purpose serverless platforms, designed primarily for cloud-native applications, fall short of the strict requirements imposed by compute-intensive domains.
\textit{Paradigms and Programming Models} ($11.5\%$) represents an area of high innovation potential. The development of data communication abstractions~\cite{pauloski_accelerating_2023, copik_process-as--service_2024}, MPI-like interfaces~\cite{copik_fmi_2023, yuan_smpi_2022}, and workflow abstractions~\cite{ejarque_enabling_2022} suggests that the field is beginning to address the programmability challenges that may hinder widespread adoption of the serverless execution model by the HPC community. These efforts aim to bridge the gap between traditional HPC programming paradigms and the serverless one.
In contrast, \textit{Application Modeling} remains significantly underexplored ($1.6\%$), despite its importance for both developer adoption and system optimization.

With regard to existing solutions, as presented in Section~\ref{subsec:research_directions}, current research efforts suggest this research field is maturing beyond preliminary studies.
Common challenges such as cold start latency, data communication overhead, and hardware heterogeneity remain persistent barriers across directions, though contributions on direct inter-function communication and invocation~\cite{chen_yuanrong_2024, copik_rfaas_2023} and abstractions such as PraaS~\cite{copik_process-as--service_2024} show promise.
Research directions maturity varies, as research platforms like Globus Compute~\cite{li_funcx_2022, chard_funcx_2020}, Faasm~\cite{shillaker_faasm_2020}, and rFaaS~\cite{copik_rfaas_2023} demonstrate the feasibility of serverless computing for scientific and compute-intensive workloads.
In contrast, research on scheduling has focused mainly on serverless AI/ML inference~\cite{yang_infless_2022, zhao_gpu-enabled_2023, bhasi_paldia_2024}, with a limited focus on general-purpose HPC workloads~\cite{carver_wukong_2020, roy_daydream_2022}.
Challenges also persist in cloud–HPC convergence and hybrid environments, where serverless elasticity complements but does not replace traditional HPC infrastructure. Several works have explored hybrid execution strategies that combine on-premises HPC resources with cloud-based serverless ones~\cite{finol_exploiting_2024, roy_mashup_2022, chahal_isesa_2022, 10.1145/3452413.3464789, aubin_helastic_2021}, exploring dynamic workload offloading during peak demand from on-premises resources to serverless ones, and vice versa.
Notably, only one study addresses quantum computing integration~\cite{nguyen_qfaas_2024}, representing a significant research gap given the growing interest in quantum machine learning, quantum-classical hybrid workflows, and the emerging Quantum-as-a-Service (QaaS) paradigm. This signals an opportunity for future work as quantum computing transitions from experimentation to practical applications.

Regarding the main use cases addressed by the surveyed papers, as discussed in Section~\ref{subsec:use_cases}, our analysis reveals that generic parallel and compute-intensive workloads ($55.7\%$) constitute the main use case, followed by ML/DL applications ($19.7\%$) and big data processing ($10.7\%$).
The strong focus on ML/DL applications aligns with the natural fit between stateless, serverless execution and inference workloads. Domain-specific frameworks for ML inference~\cite{yang_infless_2022, ali_batch_2020, zhang_mark_2019} have demonstrated the feasibility of serverless for ML serving scenarios, especially when combined with batching and accelerator scheduling optimizations.
Further, emerging applications in health ($2.5\%$), bioinformatics ($1.6\%$), and quantum computing ($0.8\%$) underscore the versatility of the serverless paradigm, yet their limited representation in the literature highlights substantial untapped potential. As the field matures, we anticipate increased research attention to solutions tailored to these emerging applications.


The publication trend analysis, presented in Section~\ref{subsec:trends}, shows steady growth of literature on high-performance serverless computing since 2018, with a marked acceleration after 2021. This trend confirms field maturity and increasing recognition. The International Conference for High Performance Computing, Networking, Storage, and Analysis (SC) leads venue contributions with seven publications, confirming the active engagement of the HPC community with serverless computing.
The most cited work, Shillaker et al.~\cite{shillaker_faasm_2020} (239 citations), addresses direct memory sharing between stateful functions, demonstrating the strong interest of the research community in overcoming the constraint of traditional FaaS platforms. The second most cited work, Chard et al.~\cite{chard_funcx_2020} (146 citations), presents Globus Compute, a federated platform for scientific applications that enables seamless execution across heterogeneous resources. The latter shows the level of maturity high-performance serverless platforms have reached.
Together, these works establish foundational contributions that have shaped subsequent research in the field.


The co-authorship network analysis we conducted, as reported in Section~\ref{subsec:bibliometric}, reveals strong interconnections among prominent research groups, with Ian Foster, Kyle Chard, and Dejan Milojicic serving as primary hubs within the collaboration network.
The concentration of expertise within a relatively small number of interconnected research groups indicates both the emerging nature of the field and its potential for broader expansion. The presence of prominent research groups with strong histories in HPC and distributed systems—such as SPCL at ETH Z\"urich and Globus Lab at University of Chicago—signals that serverless computing has attracted established HPC researchers.

Overall, our analysis of the literature indicates that high-performance serverless computing has transitioned from a set of exploratory studies to a more mature and structured research area. While significant research gaps remain, which must be addressed to enable broader adoption, we expect the field to attract increasing attention from both researchers and practitioners in the near future.

\subsection{Open Problems and Limitations}
Despite progress in high-performance serverless computing, several fundamental challenges persist. These challenges stem from inherent architectural characteristics of the serverless paradigm and, in certain cases, can hinder widespread adoption in HPC, AI, and big data domains.
Across the surveyed papers, three challenges consistently emerge: cold start latency, data communication overhead, and hardware heterogeneity.
The following provides an overview of these open problems.

\subsubsection{Cold Start Latency}
Cold start latency is a fundamental challenge for both general-purpose and compute-intensive serverless applications. While general-purpose FaaS platforms typically experience cold starts ranging from hundreds of milliseconds to several seconds, they often leverage heuristics and pre-warming techniques to reduce startup latency in subsequent function invocations.
In the case of serverless AI/ML inference, several of these techniques can be applied as well, for instance by using time-series prediction models combined with pre-warming strategies, as shown in the surveyed literature~\cite{yang_infless_2022, sankaranarayanan_pulse_2025, lu_smiless_2024}.
However, in HPC applications, heuristics based on time-series or historical data may not always be feasible, as such applications are typically executed once or repeated only during defined experimental periods. To address this, some works have focused on mitigating cold starts at the platform level~\cite{shillaker_faasm_2020,fuerst_iluvatar_2023} or through tailored prediction techniques~\cite{roy_daydream_2022}.
Yet, a one-size-fits-all solution does not exist, and different classes of applications (e.g., AI/ML, HPC) may have specific requirements and execution patterns. Consequently, cold start mitigation remains an open and ongoing challenge, representing a promising area for further research.

\subsubsection{Data Communication and State Management} Data communication overhead represents another critical bottleneck. Traditionally, inter-function communication and data access is achieved through mediated communication channels through object storage or message queues, introducing unacceptable latency for tightly coupled parallel applications. 
Current literature addresses this by introducing abstractions and providing interfaces that hide the third-party communication channel~\cite{pauloski_accelerating_2023}, however, these do not resolve the underlying performance problem. Several efforts have focused on direct function invocation using RDMA~\cite{copik_rfaas_2023}, direct inter-function communication through TCP hole punching~\cite{copik_fmi_2023}, or dedicated runtime support~\cite{chen_yuanrong_2024}. 
Direct communication channels represent the natural and more promising solution, yet few studies have partially explored this. This makes it a promising yet challenging research direction, especially due to the implications of the \textit{Serverless Trilemma}~\cite{10.1145/3133850.3133855} in the context of function composition and complex workflows. 
Similarly, state management and data access remain significant challenges that may hinder the adoption of high-performance serverless computing. In serverless service models such as FaaS, functions are inherently stateless. As a result, third-party services such as object storage or key–value stores are typically employed to persist state across function invocations. In the surveyed literature, Faasm~\cite{shillaker_faasm_2020} introduces direct memory sharing between function instances, also enabling state persistence. 
While a promising research direction, only a few works have addressed this problem in the context of compute-intensive, parallel applications. Therefore, achieving scalable stateful execution while preserving the elasticity and fault isolation intrinsic to serverless remains an open challenge.

\subsubsection{Hardware Heterogeneity and Accelerator Integration}
Support for heterogeneous hardware and accelerator integration in serverless platforms has emerged as a critical challenge for high-performance serverless computing.
Current literature presents various approaches to GPU support for serverless applications~\cite{naranjo_accelerated_2020,naranjo_delgado_acceleration_2023,yang_infless_2022,fingler_dgsf_2022,fingler_disaggregated_2023,kim_gpu_2018}. Although these contributions enable GPU access, they lack support for disaggregated GPU resources. Techniques such as NVIDIA MIG and MPS show potential to address this limitation, as demonstrated in several preliminary studies~\cite{satzke_efficient_2021,dhakal_fine-grained_2023}, but introduce additional management complexity. Advanced scheduling strategies have been proposed to support these multiplexing mechanisms~\cite{rattihalli_fine-grained_2023,hui_esg_2024,bhasi_paldia_2024,yang_infless_2022}, and dynamic kernel slicing has also been explored to reduce GPU under-utilization and meet task deadlines~\cite{prakash_optimizing_2021,garg_faaster_2021}. Yet, the majority of these efforts primarily target GPU and AI/ML inference workloads, leaving other compute-intensive workloads largely unexplored.
Recent studies have investigated the integration of alternative accelerators, such as FPGAs~\cite{ringlein_case_2021,bacis_blastfunction_2020}, DPUs~\cite{liu_fuyao_2024,du_serverless_2022}, and NPUs~\cite{ma_widepipe_2021}, into serverless architectures. Although domain-specific accelerators can significantly enhance performance for specialized workloads, current research largely remains at the proof-of-concept stage rather than achieving general-purpose integration within FaaS platforms.
Despite existing efforts, hardware heterogeneity abstraction and accelerator integration remain open challenges that have yet to be fully addressed. Effectively addressing these challenges could enable a wide range of compute-intensive applications to benefit from both accelerators and the serverless execution model.

\subsection{Research Gaps and Opportunities}
While research in high-performance serverless computing is evolving from preliminary studies to a more structured and mature discipline, several critical aspects remain unaddressed, creating notable research gaps in the current literature. The following provides an overview of these gaps and highlights opportunities for future investigation.

\subsubsection{Sustainability and Energy Efficiency}
While serverless computing can improve resource utilization through fine-grained function execution and dynamic scaling, sustainability and energy efficiency in high-performance serverless environments remain largely unexplored. Only a limited number of studies have addressed energy-aware scheduling for heterogeneous accelerators~\cite{lannurien_herofake_2023,rattihalli_fine-grained_2023,bruel_predicting_2024}. These initial efforts demonstrate the potential for reducing energy consumption through profiling-based scheduling policies. However, research in this area is still in its early stages, with most studies focusing on proof-of-concept implementations. Critical questions remain unanswered, including how to balance energy efficiency with performance guarantees under varying workloads, how to optimize energy consumption across heterogeneous hardware configurations, and how to integrate sustainability metrics into resource management policies. Furthermore, the carbon footprint of serverless execution across distributed infrastructures, particularly in the context of large-scale scientific computing and AI workloads, requires systematic investigation. As energy costs and environmental concerns continue to grow, developing energy-aware serverless platforms and scheduling strategies tailored to compute-intensive applications represents a crucial research direction.

\subsubsection{Security}
Security challenges in high-performance serverless computing present unique concerns that differ from traditional cloud environments. The nature of serverless introduces attack surfaces related to function isolation, shared resources, and inter-function communication. While commercial FaaS platforms employ various isolation mechanisms, the surveyed literature reveals a notable absence of security-focused research in the context of HPC and scientific-oriented serverless deployments. Specific challenges include protecting sensitive data in scientific workflows that span heterogeneous and federated infrastructures and ensuring secure accelerator sharing in multi-tenant environments. Moreover, stateful serverless architectures and direct memory sharing mechanisms raise questions about data leakage and side-channel attacks. The lack of security frameworks tailored to high-performance serverless environments highlights the urgent need for research in this domain. Future work must address authentication and authorization in federated serverless platforms, secure data access and communication, and threat models specific to high-performance serverless computing.

\subsubsection{Evaluation and Benchmarks}
The absence of standardized evaluation methodologies and comprehensive benchmarking frameworks represents a significant gap in high-performance serverless computing research. The surveyed literature reveals that most studies employ custom evaluation setups, making it difficult to compare approaches and assess their applicability across different contexts. While some works in the serverless literature provide benchmarking tools and simulation environments~\cite{10.1145/3464298.3476133,10.1145/3689031.3717465,
9610428}, these efforts remain fragmented and target simple general-purpose workloads. The lack of representative benchmark suites for HPC, AI, and big data workloads on serverless platforms hinders both research progress and practical adoption. Critical evaluation dimensions remain underexplored, including performance characterization across diverse hardware configurations, cost-performance trade-offs under realistic workload patterns, and scalability limits of different architectural approaches. The heterogeneity of serverless platforms, varying execution environments, and diverse programming models complicate the establishment of fair comparison baselines. To advance the field, the research community requires standardized benchmarks that capture the characteristics of compute-intensive workloads, reproducible evaluation methodologies that account for platform-specific optimizations, and metrics that extend beyond execution time to include energy consumption, cost-efficiency, and resource utilization. Such advancements would enable researchers to systematically assess trade-offs, identify performance bottlenecks, and guide the design of future high-performance serverless platforms and frameworks.

\section{Conclusion}
\label{sec:conclusion}
Serverless computing is gaining traction as a promising execution model not only for general-purpose cloud workloads but also for highly parallel and distributed workloads typical of high-performance computing environments. 
This has led to the emergence of the research area we term \textit{High-Performance Serverless Computing}, which explores how the serverless paradigm can be exploited to deploy HPC, AI, and big data applications across cloud, HPC, and hybrid infrastructures.

In this work, we have presented a comprehensive systematic literature review aimed at summarizing the current state of research on the use of serverless computing to support HPC, AI, and big data applications. To this end, we systematically collected and analyzed 122 research articles, proposing a taxonomy of eight research directions and nine targeted use case domains.
%
Our analysis reveals that research efforts are primarily focused on platform development and resource management techniques tailored to high-performance workloads, with emerging contributions in accelerator support.
Moreover, we examined the use cases and domains targeted by the surveyed works. Our findings indicate that serverless computing can effectively support parallel workloads, ML/DL, and big data applications, with emerging use cases such as health, bioinformatics, and quantum computing, which point to substantial opportunities for future research.
%
%
Regarding publication trends, our data indicate a steady increase in the number of contributions since 2018, along with a shift from preliminary feasibility studies toward more mature system architectures and optimization techniques. Based on our findings, we expect research in this area to continue growing in the coming years, potentially leading to broader adoption of serverless computing for parallel and compute-intensive applications.
Finally, the analysis of the research community structure revealed strong interconnections among leading research groups, emphasizing both the field’s emerging nature and its potential for broader expansion.

Overall, we believe that our analysis, including the taxonomy of research directions and use cases, the classification of existing solutions, the publication trend analysis, and the examination of the research community structure, provides a valuable foundation for both new researchers and experienced practitioners. In particular, it provides insight into existing architectural solutions, resource management strategies, and programming models that may support future innovations in high-performance serverless computing, and more generally in both serverless and high-performance computing.

\section*{CRediT authorship contribution statement}
\textbf{Valerio Besozzi:} Conceptualization, Methodology, Formal analysis, Writing – original draft, Writing – review \& editing.
\textbf{Matteo Della Bartola:} Data curation, Visualization, Investigation, Writing – review \& editing.
\textbf{Patrizio Dazzi:} Supervision, Validation, Writing – review \& editing.
\textbf{Marco Danelutto:} Supervision, Project administration, Writing – review \& editing.

\section*{Acknowledgment}
The authors would like to thank the editor and the anonymous reviewers for their valuable feedback and suggestions, which helped improve the quality of this work. We also extend our sincere thanks to our dear colleague and friend Alberto Ottimo for his insightful discussions and support during this research.

\bibliographystyle{unsrt}
\bibliography{sample-base}

\phantomsection
\begin{IEEEbiography}[{\includegraphics[width=1in,height=1.25in,clip,keepaspectratio]{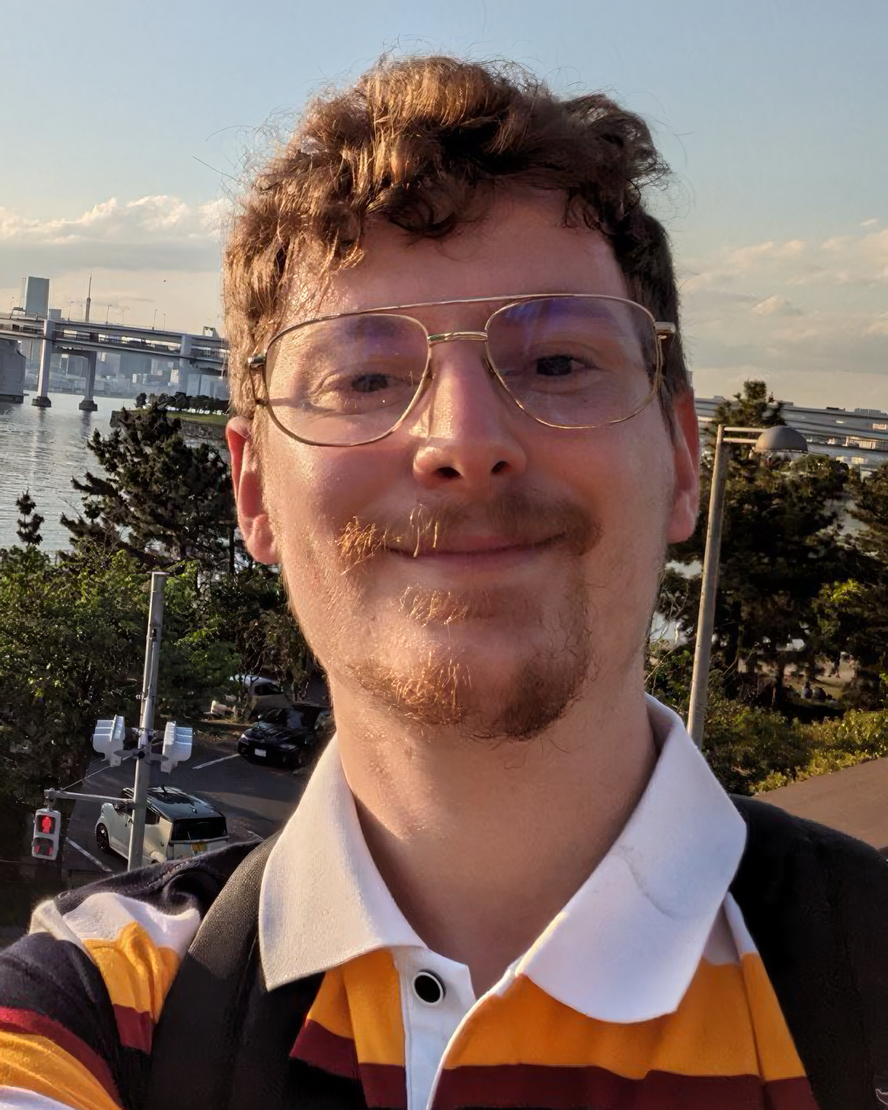}}]{Valerio Besozzi}
received the B.Sc.\ degree in Computer Science in 2021 and the M.Sc.\ degree in Computer Science and Networking in 2023, both from the University of Pisa, Italy. He is currently pursuing the Ph.D.\ degree in Computer Science at the University of Pisa, where he is a member of the Parallel Programming Models (PPM) group. 

He is the main author of the Parallelo Parallel Library (PPL), a skeletal parallel programming library targeting the Rust programming language, and SPARE, a serverless framework under development for urgent edge computing scenarios, integrating unikernels and WebAssembly support. His research interests include serverless computing, structured parallel programming, lightweight virtualization, and accelerator~integration.

Mr.\ Besozzi has served as a Conference Assistant for ACM HPDC 2024, HLPP 2024, and QUATIC 2024. He has been a Program Committee Member for the Artifact Evaluation tracks of Euro-Par 2024 and 2025, and serves as Co-Program Chair and Workshop Organizer for the AHPC3 workshop at PDP 2025 and IC2E 2025.
\end{IEEEbiography}

\begin{IEEEbiography}[{\includegraphics[width=1in,height=1.25in,clip,keepaspectratio]{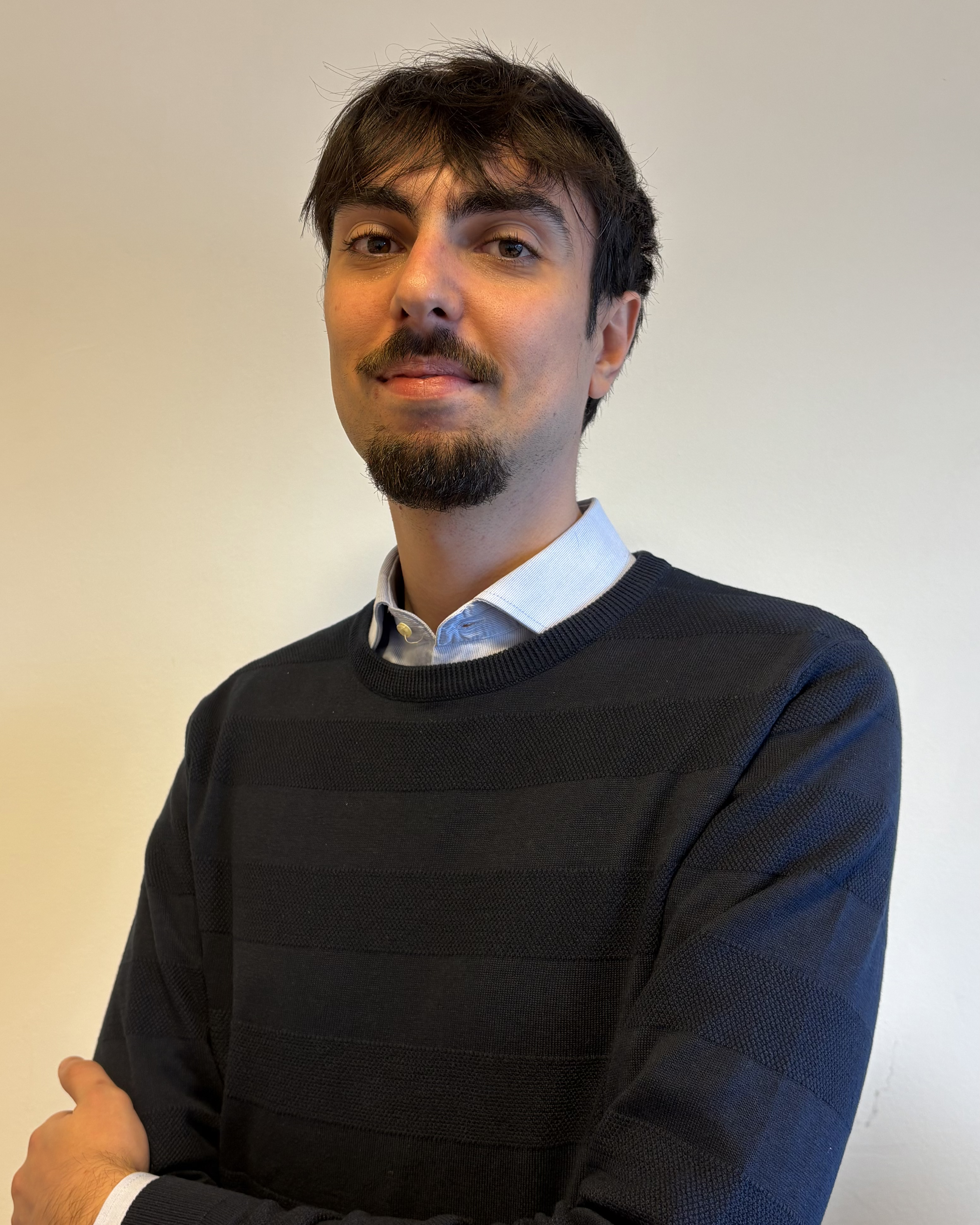}}]{Matteo Della Bartola}
received the B.Sc.\ degree in Computer Science in 2022 and the M.Sc.\ degree in Computer Science and Networking in 2024, both from the University of Pisa, Pisa, Italy. He is currently pursuing the Ph.D.\ degree in Computer Science at the University of Pisa, where he is a member of the Parallel Programming Models (PPM) group.

His research interests include parallel programming models and algorithmic skeletons, cloud computing, high-performance computing, and serverless computing.

Mr.\ Della Bartola has served as a Conference Assistant for QUATIC 2024. He is also a participant representing the University of Pisa in the EU-funded project NOUS, A catalyst for EuropeaN ClOUd Services in the era of data spaces, high-performance and edge computing.
\end{IEEEbiography}

\begin{IEEEbiography}[{\includegraphics[width=1in,height=1.25in,clip,keepaspectratio]{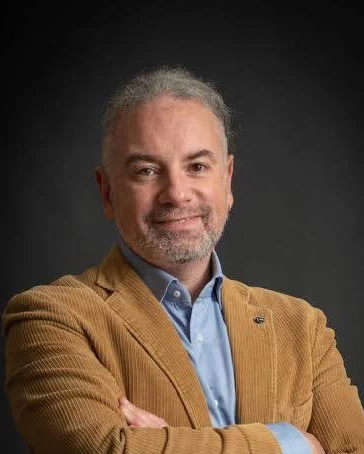}}]{Patrizio Dazzi}
received the B.Sc. degree in Computer Science in 2003 and the M.Sc. degree in Computer Technologies in 2004 from the University of Pisa, Italy. He earned his Ph.D. in Computer Science and Engineering from the IMT School for Advanced Studies Lucca in 2008.

He is an Associate Professor at the Department of Computer Science, University of Pisa. From 2005 to 2022, he was a member of the High-Performance Computing Laboratory at the Institute of Information Science and Technologies “Alessandro Faedo” of the National Research Council of Italy. Since 2020, he has been Co-Founder and Co-Director of the Pervasive Artificial Intelligence Lab, a joint initiative between the same institute and the University of Pisa. His research interests include distributed systems, edge and cloud computing, and high-performance computing.

Prof. Dazzi has coordinated several EU H2020 projects, including BASMATI and ACCORDION, and served as Innovation Coordinator for the TEACHING project. He has authored around 150 scientific publications and serves on the Executive Committee of the IEEE Technical Committee on Cloud Computing. He has served as Program Chair for ParCo 2017, IEEE JointCloud 2023, ACM IoT 2025, and as General Co-Chair for ACM HPDC 2024. He is a member of the editorial boards of ETRI Journal and Springer Computing, and currently Editor-in-Chief of the International Journal of Networked and Distributed Computing (Springer Nature).
\end{IEEEbiography}

\begin{IEEEbiography}[{\includegraphics[width=1in,height=1.25in,clip,keepaspectratio]{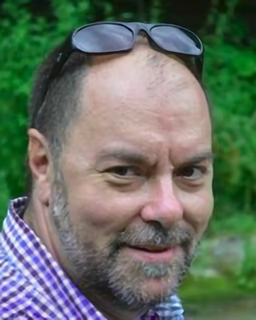}}]{Marco Danelutto}
received the Ph.D.\ degree in Computer Science from the University of Pisa, Pisa, Italy, in 1990. 

He is currently a Full Professor at the Department of Computer Science, University of Pisa, where he leads the Parallel Programming Models (PPM) group. His research interests include high-performance computing, with a focus on parallel programming models, algorithmic skeletons, autonomic computing, and formal methods.

Prof.\ Danelutto has participated in several EU-funded projects including CoreGRID, GRIDcomp, ParaPhrase, REPARA, and RePhrase. He has co-authored more than 180 refereed publications and participated in the design of structured parallel frameworks such as P3L, SkIE, Muskel, and FastFlow. He contributed to the definition of ETSI GCM and introduced the Behavioral Skeletons component model.
He has served as organizer or program chair of multiple international conferences and workshops, including Euro-Par, Euromicro PDP, HLPP, and PARCO, and participated on numerous program committees in high-performance computing.
\end{IEEEbiography}

\EOD

\end{document}